\documentclass[
reprint,
longbibliography,
aps,
twoside,
bibnotes,
nofootinbib
]{revtex4-2}
\usepackage{amsmath, amssymb, graphicx, bm}
\usepackage{bbm}
\usepackage{xcolor}
\usepackage{appendix}
\usepackage{braket}
\usepackage{multirow}
\usepackage{tikz}
\usepackage[hidelinks]{hyperref}
\usepackage{cleveref}
\usepackage{svg}
\usepackage{makecell}
\usepackage{xfrac}
\usepackage{array}
\usepackage{tabularx}

\newcommand{\ketbra}[2]{\ket{#1} \! \bra{#2}}
\newcommand{\ketbrasub}[3]{\ket{#1}_{#3} \! \bra{#2}_{#3}} 

\renewcommand{\succ}{\text{success}}

\newcommand{\ii}{\mathrm{i}}
\newcommand{\bigO}{\mathcal{O}}
\newcommand{\hilbert}{\mathcal{H}}

\newcommand{\ideal}{I}

\newcommand{\inp}{\text{inp}}
\newcommand{\out}{\text{out}}
\newcommand{\detc}{\text{det}}

\newcommand{\I}{\mathbbm{1}}

\newcommand{\psm}{\beta_{\text{A}_\inp}}
\newcommand{\qb}{\bm{q}}
\newcommand{\pb}{\bm{p}}
\newcommand{\onei}{\bm{1}_i}

\newcommand{\good}{g}
\newcommand{\bad}{b}
\newcommand{\sph}{\text{sp}}
\newcommand{\Hil}{\mathcal{H}}
\newcommand{\visible}{\text{vis}}
\newcommand{\hidden}{\text{hid}}
\newcommand{\mout}{m_\text{out}}
\newcommand{\inter}{\text{int}}
\newcommand{\Id}{\text{Id}}
\newcommand{\fockidx}{q}

\DeclareMathOperator{\tr}{tr}
\begin{document}

\title{Beyond Fidelity: Resolving Errors in Photonic Seed State Generation}

\author{Sara Bartolucci}
\author{Patrick Birchall}
\author{Jacob Bulmer}
\email{jbulmer@psiquantum.com}
\author{Sam Gartenstein}
\email{sgartenstein@psiquantum.com}
\author{Mihir Pant}
\author{Terry Rudolph}
\author{Chris Sparrow}
\author{Andrzej Veitia}
\affiliation{PsiQuantum, Palo Alto}

\begin{abstract}
An important aspect of many linear optical quantum computing architectures is probabilistically generating small dual-rail qubit entangled states, or \textit{seed states}. 
While many schemes to achieve this have been discovered, attention has been primarily focused on minimizing input resources and maximizing success probability rather than establishing a detailed understanding of linear-optical error performance.
In this work, we develop a framework for resolving sensitivities to physically distinct errors in linear optical state generation, allowing us to identify precise ways in which these errors affect the properties of the output state.
We use this framework to identify seed state generators in which destructive many-photon interference prevents detector loss and detector noise from causing a falsely successful heralding event.
We show that this effect suppresses errors, for example cutting the effective loss on the output of particular Bell state generators in half relative to a standard scheme.
We also present a new Bell state generator in which both detector loss and detector noise do not affect the quality of the output Bell state to first order, and a 4GHZ generator with $>$75\% higher success probability than a standard circuit.
In addition we provide some useful identities, bounds and invariants for studying linear optical seed state generation. 
\end{abstract}
\maketitle

\section{Introduction}
The creation of small-scale photonic entangled states, or \textit{seed states}, is a vital step in many different quantum photonic technologies.
In particular, seed state generation is an important step in a fusion-based quantum computing (FBQC) approach to building a photonic fault-tolerant quantum computer~\cite{bartolucci2023fusion}, and the source of many errors in such approaches.
Realizing these systems in the near future will require consideration of the real-world error mechanisms that affect seed state generation as well as carefully designed architectures that minimize the impact of such errors.
Certain methods for increasing error robustness are already known, such as using a dual-rail encoding~\cite{chuang1995simple} to detect loss and multiphoton errors or using purification schemes to reduce the impact of photon distinguishability~\cite{birchall2018fundamental, sparrow2018quantum,marshall2022distillation, saied2024general, somhorst2024photon,faurby2024purifying,somhorst2026below} or Pauli~\cite{bennett1996purification,pan2001entanglement, pan2003experimental,lobl2026purification} errors.
However, prior development of seed state generation has been mostly limited to finding examples of seed state generators (SSGs) with high success probabilities and low resource requirements~\cite{browne2005resource, zou2005generation, zhang2008demonstration, varnava2008good,gimeno2016towards, gubarev2020improved,fldzhyan2021compact,  paesani2021scheme, ruiz2023digital, bhatti2025heralding}, and then evaluating their error performance numerically~\cite{rohde2005frequency,rohde2006error, barbieri2009parametric,takeoka2015full, thomas2021general,shaw2023errors, wiesner2024influence, steinmetz2024simulating,randles2026limits}. 

The fundamental ingredients of an FBQC architecture are \textit{resource states}, entangled states of a fixed size, and \textit{fusions}, entangling measurements which link them together into a larger \textit{fusion network} used as the computational substrate.
One way to build resource states is by performing fusions on smaller seed states, each of which can be created by putting single photon Fock states through a linear optical interferometer and measuring some subset of the output modes which probabilistically project the photons in the remaining modes into a seed state.
Multiplexing \cite{pittman2002single,migdall2002tailoring,bartolucci2021switch} seed state generation and fusion can compensate for probabilistic success, but physical errors that corrupt the seed state can propagate into the larger resource state.
It is in this context that we consider the error performance of SSGs.

In this paper we introduce a framework for analyzing seed state generation in which errors in the photonic system can be modeled as resulting from a small number of \textit{error mechanisms} acting either on the single photons input to the SSG or the detected modes of the SSG.
Rather than reducing performance to a single fidelity metric, our framework characterizes the resulting heralded quantum state by its probability distribution over physically meaningful \textit{output state subspaces}.
These output subspaces correspond to different types of errors on the qubits of the output state, including both Pauli errors within the code subspace and errors which result in leakage outside the code space. 
The probability distribution over the output subspaces also captures classical correlations between the errors on the qubits in the output state.
By analyzing this probability distribution, we are able to achieve a richer understanding of the errors occurring during seed state generation.
This allows us to not only see how much infidelity physical error mechanisms can cause, but also to resolve the kind of impacts that this infidelity will have for a photonic architecture.

We then look at the leading order responses to a handful of physically relevant error mechanisms in seed state generation.
In particular, we use the response of the output subspace probability distribution to the different error mechanisms to show that destructive many-photon interference can prevent first-order detector errors from falsely heralding success.
We identify two Bell state generators with exactly zero first-order response to both detector loss and detector noise, and two others with zero first-order response to just detector loss.
We also show 3- and 4-GHZ state generators for which first-order detector errors result in correlated leakage errors which can be heralded out with high confidence at subsequent fusions.
Finally, we show how these properties enable implementations of these seed state generators which natively emit time-bin encoded qubits, and in which loss from switches and delays primarily reduces the heralding rate rather than the quality of the heralded state.

The leading order analysis that we perform here is aiming to elucidate as simply as possible the key mechanisms for errors and error mitigation in seed state generation.
The framework is generalizable to higher order errors and other error mechanisms than the ones we include in this paper.
We would like to highlight some recent works taking similar approaches to us for aspects of the framework we introduce here, finding low-order analytic expressions for SSG state fidelities under loss~\cite{randles2026limits} and Pauli error probabilities on linear optical entangling operations between quantum dots~\cite{karimi2026characterization}.

Although the notation in this paper is all defined within the text, please also see \Cref{tab:notation} and \Cref{fig:general_ssg} for a summary of the notation used.

\section{Primer on seed state generation}\label{sec:ssg_primer}
For the purposes of this paper, we assume that seed state generation consists of an $m$ mode linear optical unitary transformation in which $n_\inp \leq m$ of the input modes are occupied by single photons and the remaining $m - n_\inp$ are vacuum.
Following the unitary transformation, $m_\detc$ of these modes are sent to photon detectors, heralding the creation of a photonic entangled state on the remaining $m - m_\detc = \mout$ modes.
As we show in \Cref{app:photon_proof}, the generation of any \textit{mode irreducible} $\mout$ mode state will require at least $n_\inp=\mout$ single photons.

The ideal non-vacuum part of the input state of the SSG is given by
\begin{equation}
    \ket{\psi_\inp} = 
    \ket{1}^{\otimes n_\inp}
    = \left( \prod_{i=1}^{n_\inp} \hat{a}^\dag_i \right) \ket{0}^{\otimes n_\inp} \label{eq:psi_inp}
\end{equation}
where the creation operator for the $i$th mode is $\hat{a}^\dag_i$.
We also can write the ideal input state's density matrix, $\rho_{\inp, \Id} = \ketbra{\psi_\inp}{\psi_\inp}$.
Our convention here is that kets will always refer to ideal states, whereas density matrices will have the $\Id$ subscript when they refer to ideal case states. 
This input state, together with the vacuum inputs, is then sent into a linear optical interferometer described by a unitary transformation $U$.
This acts to evolve the creation operators like 
\begin{align}
    \mathcal{U} a^\dagger_i \mathcal{U}^\dagger = \sum_j U_{ji} a^\dagger_j
    \label{eq:mathcalU}
\end{align}
where $\mathcal{U}$ is the $m$-mode Fock space unitary generated by the linear optical transformation, $U$.
Combined with the invariance $\mathcal{U}\ket{0} = \ket{0}$, this is used to evolve states $\ket{\psi} = f(a_1^\dagger, \dots, a_m^\dagger) \ket{0}$ as \cite{loudon2000quantum}
\begin{align}
    \begin{split}
    \mathcal{U} \ket{\psi} & = \mathcal{U} f(a_1^\dagger, \dots, a_m^\dagger) \mathcal{U}^\dagger \mathcal{U} \ket{0} \\
    & = f(\mathcal{U}a_1^\dagger \mathcal{U}^\dagger, \dots, \mathcal{U} a_m^\dagger \mathcal{U}^\dagger) \ket{0}.
    \end{split}
\end{align}
This means that the state prior to detection, which we call the intermediate state (int), in the ideal case can be written as 
\begin{align}
    \ket{\psi_{\inter}} = \mathcal{U} \left(\ket{\psi_{\inp}} \otimes \ket{0}^{\otimes (m-n_\inp)}\right)
    \label{eq:psi_int}
\end{align}
with corresponding density matrix $\rho_{\inter, \Id} = \ketbra{\psi_{\inter}}{\psi_{\inter}}$.

For convenience, we define the detectors as being placed on the first $m_\detc$ modes, and the set of vectors $\{\bm{p}\}$ as the possible detector outcomes which herald success, $\bm{p} = (p_1, \dots, p_{m_\detc})$.
The successful measurement operators are given by 
\begin{align}
    \Pi^{(\pb)}_{\Id} = \ketbra{\pb}{\pb} \label{eq:DI}
\end{align}
where we use $\ket{\pb} = \bigotimes_{i=1}^{m_\detc} \ket{p_i}$. 
Later, we will use $n_\detc$ to denote the total number of detected photons $\sum_{i=1}^{m_\detc} p_i$.

The state after a successful heralding measurement will depend on the particular success outcome we see and can be written as 
\begin{align}
    \ket{\psi_\out^{(\pb)}}= \bra{\pb} \frac{\Pi^{(\pb)}_{\Id} \ket{\psi_\inter}}{\sqrt{ \Pr(\bm{p} ; \Id)}} 
    \label{eq:psi_out}
\end{align}
with corresponding density matrix $\rho_{\out, \Id}^{(\bm{p})} = \ketbra{\psi_{\out}^{(\bm{p})}}{\psi_{\out}^{(\bm{p})}}$.
The probability of outcome $\bm{p}$ is 
\begin{align}
    \Pr(\bm{p} ; \Id) = \tr\left(\Pi^{(\bm{p})}_{\Id} \cdot \rho_{\inter, \Id}\right)
\end{align}
and so the total success probability in the ideal case is given by
\begin{align}
    \Pr(\text{success}; \Id) = \sum_{\bm{p}}  \Pr(\bm{p} ; \Id).
\end{align}
These probability distributions are defined by considering the ideal operation of the circuit, so we add the $\Id$ subscript to label these distributions explicitly.
This indicates that we are calculating these quantities with no errors occurring at any point in the seed state generation process.

We aim to herald $n$-photon states where $n$ can be 2 or higher, though typically not higher than $n=4$.
The SSGs are designed to generate maximally entangled stabilizer states of dual-rail qubits, so $n$-photon states form $n$-qubit states on $m_\out=2n$ rails.
This paper focuses on GHZ states, up to local Pauli operations, although other states such as 4 qubit line graph states may also be of interest in FBQC architectures.

When we have an SSG with multiple valid $\bm{p}$ each one may produce a different $\ket{\psi_{\out}^{(\bm{p})}}$.
In all circuits considered here, these states will be equivalent up to local Pauli operations. 
In typical approaches for fault tolerant FBQC, we do not need to do any adaptive physical corrections to make up for these differences.
Instead we can simply keep track of them in the \textit{Pauli-frame}~\cite{knill2005quantum} and account for them entirely with classical post-processing of the measurement outcomes when computing checks. 
Because of this, we treat all ideal states $\ket{\psi_{\out}^{(\bm{p})}}$ as equally useful.
So, unless we specify otherwise, discussion of \textit{success} in the remainder of the paper is implicitly using an appropriately weighted sum over all possible success patterns and all valid heralded states. 

\section{Error Modeling}\label{sec:error_modeling}

Accurately modeling the state of a large and noisy quantum system can be difficult because the state space grows quickly with each additional qubit and often with each new error mechanism considered.
For this reason, many previous descriptions of errors in linear optical systems have either focused on one particular type of error such as loss \cite{varnava2008good} or photon distinguishability \cite{rohde2005frequency,rohde2006error,shaw2023errors}, and/or evaluated the system under some catchall metric such as state fidelity~\cite{wiesner2024influence}.
Two states with the same fidelity will not necessarily have the same utility in realizing a photonic quantum computer, so to best design an architecture it is worthwhile to gain a more fine-grained understanding of how the different error processes impact the rate and structure of the errors on the output states. 

\subsection{Error mechanisms}\label{sec:error_channels}

\begin{figure*}
    \centering
    \includegraphics[trim={0 0 4cm 0},width=0.85\textwidth]{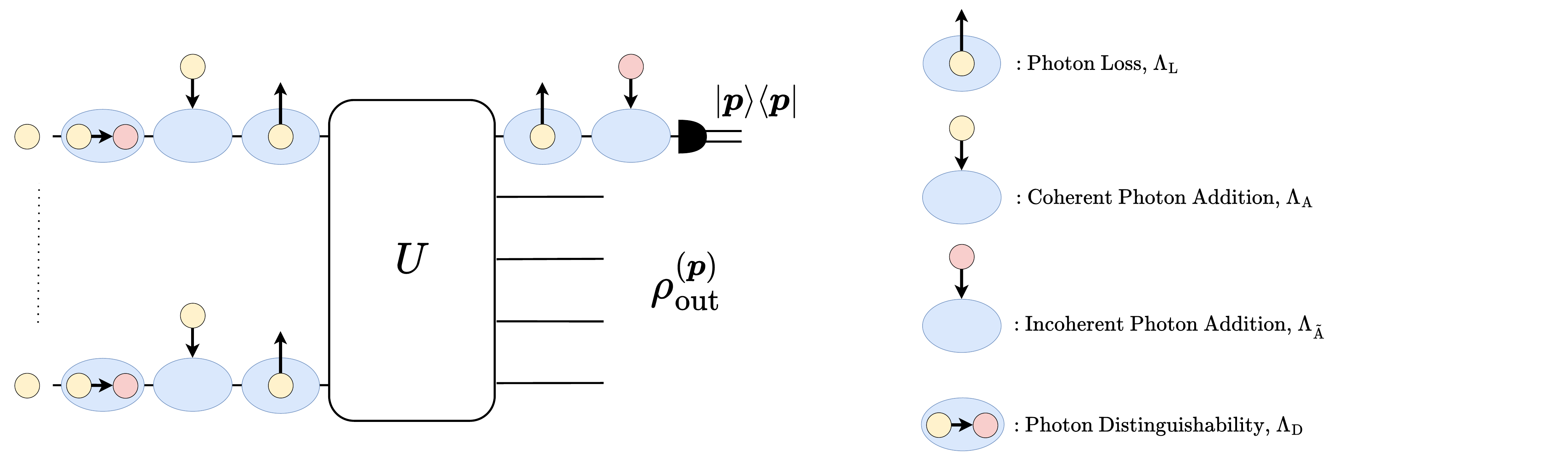}
    \caption{Schematic for the error mechanisms that we consider in this work. 
    Distinguishability, multiphoton (coherent photon addition), and loss at the input, and loss and detector noise (incoherent photon addition) at the detectors.
    Yellow circles are photons in the ``good'' mode, red circles are photons in the ``bad'' mode.
    Here shown as an example for some $n_\inp=m=5$ Bell state generation circuit, e.g.\ the 5DFT in \Cref{fig:ssg_circuits}a. 
    These error mechanisms lead to a noisy output state, $\rho_\out^{(\pb)}$, which we can then use to calculate output subspace probabilities $\Pr(\bm{Q} | \pb)$ as discussed in \Cref{sec:subspaces}.
    \label{fig:channels}}
\end{figure*}

We use an error model where instead of considering all error mechanisms being applied to the SSG simultaneously, our model assumes that each error has its own single-mode error channel that is applied to the state one at a time.
Another assumption we make is to consider all channels to be applied either before or after the interferometer.
Although errors within the interferometer, such as imbalanced losses or imperfect splitting ratios and phase shifts can lead to errors on the output seed state~\cite{somhorst2026mitigating}, we do not investigate these errors in this work.
The performance already demonstrated for passive photonic components~\cite{alexander2024manufacturable} means that the impact of errors associated with imperfect unitary transformations should be a minor concern compared to the error mechanisms from the inputs and detectors.

These assumptions allow us to categorize the error mechanisms into two sets: $\mathcal{E}_\inp$ and $\mathcal{E}_\detc$, which occur on the SSG input and detected modes, respectively. 
We will use $\mathcal{E}$ (with no subscript) to denote the combination of the input and detected mode error mechanisms.
Each error mechanism $\text{E} \in \mathcal{E}$ has an associated channel $\Lambda_\text{E}$ which describes the action of the error mechanism on the single mode to which it is applied.
A sketch of this is shown in \Cref{fig:channels}.
Each channel is parameterized by a probability $\beta_\text{E}$ such that when it acts on a single photon we see 
\begin{align}
    \begin{split}
    \Lambda_\text{E}(\ketbra{1}{1}) & = (1 - \beta_\text{E})\ketbra{1}{1} + \beta_\text{E}\sigma \\ 
    \text{where:} \ \braket{1 | \sigma | 1} & = 0 
    \end{split}
    \label{eq:channel_param}
\end{align}
i.e.\ that it will transform a single photon input state into some orthogonal state, $\sigma$, with probability $\beta_\text{E}$.
For example, for the linear loss channel $\text{L}$, we have $\beta_\text{L}$ being the single photon loss probability: the probability a single photon ($\ketbra{1}{1}$) gets mapped to the vacuum state ($\ketbra{0}{0}$).

Let $\Lambda_{\text{E}, j}$ be the channel acting on mode $j$.
We model the realistic, non-ideal input state as
\begin{align}
    \rho_\inp = \left[\bigotimes_{j=1}^{n_\inp}  \bigcirc_{\text{E}\in \mathcal{E}_\inp}\Lambda_{\text{E}, j} \right] \left( \rho_{\inp, \Id} \right) \label{eq:rho_inp}
\end{align}
where $\bigcirc_{\text{E} } \Lambda_\text{E}$ represents the channel composition ${\left[\Lambda_{\text{E}_1} \circ \Lambda_{\text{E}_2} \circ \dots \right] (\rho)} = \dots \left(\Lambda_{\text{E}_1} \left( \Lambda_{\text{E}_2}(\rho) \right) \right)$.
For small error mechanism parameters, $\beta_\text{E} \ll 1$ for all $\text{E} \in \mathcal{E}_\inp$, we can expand \Cref{eq:rho_inp} to first order:
\begin{align}
    \begin{split}
    \rho_\inp = \beta_{\inp, \Id} \rho_{\inp, \Id}
    & + n_\inp \sum_{\text{E}\in\mathcal{E}_\inp} \beta_\text{E} \rho_{\inp,\text{E}} \\
    & + \bigO\left(\sum_{\text{E}_{1}, \text{E}_{2}} \beta_{\text{E}_{1}}\beta_{\text{E}_{2}}\right)
    \end{split}\label{eq:decomp_input}
\end{align}
where $\beta_{\inp, \Id} = 1 - n_\inp \sum_{\text{E} \in \mathcal{E}_\inp}  \beta_\text{E}$.
Here, $\rho_{\inp,\text{E}}$ is the normalized first-order error component obtained by averaging over the $n_\inp$ ways that one ideal input photon can be transformed by the error mechanism $\text{E}$ to some other state.
After evolving this through the interferometer, we have $\rho_\inter$ which we define as
\begin{align}
    \rho_\inter = \mathcal{U} \left(\rho_\inp \otimes \ketbra{0}{0}^{\otimes (m-n_\inp)} \right)\mathcal{U}^\dagger.
    \label{eq:rho_int}
\end{align}
With this, we can model the output state as
\begin{align}
    \rho_\out^{(\bm{p})} = \frac{1}{\Pr(\bm{p})} \tr_{[1,m_\detc]} \left( \left[ \bigotimes_{j=1}^{m_\detc} \bigcirc_{\text{E} \in \mathcal{E}_\detc}\Lambda_{\text{E}, j}\right] \left( \rho_\inter \right) \cdot \Pi^{(\bm{p})}_{\Id} \right) 
    \label{eq:rho_h_p}
\end{align}
where $\tr_{[1,m_\detc]}$ is a trace over the detected modes, which run from index 1 to $m_\detc$ and $\Pr(\bm{p})$ is the probability that the pattern $\bm{p}$ is detected.

If we apply the detection error mechanism channels to the measurement operators instead of to the state, we can rewrite \Cref{eq:rho_h_p} as 
\begin{align}
    \rho_{\out}^{(\bm{p})} = \frac{1}{\Pr(\bm{p})} \tr_{[1,m_\detc]}\left( \rho_\inter \cdot \Pi^{(\bm{p})} \right)
    \label{eq:noisy_measurement}
\end{align}
where $\Pi^{(\bm{p})}$ is the noisy measurement operator which captures the impacts of the detector error mechanisms.

We can also write down a decomposition of the noisy positive operator-valued measure (POVM) element $\Pi^{(\pb)}$ to first order in $\beta_\text{E}$:
\begin{align}
    \begin{split}
        \Pi^{(\pb)} = \beta_{\detc, \Id} \Pi_{\Id}^{(\pb)}
        & + \sum_{\text{E} \in \mathcal{E}_\detc}\nu_\text{E} \beta_\text{E} \Pi_\text{E}^{(\pb)}\\
        & + \bigO\left(\sum_{\text{E}_{1}, \text{E}_{2}} \beta_{\text{E}_{1}}\beta_{\text{E}_{2}}\right)
    \end{split}\label{eq:decomp_POVM}
\end{align}
where $\nu_\text{E}$ is a constant for each error mechanism which accounts for how its impact scales with the number of photons/modes.
Derivations for the noisy operators $\Pi_\text{E}^{(\bm{p})}$ and their corresponding coefficients $\nu_\text{E}$, as well as the expression for $\beta_{\detc, \Id}$ are given in \Cref{app:error_channels}, with the resulting values summarized in \Cref{tab:nu_values}.

Using \Cref{eq:decomp_input} and \Cref{eq:decomp_POVM} we can expand \Cref{eq:noisy_measurement} as the weighted sum of the ideal output state and many other states, each of which can be associated with a particular error.
Because $\rho_{\out,\text{E}}^{(\bm{p})}$ is defined below as a normalized state conditioned on the detector pattern $\bm{p}$, the contribution from each error mechanism must be weighted by the corresponding probability $\Pr(\bm{p};\text{E})$ of producing that detector pattern:
\begin{align}
    \begin{split}
    \rho_\out^{(\bm{p})} \approx \frac{1}{\Pr(\bm{p})} \biggl( &\beta_{\Id}\Pr(\bm{p};\Id) \rho_{\out,\Id}^{(\bm{p})} \\
    & + \sum_{\text{E} \in \mathcal{E}} \nu_{\text{E}}\beta_\text{E} \Pr(\bm{p};\text{E}) \rho_{\out,\text{E}}^{(\bm{p})} \biggr)
    \end{split}
    \label{eq:decomp_output}
\end{align}
where this sum now runs over both the input and detector error mechanisms, $\beta_\Id = \beta_{\Id, \detc} - n_\inp \sum_{\text{E} \in \mathcal{E}_\inp} \beta_\text{E}$ and $\Pr(\pb)$ is the (non-idealised) probability of measuring pattern $\bm{p}$.
We can see that $\nu_\text{E} = n_\inp$ for all $\text{E}\in\mathcal{E}_\inp$ from \Cref{eq:decomp_input}.

The output state associated with the error mechanism $\text{E}$ can be calculated as
\begin{equation}
\rho^{(\bm{p})}_{\out,\text{E}} = 
    \begin{cases}
        \frac{1}{\Pr(\bm{p} ; \text{E})} \tr_{[1,m_\detc]} \left( \rho_{\inter, \text{E}}   \cdot \Pi_\Id^{(\bm{p})}\right), \quad \ \text{E} \in \mathcal{E}_\inp\\
        \frac{1}{\Pr(\bm{p} ; \text{E})}  \tr_{[1,m_\detc]} \left(  \rho_{\inter, \Id} \cdot \Pi^{(\bm{p})}_{\text{E}}  \right), \quad \ \text{E} \in \mathcal{E}_\detc 
    \end{cases}
\end{equation}
where $\rho_{\inter, \text{E}}$ is found by evolving $\rho_{\inp, \text{E}}$ as in \Cref{eq:rho_int} and $\Pr(\pb ; \text{E})$ is the probability distribution on outcomes $\pb$ for the state $\rho_{\inter, \text{E}}$ for the input error mechanisms and when $\rho_{\inter, \Id}$ is measured with $\Pi_\text{E}^{(\bm{p})}$ for the detected mode error mechanisms.

Equations \ref{eq:decomp_input} and \ref{eq:decomp_output} give us the input and heralded output states of the SSG as sums of an ideal state and terms which each correspond to the action of a particular error mechanism.
These decompositions allow us to pinpoint the first order effects of individual error mechanisms on the quality of the SSG output.

\subsection{Output state subspaces}\label{sec:subspaces}
To categorize the errors on the output state of an SSG, we can decompose its Hilbert space into different subspaces.
We tailor this decomposition such that we can interpret each subspace as corresponding to states which we associate with the absence or presence of different kinds of errors.
We begin by defining the Hilbert space in which the output state lives.
Let the state space for a single optical, Bosonic mode be
\begin{equation}
\hilbert_{\text{mode}} = \left\{\sum_\fockidx \alpha_\fockidx \ket{\fockidx}: \alpha_\fockidx \in \mathbb{C}, \fockidx \in \mathbb{N}_0\right\}.
\end{equation}
A single rail is defined as a collection of $k$ modes $\hilbert_{\text{rail}} = \hilbert_{\text{mode}}^{\otimes{k}}$. 
This is typically interpreted as all possible optical modes within a particular single-spatial-mode waveguide.

In this work, we make an approximation which sets $k=2$.
We treat one of these modes as the desired, or ``good'' one, $g$, and the other as an orthogonal ``bad'' mode, $b$.
So, we write 
\begin{equation}
\hilbert_{\text{rail}} = \hilbert_g \otimes \hilbert_b. \label{eq:H_rail}
\end{equation}
We only analyze first order impacts of distinguishability, so we will have at most one photon in the ``bad'' mode in any of the rails. 
Motivated by the observed temporal/spectral response of the hardware~\cite{alexander2024manufacturable} over the relevant modes, we also assume that all interferometers and detectors act identically on photons in the good mode or the bad mode.

The space of a dual-rail qubit is then $\hilbert_{\text{dr}} = \hilbert_{\text{rail}}^{\otimes 2}$.
We decompose this dual-rail subspace by the total photon number, splitting into vacuum, single photon, and multiphoton subspaces.
\begin{align}
    \Hil_\text{dr} = \Hil_V \oplus \Hil_{\sph} \oplus \Hil_M 
\end{align}
We call the $\Hil_V$ and $\Hil_M$ subspaces the \textit{leakage} subspaces because they can be discriminated using a mode insensitive photon number resolving measurement, i.e.\ they have \textit{leaked} outside of the dual-rail logical subspace which we ideally compute in.

It is important to add further structure to the single photon subspace $\hilbert_\sph$ because, unlike in the leakage subspaces, deviations from the ideal state here cannot be immediately detected by photon number measurements.
Specifically, we first map $\Hil_\sph$ to an isomorphic space 
\begin{align}
    \Hil_\sph \cong \Hil_\visible \otimes \Hil_\hidden
\end{align}
where the $\Hil_\visible$ space contains the \textit{visible} which-rail information~\cite{geller2026measuring} and is spanned by the qubit basis states $\ket{0}_\visible, \ket{1}_\visible$.
The $\Hil_\hidden$ space is spanned by the single photon's \textit{hidden} degree of freedom, which mode it occupies within the rail. 
In our model, this is spanned by the states $\ket{g}_\hidden$ and $\ket{b}_\hidden$, corresponding to the photon being in the ``good'' or ``bad'' mode.
This mapping is motivated by the insensitivity of our interferometers and detectors in our model to the \textit{hidden} information.

To resolve errors within the $\Hil_\visible$ subspace of all the non-leaked qubits, we first look at the subspace defined by the reduced state of the ideal state with any leaked qubits traced out.
We think of this as the ideal subspace of the unleaked qubits.
We can then specify a set of Pauli strings which take us from this ideal subspace to all other subspaces of the unleaked qubits.
Each subspace is then labeled by its corresponding Pauli string.
Definitions of these subspaces are given in \Cref{app:subspaces}.

To label all the subspaces of the Hilbert space for the state, we use $\bm{Q}$, an $n$-tuple of qubit-sub-labels, $Q_i$, so: $\bm{Q} = (Q_1, Q_2, \dots, Q_n)$. 
Our Hilbert space decomposition procedure gives us a set of 10 possible values $Q_i$ for each of the $n$ qubits.
These $Q_i$ can take the following values:
\begin{itemize}
    \itemsep0em 
    \item[-] Vacuum ($V$)
    \item[-] Multiphoton ($M$)
    \item[-] Single photon, good mode, no Pauli error ($\ideal$)
    \item[-] Single photon, good mode, Pauli error ($X, Y, Z$)
    \item[-] Single photon, bad mode, no Pauli error ($B$)
    \item[-] Single photon, bad mode, Pauli error ($BX, BY, BZ$).
\end{itemize}

To characterize an $n$ qubit SSG output in the presence of errors, we use these labels to define probability distributions $\Pr(\bm{Q})$ for the probabilities that the state ends up in the subspace labeled by $\bm{Q}$.
We can view each $Q_i$ as specifying the error on each qubit $i$ when the state lives in this corresponding subspace. 

The quantity $\Pr(\bm{Q} | \bm{p})$ is the probability that for the state heralded by detector pattern $\bm{p}$, if we were able to measure all qubits in a basis which resolved the subspaces, we would observe a state within the subspace labeled by $\bm{Q}$.
We note that not all combinations of $Q_i$ correspond to a valid subspace label, and therefore there are fewer than $10^{n}$ valid subspace labels for an $n$-qubit state, even though we have $10$ options for each $Q_i$. 
Also, some possible subspace labels will refer to the same subspace, e.g. the subspaces labeled $(X, I, I)$ and $(I, X, X)$ for a 3GHZ state.
More details on the procedure for decomposing the Hilbert space and labeling the subspaces are shown in \Cref{app:subspaces}, as well as a description of which $\bm{Q}$ form valid labels for a given ideal-case state.

When viewed together, the conditional heralded subspace probability distribution $\Pr(\bm{Q} | \bm{p})$ provides useful information about the state created by the SSG.
It is not a complete representation of the state as we are missing information within the subspaces that are larger than 1-dimensional, such as the multiphoton space, and we are missing the coherences between the subspaces.
However, as we argue in \Cref{app:error_levels}, ignoring these intra-subspace structures and inter-subspace coherences 
can still give reasonable accuracy 
when modeling seed states in FBQC architectures, given some motivated assumptions.

\subsubsection*{Examples}
This decomposition of the output state into terms contained in distinct subspaces of interest is a key step in understanding the error properties of SSGs.
We provide some example states in these subspaces.

First, assume the ideal state is a dual-rail Bell state in the ``good'' mode:
\begin{align}
     %\ket{\psi_h^{(\bm{p})}} = 
& \frac{1}{\sqrt{2}}\left( \ket{0101}_g  + \ket{1010}_g \right) \otimes \ket{0000}_b \\
& = \frac{1}{\sqrt{2}}\left( \ket{00}_\visible + \ket{11}_\visible \right) \otimes \ket{gg}_\hidden.
\end{align}
% written before and after mapping $\Hil_\sph$ to $\Hil_\visible \otimes \Hil_\hidden$.
For our first example we consider the tuple $\bm{Q} = (V, \ideal)$ which labels a subspace corresponding to the state: 
\begin{align}
   & \frac{1}{2}\left( \ketbra{00}{00}_g \otimes \ketbra{00}{00}_b \right) \\  
   & \ \quad \quad \quad \otimes  \left( \bigl[ \ketbra{0}{0}_\visible + \ketbra{1}{1}_\visible \bigr] \otimes \ketbra{g}{g}_\hidden \right).
\end{align}
We can see that the modes of the first qubit are in the vacuum state and the second qubit is in the maximally mixed state, which is what we would expect to find after tracing out the leaked qubit from the initial Bell state.

The tuple $\bm{Q} = (\ideal, B)$ labels a subspace which is given by the span of the state 
\begin{align}
        \frac{1}{\sqrt{2}}\left( \ket{00}_\visible + \ket{11}_\visible \right) \otimes \ket{gb}_\hidden,
\end{align}
and the subspace $\bm{Q} = (\ideal, X)$ corresponds to the span of the state
\begin{align}
    \frac{1}{\sqrt{2}}\left( \ket{01}_\visible + \ket{10}_\visible \right) \otimes \ket{gg}_\hidden.
\end{align}
Note that this can also be labeled with the tuple $\bm{Q} = (X, \ideal)$.

\subsection{Error Response Coefficients}\label{sec:response_coefficients}

In this section, we introduce the concept of \textit{error response coefficients} which help us to understand how the different error mechanisms we consider acting in the SSG can lead to the output state being found in different output subspaces.
To begin, we note that an ideal SSG will always produce states which are in the entirely ideal subspace, i.e.\
\begin{equation}
    \Pr(\bm{Q} | \bm{p} ; \Id) = 0
\end{equation}
for any $\bm{Q}\neq \bm{Q}_\ideal := (\ideal, \ideal,\ldots\ideal)$.

With this in mind, if $\beta_\text{E} \ll 1$ for all $\text{E} \in \mathcal{E}$, we can use the expansion from \Cref{eq:decomp_output} to write
\begin{align}
    \Pr(\bm{Q} | \bm{p}) \approx \sum_{\text{E} \in \mathcal{E}} \nu_\text{E} \frac{\Pr(\bm{p};\text{E})}{\Pr(\bm{p};\Id)} \Pr(\bm{Q} | \bm{p} ; \text{E}) \beta_\text{E}
    \label{eq:expanded_cond_prob}
\end{align}
for any $\bm{Q}\neq \bm{Q}_\ideal$.
Here, the factor $\Pr(\bm{p};\text{E})/\Pr(\bm{p};\Id)$ accounts for the fact that a given error mechanism can change the probability of heralding the detector pattern $\bm{p}$.
In practice, the probabilities $\Pr(\bm{Q} | \bm{p} ; \text{E})$ can be found by numerically or symbolically computing $\rho_{\out,\text{E}}^{(\bm{p})}$ then computing the probabilities for each of the subspaces $\bm{Q}$. 
More details are in \Cref{app:sub_probs} and an example is walked through in \Cref{app:example}.

This motivates us to define
\begin{equation}
    c_{{\text{E}}\to \bm{Q}}^{(\bm{p})} = \nu_\text{E} \frac{\Pr(\bm{p};\text{E})}{\Pr(\bm{p};\Id)} \Pr(\bm{Q}|  \bm{p} ; \text{E})\label{error_coeff}
\end{equation}
as the first-order \textit{error response coefficient}.
It is a non-negative constant which, to first order in the error mechanism probabilities, relates the error mechanism $\text{E}$ to the probability of the state being in subspace $\bm{Q}$ given the measurement of the successful detector pattern $\bm{p}$.
We then rewrite \Cref{eq:expanded_cond_prob} in a more compact form
\begin{align}\label{eq:coefs_to_error_probs}
    \Pr(\bm{Q}|\bm{p}) \approx  \sum_{{\text{E}\in\mathcal{E}}}  c_{{\text{E}}\to \bm{Q}}^{(\bm{p})} \beta_\text{E}.
\end{align}
We see from the above equation that, to first order, the conditional probability of finding the state in subspace $\bm{Q}$ is a sum of contributions from different error mechanisms.
Each term in the sum is the product of the probability which parameterizes the error mechanism $\beta_\text{E}$ with the first order error response coefficient $c_{{\text{E}}\to \bm{Q}}^{(\bm{p})}$.
We can also average the coefficients over all $\bm{p}$
\begin{align}
    c_{\text{E}\to \bm{Q}} = \frac{\sum_{\bm{p}} \Pr(\bm{p} ; \Id)c_{\text{E}\to\bm{Q}}^{(\bm{p})}}{\Pr(\text{success} ; \Id)}\label{avg_error_coeff}.
\end{align}
This then gives a version of \Cref{eq:coefs_to_error_probs} which is averaged over different success patterns
\begin{align}
    \Pr(\bm{Q}|\text{success}) \approx  \sum_{\text{E} \in \mathcal{E}}  c_{\text{E} \to \bm{Q}} \beta_\text{E} . \label{eq:response_use}
\end{align}
It should be noted that there can be differences in the errors on the state conditional on the different outcomes $\pb$.
\Cref{app:det_pattern_error_dependence} discusses how we sometimes can benefit from analyzing error responses for each $\pb$, without averaging.
However, the subsequent sections of this paper deal with the averaged values.

To build a more complete understanding of how error mechanisms impact an SSG beyond these first order errors, we can also investigate error coefficients for simultaneous error channels, e.g.\ the loss of more than one photon or simultaneous multiphoton and photon loss events from different inputs.
We can look at the next-to-leading order terms of \Cref{eq:decomp_output} and follow similar steps to what we have shown for the first order terms.
Doing so would give us an expanded version of \Cref{eq:coefs_to_error_probs}, where the additional terms look like:
\begin{align}
    +\sum_{\text{E}_1, \text{E}_2 } c_{(\text{E}_1, \text{E}_2)\to \bm{Q}}^{(\bm{p})} \beta_{\text{E}_1} \beta_{\text{E}_2} .
    \label{eq:2nd_order}
\end{align}

For second order effects, care should be taken when applying errors which do not commute, for example simultaneous detector loss and detector photon addition.
In our calculations of second order error mechanisms we use the ordering shown in \Cref{fig:channels}.

The procedure for calculating error response coefficients is summarized in \Cref{tab:coeff_calc_method}.

\begin{table}[t]
    \centering
    \small
    \setlength{\tabcolsep}{4pt}
    \renewcommand{\arraystretch}{1.35}
    \begin{tabular}{@{}cl@{}}
        \hline \hline
        \textbf{Step} & \textbf{Operation} \\ \hline
        \textbf{1} & \parbox[t]{0.78\columnwidth}{\raggedright\textbf{Propagate.} Choose an error mechanism $\text{E}$ and propagate the relevant input state, either the error-modified state for input errors or the ideal one for detector errors, through the SSG unitary, giving the intermediate state $\rho_{\inter, \text{E}}$.}\\[2pt]
        \textbf{2} & \parbox[t]{0.78\columnwidth}{\raggedright\textbf{Herald.} Apply a success-pattern measurement $\bm{p}$ on the detected modes, using the ideal POVM element for input errors or the error-modified POVM element for detector errors, then trace out those modes to obtain the heralded output $\rho^{(\bm{p})}_{\out, \text{E}}$.}\\[2pt]
        \textbf{3} & \parbox[t]{0.78\columnwidth}{\raggedright\textbf{Project.} Resolve the heralded output into output subspaces by projecting onto each subspace of interest, $\bm{Q}$, giving the conditional probabilities $\Pr(\bm{Q}|\bm{p};\text{E})$.}\\[2pt]
        \textbf{4} & \parbox[t]{0.78\columnwidth}{\raggedright\textbf{Extract.} Using the appropriate combinatorial prefactor, $\nu_\text{E}$, ideal success probability $\Pr(\bm{p} ; \Id)$ and noisy success probability $\Pr(\bm{p} ; \text{E})$  determine $c^{(\bm{p})}_{\text{E}\to\bm{Q}}$, and average over successful detector patterns if desired. }\\
        \hline \hline
    \end{tabular}
    \caption{Procedure for calculating first-order error response coefficients, $c_{\text{E}\to\bm{Q}}$. 
    This gives the proportionality between the probability of an error mechanism $\text{E}$ occurring on a measured or input photon and the probability of heralding an output state with some error causing it to be in subspace $\bm{Q}$, as stated in \Cref{eq:response_use}.}
    \label{tab:coeff_calc_method}
\end{table}

\subsection{Interpreting the error response coefficients}
As shown in \Cref{eq:response_use}, the response coefficients $c_{\text{E} \to \bm{Q}}$ are first-order sensitivities of the heralded output state.
These coefficients are defined after conditioning on successful heralding, so they include both the subspace occupied by the state after the error and any change in the probability that the corresponding detector pattern is heralded, through the factor $\Pr(\bm{p};\text{E})/\Pr(\bm{p};\Id)$ in \Cref{error_coeff}.
Thus, a zero response coefficient means that the error mechanism gives no first-order contribution to that output subspace after conditioning on success; this can happen either because the erroneous state never heralds success, or because it heralds only into other subspaces.
For response coefficients which we know cannot be zero, we are interested in bounding their size, especially for output subspaces which correspond to errors which are harder to correct in an architecture.

\section{Connecting and Bounding Error Responses}\label{sec:loss_and_multi}

To this point, we have dealt with a general set of error mechanisms $\text{E} \in \mathcal{E}$ which can be applied to the input or detected modes of an SSG.
Such a description is valid for a wide range of error analyses.
For the remainder of this analysis however, we will concern ourselves with a small set of possible errors which we consider the dominant error mechanisms in realistic SSGs.

The errors in our simplified model are described by the application of one of four possible channels: linear photon loss (L), photon distinguishability (D), coherent (mode-matched) photon addition (A), and incoherent photon addition ($\tilde{\text{A}}$).
These channels are defined in \Cref{app:error_channels} where their effects on single photon inputs and ideal POVM elements are derived.
Table \ref{tab:error_channels} gives a low order approximation of each channel's action on diagonal Fock density matrix elements.
In \Cref{fig:channels} we show where these channels are applied when we model SSG circuits.

At the input we consider loss, multiphoton contamination, and photon distinguishability which correspond to the loss, coherent photon addition, and distinguishability channels acting on the ideal input state.
On the detected modes we consider loss and dark counts, which correspond to the loss and incoherent photon addition channels, respectively.
% We label input error mechanisms with the subscript $\inp$ and detected channels with the subscript $\detc$.
As noted previously, although these errors are not a full model of the system, we expect them to capture the most impactful effects.

\begin{table*}
    \centering
    \setlength{\tabcolsep}{8pt}
    \renewcommand{\arraystretch}{1.8}
    \begin{tabular}{c|c|l}
        \hline
        \textbf{Error Channel} & \textbf{ Symbol} & \multicolumn{1}{c}{\textbf{Channel Map (First Order)}} \\ \hline \hline
        Loss & $\Lambda_\text{L}$ & $\ketbra{\fockidx}{\fockidx} \to \left(1 - \beta_\text{L} \fockidx\right)\ketbra{\fockidx}{\fockidx} + \beta_\text{L}\fockidx\ketbra{\fockidx - 1}{\fockidx-1} + \dots$ \\ 
        Coherent photon addition & $\Lambda_\text{A}$ & $\begin{aligned}\ketbra{\fockidx}{\fockidx} \to \bigl[ 1  & - \frac{1}{2}\beta_\text{A}(\fockidx+1)\bigr] \ketbra{\fockidx}{\fockidx} + \frac12 \beta_\text{A}(\fockidx+1)\ketbra{\fockidx+1}{\fockidx+1} + \dots \end{aligned}$ \\
        Incoherent photon addition & $\Lambda_{\tilde{\text{A}}}$ & $\ketbra{\fockidx}{\fockidx} \to \bigl(1-\beta_{\tilde{\text{A}}}\bigr)\ketbra{\fockidx}{\fockidx} + \beta_{\tilde{\text{A}}}\ketbra{\fockidx+1}{\fockidx+1} + \dots$ \\ 
        Distinguishability & $\Lambda_{\text{D}}$ & $\ketbra{10}{10}_{gb} \to (1 - \beta_\text{D})\ketbra{10}{10}_{gb} + \beta_\text{D}\ketbra{01}{01}_{gb}$ \\ \hline 
    \end{tabular}
    \caption{    
    Channels which can occur on photonic rails during seed state generation, along with their action on diagonal Fock basis density matrix elements, $\ketbra{\fockidx}{\fockidx}$.
    Note that for $q=1$ these expressions are all of the form outlined in \Cref{eq:channel_param}, where with probability $\beta_\text{E}$ the mechanism will map a single photon state to a particular orthogonal state.
    The distinguishability channel is only required to be defined for $\ketbra{10}{10}_{gb}$ state shown here.
    The $g$ and $b$ subscripts are used to label the modes of an indistinguishable (good) photon and a distinguishable (bad) photon respectively.
    Each channel map is parameterized by a corresponding probability $\beta_\text{E}$, and the first order expansion is valid when $\beta_\text{E} \ll 1$.
    More detail, including the application of these channels to SSG detection measurement operators, can be found in \Cref{app:error_channels}. \label{tab:error_channels}}
\end{table*}

\subsection{Vacuum Subspace Responses}

\begin{figure*}
    \centering
    \includegraphics[width=0.9\textwidth]{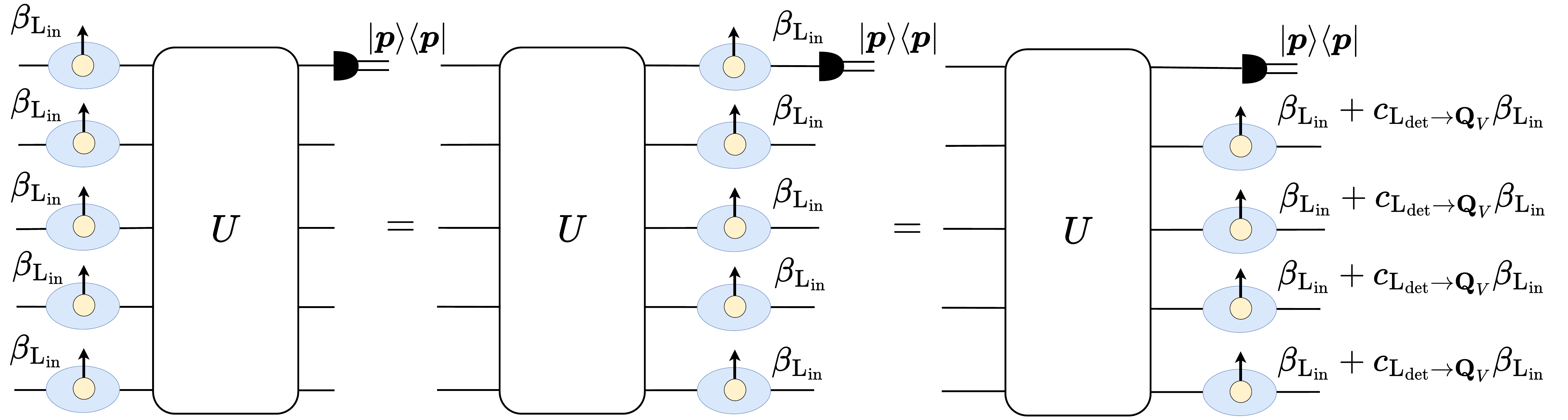}
    \caption{
        Commuting of the loss channels at the input to loss channels on the SSG output modes.
        The quantities next to the ellipses give the probability which parameterizes the loss channel, as defined in \Cref{app:error_channels}.
        The first step shows how the balanced loss channels commute across a unitary matrix to loss channels on the output and detected modes, while the second step shows the transformation of loss on the detected mode to effective loss on the SSG outputs.
        The resulting effective loss can be seen to be minimized when the error response coefficient $c_{\text{L}_\detc\rightarrow\textbf{Q}_V} = 0$.
        The remainder of the loss on the outputs is unavoidable.}
    \label{fig:source_to_output_loss}
\end{figure*}

It is well known that the balanced loss channel $\Lambda_{\text{L}}$, commutes with linear optics~\cite{loudon2000quantum,berry2011preservation}. 
Because of this, we are able to view any balanced loss at the SSG input as equivalent to a combination of detector loss and loss on the heralded SSG output modes of equal magnitude.
This means that the effect of input loss is a combination of the unavoidable impact of loss which is commuted through the circuit and acts on the output modes, and the impact of the loss which acts on the detected modes, which depends on the properties of the particular SSG circuit.
This concept is illustrated in \Cref{fig:source_to_output_loss}.

We capture the above observations in an equation linking two error response coefficients:
\begin{align}
c_{\text{L}_\inp \rightarrow \bm{Q}_V} = 1 + c_{\text{L}_\detc \rightarrow \bm{Q}_V}. \label{eq:SL_DL}
\end{align}
where $\bm{Q}_V = (V, \ideal, \dots, \ideal)$.
In the above equation, the $+1$ term represents the effect of the commuted input loss on the output modes.
The $c_{\text{L}_\detc \rightarrow \bm{Q}_V}$ term represents the additional vacuum probability induced in the output state due to loss on the detected modes.
\Cref{eq:SL_DL}, combined with the non-negativity of the error response coefficients gives the lower bound of
\begin{equation}
c_{\text{L}_\inp \rightarrow \bm{Q}_V} \geq 1.
\end{equation}
This is commensurate with the results in Ref.~\cite{berry2011preservation}.

We take the opportunity here to note that, for the Bell state and GHZ state examples we consider later in this paper, the error distributions over the different qubits will be symmetric under permutations of qubit labels e.g.\ ${\Pr(\bm{Q}=(I, I, V)) = \Pr(\bm{Q}=(I, V, I))}$ and so on.
So, while the vector $\bm{Q}_V$ describes a subspace with the first qubit in the vacuum subspace, it also occurs with probability equal to the probability of other subspaces with a different qubit in vacuum (and remaining qubits as ideal).
This is not true in general, so care should be taken here.

\subsection{Multiphoton Subspace Responses}

In the case when all the SSG input modes contain a single photon, $m=n_\inp$, the coherent photon addition channel $\Lambda_\text{A}$ can also be commuted through the interferometer, a property which is derived in \Cref{app:multiphoton_commuting}.
For this specific case of circuits with $m=n_\inp$ we can write a relationship between input and detector error response coefficients, analogously to \Cref{eq:SL_DL}.
This is:
\begin{equation}
    \begin{split}
        c_{\text{A}_\inp \rightarrow \bm{Q}_{M}} = \frac{3}{2} + c_{\text{A}_\detc \rightarrow \bm{Q}_M}\\
        \text{if } m = n_\inp \label{eq:SM_DM}
    \end{split}
\end{equation}
where $\text{A}_\detc$ refers to the coherent photon addition channel acting on the detected modes, a mechanism which is not considered a priori in our error model defined in \Cref{fig:channels} since it has no direct physical cause, resulting only from the commutation of $\text{A}_\inp$ through the LO unitary.
$\bm{Q}_M$ labels the subspace with a single output qubit in the multiphoton subspace
\begin{align}
    \bm{Q}_M = (M, \ideal, \dots, \ideal).
\end{align}
This expression, including the factor of $3/2$ is derived in \Cref{app:multiphoton_responses}.
This error response coefficient is bounded below by
\begin{equation}
    \begin{split}
        c_{\text{A}_\inp \to \bm{Q}_M} \geq \sfrac{3}{2}\\
        \text{if } m = n_\inp
    \end{split}
\end{equation}
coming from
$$c_{\text{A}_\detc \to \bm{Q}_M} = 0$$
As in \Cref{eq:SL_DL}, additional contributions to $\bm{Q}_M$ are caused by photon addition on the detected modes which leads to an erroneously successful heralding event.

\subsection{Pauli Subspace Responses} \label{sec:pauli_response}

We now consider the output subspaces which contain a Pauli $P \in \{X, Y, Z\}$ on a single qubit
\begin{align}
    \bm{Q}_P = (P, \ideal, \dots, \ideal).
\end{align}
We first note that there is no leakage in these subspaces.
Since unitary linear optics preserves total photon number, we know that first order loss or photon addition will have zero contribution to these Pauli subspaces.
Instead, the relevant error response coefficients here are the ones for distinguishability errors and second order combinations of photon addition and loss.
Bounds for distinguishability error responses are more subtle and are discussed in \Cref{sec:dist}.
This section focuses on Pauli error responses due to second order impacts of combined multiphoton (from coherent photon addition) and loss.
Recall from \Cref{eq:2nd_order} and surrounding discussion how we define and denote these second order error response coefficients.

As with the previous response coefficients we have looked at, we can start by considering the action of the error mechanisms when they are applied to the output modes.
We can build on the results of the previous subsections to show that an SSG where all inputs are occupied by a single photon, i.e. where $m=n_\inp$, obeys the bound
\begin{equation}
    \begin{split}
        c_{(\text{A}_\text{out}, \text{L}_\text{out}) \to \bm{Q}_P} = \frac{1}{4}\\
        \text{if } m = n_\inp
    \end{split}
\end{equation}
which we can use in our expression to link input losses and photon addition to the Pauli error subspace.
This response is derived in \Cref{app:multiphoton_responses}.
Commuting these input error mechanisms to the output and accounting for all possible combinations of these errors on the output modes, gives us 
\begin{equation}
    \begin{split}
        c_{(\text{A}_\inp, \text{L}_\inp) \to \bm{Q}_P} = \frac{1}{4}
        &+ c_{(\text{A}_\detc, \text{L}_\detc) \to \bm{Q}_P}\\
        &+ c_{(\text{A}_\detc, \text{L}_\text{out}) \to \bm{Q}_P}\\
        &+ c_{(\text{A}_\text{out}, \text{L}_\detc) \to \bm{Q}_P}
        \label{eq:SM_SL_P}
    \end{split}
\end{equation}
which provides the lower bound
\begin{equation}
    \begin{split}
        c_{(\text{A}_\inp, \text{L}_\inp) \to \bm{Q}_P} \geq \sfrac14\\
        \text{if } m = n_\inp.
    \end{split}
    \label{eq:pauli_bound}
\end{equation}

\subsection{Bad Photon Subspace Responses}\label{sec:dist}

Although the photon distinguishability channel at the SSG input cannot be simply commuted to the SSG outputs and detectors, it is still possible to think about distinguishability errors in terms of photon addition and subtraction on these modes.
As we show in detail in \Cref{app:dist}, photon distinguishability at the input can cause errors on the output state through a set of distinct mechanisms, corresponding to different ways the ``good'' photons and ``bad'' photon can arrive at the detected and output modes.
Here, we focus on the processes which can lead to a qubit of the output state being projected into the bad photon subspaces: $B, BX, BY$, and $BZ$.

To start, we write the input state with a bad photon in mode $i$ as
\begin{align}
    \ket{\psi_{\inp, \text{D}_i}} = a_{i} d_{i}^\dagger \ket{\psi_{\inp, \Id}}
\end{align}
where $a_{i}$ is the annihilation operator for the good mode of input $i$ and $\hat{d}^\dagger_{i}$ is the creation operator for distinguishable photons in the bad mode of input $i$.
We can then write the pre-detection state as
\begin{equation}
\ket{\psi_{\inter, \text{D}_i}} = \mathcal{U}\ket{\psi_{\inp, \text{D}_i}} = \sum_{kj}U^*_{ki}U_{ji}a_k d^\dagger_j\ket{\psi_{\inter, \Id}}\label{eq:dist_out}.
\end{equation}
The output state of the SSG will consist of those terms in the above equation which survive detection, i.e.\ they yield a successful pattern when measured with detectors that cannot distinguish between the hidden modes of the photons.

We consider the first order responses relating input photon distinguishability to output spaces which have a single bad photon in one output qubit and a single good photon in all the others.
The response coefficients of interest are $c_{\text{D}_\inp\to \bm{T}}$ for
\begin{align}
    \bm{T} &\in\{\bm{Q}_B, \bm{Q}_{BX}, \bm{Q}_{BY}, \bm{Q}_{BZ}\}\\
    & = \{(B, I, \ldots, I), (BX, I, \ldots, I), \ldots\}.
\end{align}

The only terms in \Cref{eq:dist_out} which can contribute to these responses are ones where the distinguishable photon is created in one of the $m_\out$ output modes.
We label this mode $j \in \mu_\out$ where $\mu_\out$ is the set of output modes.
The good photon can then be annihilated in the same mode ($k=j$), a different output mode ($k\in\mu_\out \backslash \{j\}$), or a detected mode ($k\in\mu_\detc$).

As in the previous subsections, reasoning about photon addition and loss can help us to bound the distinguishability responses.
For example, if $c_{\text{L}_\detc\to\bm{Q}_V} = 0$ then a single loss on the detected modes can't produce a state in which all qubits can be measured in $\Hil_\sph$ via the addition of a single photon in the output modes, i.e.\ the $k \in \mu_\detc$ process above cannot contribute.

Continuing along this sort of reasoning, along with explicit calculations of the contributions due to the non-zero processes, it is possible in general to determine the values of the above coefficients.
For example, in \Cref{app:dist} we show that we can link and bound the distinguishability responses as

\begin{equation}
    \begin{split}
        c_{\text{D}_\inp \to \bm{Q}_B} = & \sum_{P \in \{X, Y, Z\}}c_{\text{D}_\inp \to \bm{Q}_{BP}} \geq \frac{1}{n_\inp}\\
        & \text{if } m = n_\inp \text{ + symm. conditions}
    \end{split}
\end{equation}
for SSGs which have a particular type of symmetry which allows for suppression of these outputs, as proved in \Cref{app:dist}.
These conditions are true for all of the SSGs analyzed in \Cref{sec:example_circuits}.

On the way to deriving these results, \Cref{app:dist} also contains new, more generalized versions of multiphoton interference suppression laws \cite{dittel2018totally,dittel2018totally2}.

\section{Detector error mitigation}\label{sec:det_only}
The results in \Cref{eq:SL_DL,eq:SM_DM,eq:SM_SL_P} introduced in the previous section give the responses of SSGs to certain input errors as functions of errors on the detected and output modes.
From this we can see that combining an understanding of the effects of a detector-error-only model with the unavoidable errors which commute to the output modes of the SSG is sufficient to calculate many of the first order error responses of an SSG.

This insight into the importance of understanding the detector error responses motivates us to look at detector errors in isolation.
In this section, by focusing only on loss and \textit{noise}, which refers to either of the photon addition errors, on the detected modes, we discuss how certain properties of the ideal intermediate state, after the unitary transformation but before detection, can imply different types of error mitigating properties.

\subsection{Intrinsically Removed Errors} 

The first property we define is what we refer to as \textit{intrinsic removal} of detector loss and detector noise.
This refers to the property of an SSG where the ideal intermediate state of the SSG contains a \textit{photon number gap} which prohibits the SSG from mistakenly heralding success when a single detector loss or noise event occurs.
That is, the destructive many-photon interference in the SSG yields no terms in which the state of the detected modes can be mapped to one which would result in a successful detector pattern when measured with one of the first-order error POVM elements.
So for any measurement pattern $\tilde{\bm{p}}$ which is found by adding one to any of the detector counts in a success pattern $\bm{p}$, an SSG with intrinsic removal of detector loss will have $\braket{\tilde{\bm{p}} | \psi_{\inter}} = 0$ and loss then has nothing to map back to a success pattern $\bm{p}$.
For SSGs with intrinsic removal of detector noise errors, this is true when $\tilde{\bm{p}}$ is found by taking one photon away from any success pattern.

\begin{figure*}
    \includegraphics[width=\linewidth]{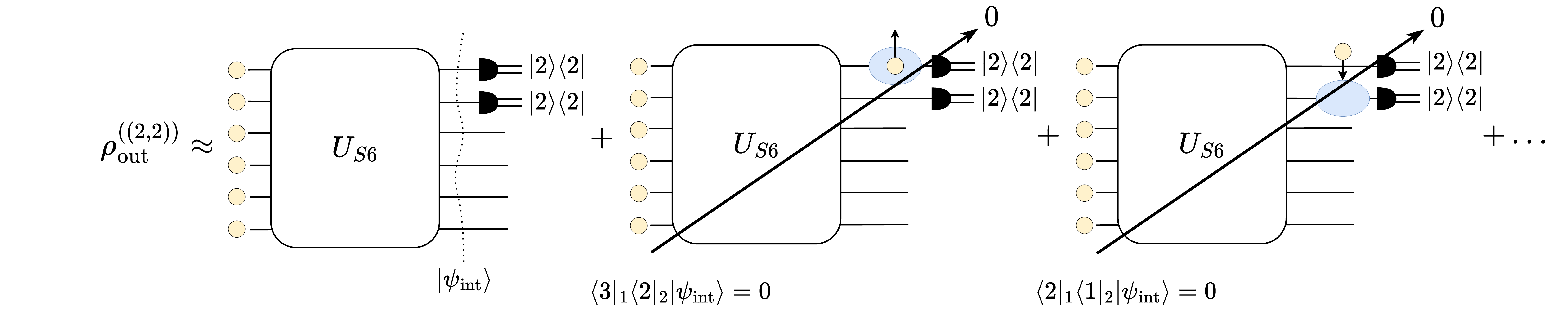}
    \caption{
    Sketch of the error processes which should be considered to model detector loss and detector noise, with Kraus operators as defined in \Cref{app:error_channels}. 
    In this example, the S6 Bell state generator, introduced in \Cref{sec:S6} with a successful detection outcome $\bm{p} = (2,2)$, is drawn. 
    When loss and multiphoton errors are commuted through the circuit, the intrinsic error removal properties of the S6 SSG mean that there is zero amplitude for any output where the states $\ket{2}\ket{3}$, $\ket{3}\ket{2}$, $\ket{1}\ket{2}$, or $\ket{2}\ket{1}$ are present in the two detected modes.
    Thus, when we condition on the successful measurement outcome, to first order there is no impact from these error channels.
    }
    \label{fig:concept}
\end{figure*}

We can rephrase these statements about intrinsic removals in terms of the error response coefficients.
An SSG intrinsically removes either detector loss ($\text{E} = \text{L}_\detc$) or detector photon addition ($\text{E} \in \{\text{A}_\detc, \tilde{\text{A}}_\detc\}$) to first order if
\begin{align}
    \Pr(\bm{p}; \text{E}) = 0
    \label{eq:p_IR}
\end{align}
for all success patterns $\pb$.
This implies $\Pr(\bm{Q}, \pb ; \text{E}) = 0$ for all subspace labels $\bm{Q}$. 
Note that this probability being zero means that the conditional probability $\Pr(\bm{Q} | \pb ; \text{E})$ is not well defined. 
However, we can see that the definition implied by \Cref{eq:p_IR}
\begin{equation}
    c_{\text{E}\to\bm{Q}} = 0
\end{equation}
for a detector error mechanism $\text{E}$ and \textit{any} set of subspaces $\bm{Q}$ is consistent with the use of these coefficients, for example in \Cref{eq:coefs_to_error_probs}, i.e.\ the error mechanism $\text{E}$ will not have any first order contributions at all to the probability of errors on the SSG output state.
Therefore, circuits with intrinsic removal properties can saturate the lower bounds given in \Cref{sec:loss_and_multi}.
A sketch of the S6 Bell state generator circuit, discussed in \Cref{sec:S6}, which intrinsically removes both detector loss and detector noise can be seen in \Cref{fig:concept}.
The intrinsic removal of detector loss property was also observed by the authors of ref.~\cite{randles2026limits}.

We show in \Cref{app:n_ghz} that the SSG circuits built from performing fusions on postselected entanglement~\cite{browne2005resource,bartolucci2021creation}, which we will refer to as the \textit{standard} SSGs, do not have intrinsic removal properties.
Seed states made with these circuits have an error response coefficient for input loss to vacuum, $c_{\text{L}_\inp \to \bm{Q}_V}=2$, twice that of an SSG which intrinsically removes detector loss.
This tells us that the standard SSG success outcomes are equally likely in the ideal case or when we condition on a photon being lost.
This is a meaningful performance difference, where SSGs which intrinsically remove detector loss errors can halve the impact that photon loss before the SSG interferometer has on the probability of having qubits leaked into the vacuum subspace $\bm{Q}_V$.

It is possible to introduce a variant of the standard circuit that trades off a lower probability of an output qubit in the $V$ subspace in exchange for worse success probability and multiphoton performance, approaching intrinsic removal in the limit of zero success probability.
We discuss this in more detail in \Cref{app:diff_down_coupling}.
In practice, we find that the circuits that have intrinsic removal properties without needing to trade-off against success probability perform better, even though they typically require more input resources, as can be seen in \Cref{tab:table}.

\subsection{Fusion Heraldable Errors}

When an SSG does not have intrinsic removal properties, it is still possible for it to have a more restricted kind of mitigation of detector errors.
Considering how seed states are used in resource state generation (RSG) provides important context to understand this restricted kind of mitigation.
During RSG, some or all of the seed state qubits undergo multiplexed fusion.
Assuming ideal linear optical fusion, if one qubit entering a fusion is in a leakage subspace, this error will be heralded by the fusion circuit's measurement outcome.
Therefore, for small error probabilities, these leakage errors on qubits fused during RSG are unlikely to falsely herald success, and therefore are unlikely to end up forming part of the resource state.

We refer to SSGs in which detector errors produce states which always have $>1$ qubits which are in leakage subspaces as having \textit{fusion heraldable} errors.
Note here that photon number conservation implies that if a photon was lost or added at detection, then at least 1 qubit of the output state will always be in a leakage subspace.
This is why we define the fusion heraldable property as applying to the SSGs which contain more than 1 qubit in a leakage subspace.
When leakage errors are correlated across more qubits in the state, the power of this type of error mitigation increases as there are more opportunities to catch these leakage terms at a fusion.

We can characterize an SSG's fusion heraldability by considering the \text{leakage weight}, $k$.
This is defined to be the smallest number of leaked qubits present in the nonzero probability subspaces of $\rho_{\out, \text{E}}^{(\pb)}$.
When we have a higher leakage weight, it means that there are more opportunities to catch the error at a fusion.

This implies a spectrum of SSG error mitigation capabilities, where having $k>1$ fusion heraldability implies some degree of additional mitigation, with $k=n$ implying that detector error mechanisms can be readily mitigated in any scheme which uses this SSG as a part of an RSG. 
We will refer to the $k=n$ case as \textit{full fusion heraldability}.
Note that this is distinct from the stronger property of intrinsic removal where errors are removed without any further processing of the state.

We can use the leakage weight of an SSG to help inform the design of a resource state generator which mitigates first-order detector error mechanisms.
To do this, we should ensure that at least $n-k+1$ of the SSG's qubits are fused and multiplexed during the RSG process to make sure that most of the fusion heraldable errors on the seed states have been removed by the time that the resource state has been created.

We can use the error response coefficients to tell us about the fusion heraldable properties of an SSG under some error mechanism, $\text{E}$.
For example, if we want to check whether an SSG has fully fusion heraldable properties, an SSG capable of full fusion heraldability of some detector error mechanism $\text{E}$ will obey the definition
\begin{align}
    \label{eq:fh}
    c_{\text{E}\to\bm{Q}} = 0,
    \quad \text{for all } \bm{Q} \text{ where } \exists \ Q_i \notin \{V,M\}.
\end{align}
for any $\text{E} \in \mathcal{E}_\detc$.
That is, any non-zero response coefficient for the error $\text{E}$ must be for a subspace tuple $\bm{Q}$ in which every qubit is in a leakage subspace.

\section{Example circuits}\label{sec:example_circuits}

\begin{figure*}
    \includegraphics[width=0.9\textwidth]{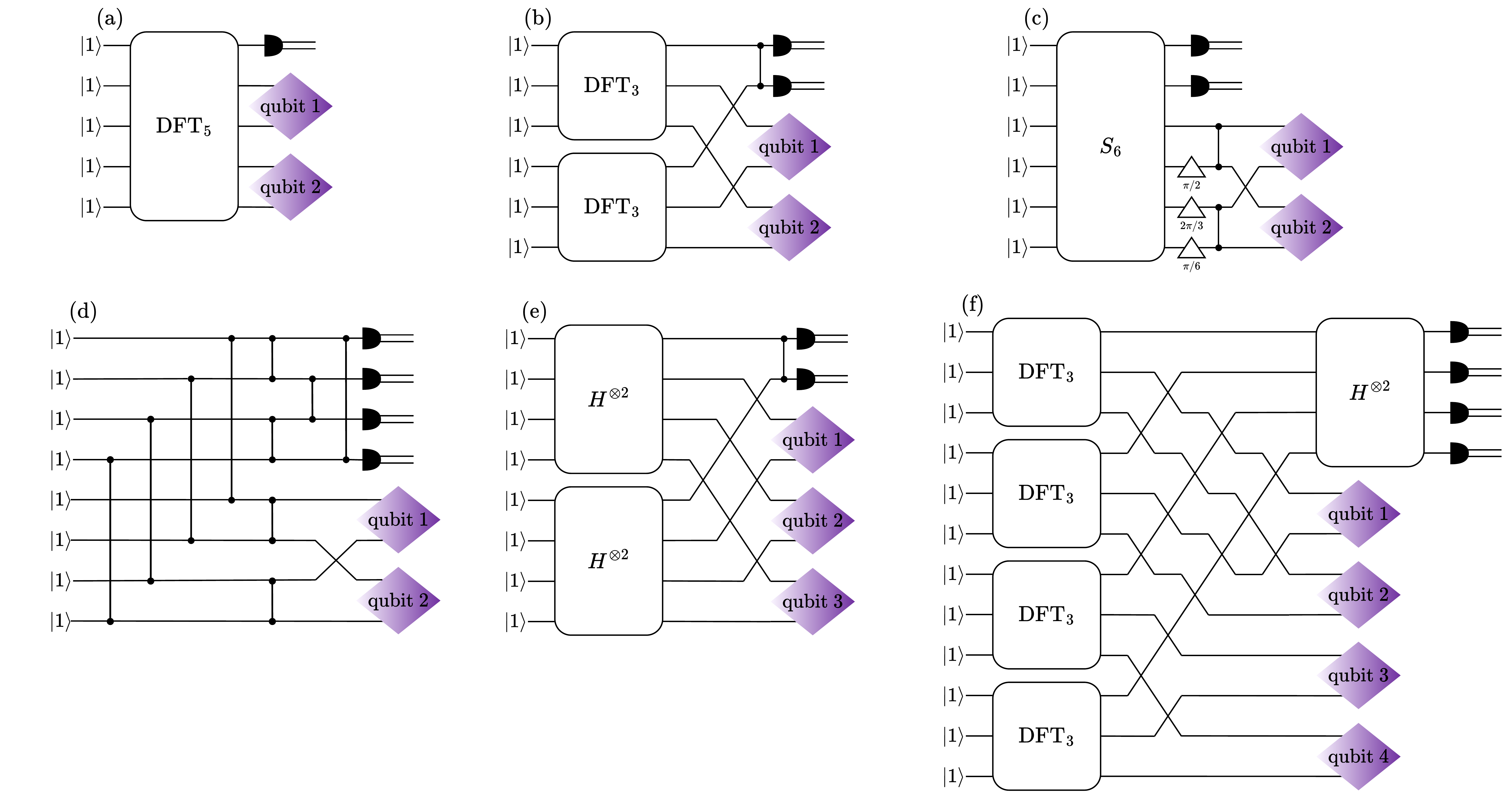}
    \caption{  
    Circuit schematic of seed state generators discussed in \Cref{sec:example_circuits}. 
    The 5DFT (a), 2-3DFT (b), S6 (c), and H8 (d) Bell state generator circuits, the 2-H4 3GHZ state generator circuit (e) and the 4-3DFT 4GHZ state generator circuit (f).
    Black horizontal and crossing lines represent optical modes.
    Rounded boxes indicate any implementation of the indicated unitary transformation. 
    Black vertical lines indicate balanced beam splitters, with transformation given by $H$.
    Triangles are passive phase shifts. 
    Circuit layouts are drawn with detectors on the first $m_\detc$ modes. 
    This is not intended to represent how these circuits should be realized in practice, which should consider practical constraints to minimize loss and enable path length matching, as well as optimize over invariants discussed in \Cref{app:pauli_to_leakage}.
    See also \Cref{sec:realization} for further discussion on realization.
    \label{fig:ssg_circuits}}
\end{figure*}

\begin{table*}[t]
\begin{tabular}{cc|cr|ccccc|cc}
    \hline \hline 
    \multicolumn{2}{c|}{\textbf{SSG}} & \multicolumn{2}{c|}{\textbf{Resource req.}} & \multicolumn{5}{c|}{\textbf{Error response coefficients}} & \multicolumn{2}{c}{\textbf{Mitigation properties}} \ \\ 
    Name & $n$-GHZ & $n_\inp$ & $\Pr(\text{success};\Id)$ & $c_{\text{L}_\detc\to\bm{Q}_V}$ & $c_{\tilde{\text{A}}_\detc\to\bm{Q}_M}$ & $c_{\text{A}_\inp\to\bm{Q}_M}$ & $c_{\text{D}_\inp\to\bm{Q}_B}$ & $\sum_P c_{\text{D}_\inp\to\bm{Q}_P}$ & $\text{L}_\detc$ & $\tilde{\text{A}}_\detc$ \\ \hline \hline
    PC  & 2 & 4 & $1/8 = 0.125$ & 1 & 1 & 3/2 & 1/4 & 1/4 & - & - \\ 
    5DFT & 2 & 5 & $12/125=0.096$ & 0 & 1/3 & 2 & 1/5 & 2/15 & IR & - \\ 
    2-3DFT & 2 & 6 & $4/27=0.148$ & 0 & 0 & 3/2 & 1/6 & 5/24 & IR & FFH \\ 
    S6 & 2 & 6 & $1/18=0.056$ & 0 & 0 & 3/2 & 5/24 & 1/12 & IR & IR \\
    H8 & 2 & 8 & $1/8=0.125$ & 0 & 0 & 3/2 &1/8 & 3/8 & IR & IR \\ \hline
    PC & 3 & 6 & $1/32=0.031$ & 1 & 1 & 3/2 & 1/4 & 1/4 & - & - \\ 
    2-H4 & 3 & 8 & $3/64=0.047$ & 0 & 0 & 3/2 & 1/8 & 5/24 & FFH & FFH \\ \hline
    PC & 4 & 8 & $1/128=0.008$ & 1 & 1 & 3/2 & 1/4 & 1/4 & - & - \\ 
    4-3DFT & 4 & 12$\ $ & $10/729=0.014$ & 0 & 0 & 3/2 & 1/6 & 71/480 & FFH & FH, $k=2$\\
    \end{tabular}
    \caption{
        Table summarizing the resource requirements, error responses and mitigation properties of the SSGs described in \Cref{sec:example_circuits}.
        Error responses are given for output subspaces which contain 1 qubit in the non-ideal subspace and remaining in ideal subspaces, e.g. $\bm{Q}_V = (V, I, I)$ for a 3GHZ state. 
        The label PC refers to the ``primate-converted'' variation of the standard n-GHZ generator explored in \Cref{app:n_ghz}. This variation is chosen as it outperforms some realisations of the standard SSGs by converting certain Pauli errors into leakage errors.
        An important observation from these results is the noticeably improved error performances of SSGs with modestly higher resource requirements.
        Error mitigation properties to detector loss (L) and incoherent detector photon addition ($\tilde{\text{A}}$) noise are specified as being intrinsically removable (IR), $k=n$ fully fusion heraldable (FFH), $1<k<n$ fusion heraldable (FH) or trivial $k=1$ fusion heraldability  (-).
    }
    \label{tab:table}
\end{table*}

In this section, we discuss several example SSG circuits which have intrinsic removal or fusion heraldability properties. 
This discussion focuses on responses to detector noise and detector loss, which using the results of \Cref{sec:loss_and_multi}, can help us understand other error mechanism responses.
Schematics for these circuits are shown in \Cref{fig:ssg_circuits}.
A summary table for the main properties of these circuits is given in \Cref{tab:table}.
Tables of the important error response coefficients for the example SSGs can be found in \Cref{app:error_responses}.
These coefficients were obtained using the method of \Cref{tab:coeff_calc_method}.

\subsection{5DFT Bell State Generator} \label{sec:5dft}
First reported in ref.~\cite{paesani2021scheme} and shown in \Cref{fig:ssg_circuits}(a), this Bell state generator takes $n_\inp=5$ input photons and has a success probability of $\Pr(\text{success} ; \Id)=12/125$.
The interferometer is a 5-mode discrete Fourier transform (DFT) circuit.
We write the DFT unitary as 
\begin{align}
    \text{DFT}_m = \frac{1}{\sqrt{m}}
    \begin{pmatrix}
        1 & 1 & 1 & \dots & 1 \\ 
        1 & \omega & \omega^2 & \dots & \omega^{m-1} \\ 
        1 & \omega^2 & \omega^4 & \dots & \omega^{2(m-1)} \\ 
        \vdots & \vdots & \vdots & \ddots & \vdots \\ 
        1 & \omega^{m-1} & \omega^{2(m-1)} & \dots & \omega^{(m-1)(m-1)}
    \end{pmatrix}
\end{align}
where $\omega=\exp(2\ii \pi / m)$.

This circuit has just one detected mode ($m_\detc=1$). 
If we look at the ideal case photon number measurement probabilities, we find that any pattern with $p_1=4$ has probability 0.
This can be understood as a consequence of the \textit{zero-transmission law} for discrete Fourier transform (DFT) interferometers~\cite{tichy2010zero}, which is an example of how multiphoton interference can cause many output patterns of a photonic circuit to be suppressed.

There is one success pattern for this circuit, $\bm{p}=(3)$. 
Therefore, the fact that $\bra{4}_1 \ket{\psi_{\inter}} = 0$ means that a first order detector loss error will not lead to a falsely heralded success outcome i.e.\ $\Pi_{\text{L}_\detc}^{((3))} \ket{\psi_{\inter}} = 0$.
This proves intrinsic removal of first order detector loss errors.
On the other hand, there is no such suppression of $\bra{2}_1 \ket{\psi_{\inter}}$ terms and thus, no intrinsic removal of detector noise errors.

As we show in \Cref{app:5dft_variants} there are additional variants of this SSG which also produce Bell states.

\subsection{2-3DFT Bell State Generator}
This circuit, shown in \Cref{fig:ssg_circuits}(b), combines two copies of a 3-mode discrete Fourier transform (DFT) interferometer and couples one output mode from each at a balanced beam splitter.
This scheme is reported in ref.~\cite{bhatti2025heralding}.
It takes $n_\inp=6$ input photons and has a success probability of $\Pr(\text{success};\Id) = 4/27$.
The successful detector outcomes are $\bm{p} \in \{(4,0),(3,1),(1,3),(0,4)\}$.

The multiphoton interference in the circuit means that all the patterns found by adding a photon to those patterns in $\{\bm{p}\}$ are suppressed, i.e.\ 
\begin{align}
    \bra{4}_1\bra{1}_2 \ket{\psi_{\inter}} & = 0 \\ 
    \bra{5}_1\bra{0}_2 \ket{\psi_{\inter}} & = 0 \\
    \bra{3}_1 \bra{2}_2 \ket{\psi_{\inter}} & = 0 \\ 
    \dots \text{etc.} 
\end{align}
This means the 2-3DFT SSG also has intrinsic removal of detector loss errors, like the 5DFT SSG.

If we consider the patterns which result from subtracting a photon from a successful detector pattern, which are projected on by the detector noise measurement operators, $\Pi_{\text{A}_\detc}^{(\bm{p})}$ and $\Pi_{\tilde{\text{A}}_\detc}^{(\bm{p})}$, we see that they always result in states with both qubits in non-dual-rail error levels, either $(M, V)$ or $(V, M)$.
In this scenario, because both qubits in the output are outside the dual-rail subspace, we can say that detector noise in this SSG is fully fusion heraldable.
Like in the intrinsic removal case, the fact that we see no dual-rail qubit terms in the state after a detector noise error can be understood as being a result of many-photon interference occurring in the circuit.

\subsection{S6 Bell State Generator}\label{sec:S6}
A new circuit which we introduce in this work was inspired by the catalog of complex Hadamards~\cite{bruzda2006complex}.
The catalog lists a complex Hadamard matrix, $S_6$, defined as 
\begin{align}
    S_6 = \frac{1}{\sqrt{6}}\begin{pmatrix}
        1 & 1 & 1 & 1 & 1 & 1 \\ 
        1 & 1 & \omega & \omega & \omega^2 & \omega^2 \\ 
        1 & \omega & 1 & \omega^2 & \omega^2 & \omega \\ 
        1 & \omega & \omega^2 & 1 & \omega & \omega^2 \\
        1 & \omega^2 & \omega^2 & \omega & 1 & \omega \\ 
        1 & \omega^2 & \omega & \omega^2 & \omega & 1
    \end{pmatrix}
\end{align}
with $\omega=\exp(2\ii \pi / 3)$.
If $n_\inp=6$ photons are injected into this circuit, then the output state contains terms which are equivalent up to mode permutations of only the following Fock states
\begin{align}
    \ket{600000}\!, \ket{411000}\!, \ket{330000}\!, \ket{311100}\!, \ket{222000}\!, \ket{111111}\!.
\end{align}
If we set $m_\detc=2$ and herald success on the pattern $\bm{p}=(2,2)$, we can project onto a state which, up to passive phase shifts, looks like 
\begin{align}
    (\ket{2000} + \ket{0200} + \ket{0020} + \ket{0002})/2 ,
\end{align}
which can be deterministically converted into a dual-rail Bell pair using linear optics. 
A circuit for this is shown in \Cref{fig:ssg_circuits}(c).
This success outcome occurs with probability $\Pr(\text{success};\Id) = 1/18$.

The absence of terms with $\ket{21}, \ket{12}, \ket{32}$ and $\ket{23}$ in the detected modes means that the S6 Bell state generator has intrinsic removal of both detector loss and detector noise errors.
We have drawn \Cref{fig:concept} to diagrammatically show this.

We note also that this BSG can be seen as equivalent (in that it is the same up to a unitary transformation on the output modes) as one which consists of 6 photons entering the 6-mode DFT, then placing 2 detectors separated by 1 mode, e.g. on modes 4 and 6.

\subsection{H8 Bell State Generator}\label{sec:H8}
The H8 Bell state generator, previously mentioned in \cite{bartolucci2021creation} (section IV.B therein) and shown in \Cref{fig:ssg_circuits}(d), has a circuit which resembles the standard Bell state generator shown in \Cref{app:n_ghz}, but where all eight input modes are occupied by a single photon, $m=n_\inp=8$.
Like the standard Bell state generator it succeeds with probability $1/8$, absent any output mode reconfiguration, but has significantly better error properties including intrinsic removal of both detector loss and detector noise.
The Hong-Ou-Mandel interference at the input beam splitters makes it so that all terms in $\ket{\psi_{\inter, \Id}}$ have an even number of photons in the detected modes, which leads to intrinsic removal of both first order detector loss and detector noise error mechanisms.

Though it is beyond the scope of this paper where we restrict ourselves to only ballistic SSGs, we note that adding some adaptive phase shifts in the circuit can increase the ideal success probability to $1/4$.

\subsection{2-H4 3GHZ Generator}\label{sec:2-H4}
Another circuit which was reported in ref.~\cite{bhatti2025heralding} creates 3GHZ states using 2 copies of a 4-mode Hadamard interferometer, $H^{\otimes 2}$, with 
\begin{align}
    H = \frac{1}{\sqrt{2}} \begin{pmatrix}
        1 & 1 \\ 1 & -1
    \end{pmatrix} .
\end{align}
This circuit, shown in \Cref{fig:ssg_circuits}(e), uses $n_\inp=8$ photons and succeeds with $\Pr(\text{success};\Id) = 3/64$.
All 5 photon measurements are successful detector outcomes, $\bm{p} \in \{(5,0),(4,1),(3,2),(2,3),(1,4),(0,5)\}$.

\begin{figure*}
    \includegraphics[width=\linewidth, trim={20 50 20 20}]{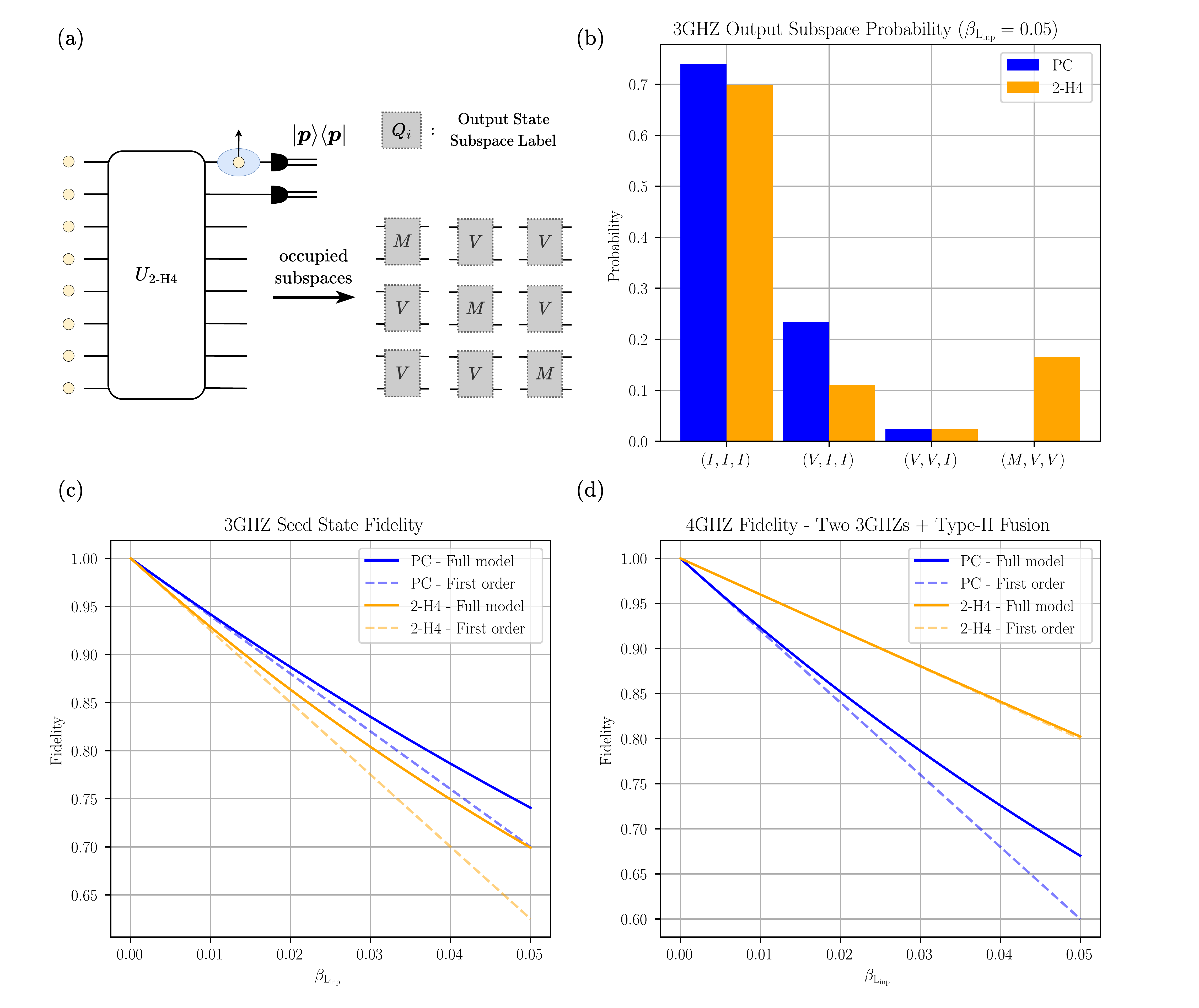}
    \caption{
    Certain error mechanisms in SSG produce states in which multiple qubits are simultaneously in leakage subspaces.
    Such leakage is likely to cause subsequent fusions to fail, preventing the affected states from entering the resource state. 
    We call these errors \textit{fusion heraldable}.
    \textbf{Panel (a)} sketches one way in which these fusion heraldable states can be generated.
    Here detector loss in the 2-H4 3GHZ generator results in a state which lives in the $(M, V, V), (V, M, V),$ or $(V, V, M)$ output subspaces.
    \textbf{Panel (b)} gives the output subspace probabilities for the 2-H4 and ``standard'' PC 3GHZ generators with input loss, $\text{L}_\inp$, of $\beta_{\text{L}_\inp} = 0.05$ and no other errors.
    For both generators, permutations of a given error label pattern have equal probability; the plotted probability is the total across all the symmetric configurations.
    The $(V,V,V)$ term is omitted due to its small probability for both SSGs.
    \textbf{Panel (c)} shows the fidelity of these output states, $\bra{\psi_\out^{(\pb)}} \rho_\out^{(\pb)} \ket{\psi_\out^{(\pb)}}$, averaged over all success outcomes $\pb$, shown both to first order in $\beta_{\text{L}_\inp}$ and exactly. 
    It is clear that a first order approximation holds only for small values of $\beta_{\text{L}_\inp}$.
    However, the first order approximations give the correct indication of which SSG has the highest fidelity.
    \textbf{Panel (d)} shows the fidelity of the 4GHZ created when two copies of the 3GHZ are fused with a type-II fusion that fails in the Z basis.
    This shows us that if we were to use seed state fidelity as the sole metric, we would incorrectly conclude that the PC 3GHZ generator is better than the 2-H4.
    However, much of the 2-H4 infidelity consists of fusion heraldable errors.
    As Panel (d) shows, this means states produced by fusing states from 2-H4 SSG are more robust to loss when used as part of a resource state generation scheme.
    The first-order and full-model fidelity curves in panel (d) for 2-H4 nearly coincide because the fidelity expansion contains no quadratic term in the input-loss probability.
    \label{fig:h4_errors}}
\end{figure*}

The circuit does not have any intrinsic removal properties.
However, if we look at the state after applying the detector loss operator, $\Pi_{\text{L}_\detc}^{(\bm{p})}$, we can see that all terms in the state correspond to error levels $(V, V, M)$, $(V, M, V)$, $(M, V, V)$.
This is shown in \Cref{fig:h4_errors}a.
Similarly, if we look at the state after applying a detector noise operator, e.g.\ $\Pi_{\text{A}_\detc}^{(\bm{p})}$, we can see that all terms in the state correspond to error levels which are qubit relabelings of $(M, M, V)$ or $(M, V, V)$. 
Thus, we can see that detector loss and detector noise errors in the 2-H4 3GHZ generator are both fully fusion heraldable.

To help demonstrate the importance of the full fusion heraldable property, we compare the 2-H4 3GHZ generator to an implementation of the ``standard'' 3GHZ SSG~\cite{browne2005resource,bartolucci2021creation} which we call \textit{primate-converted} (PC) when we add the input loss ($\text{L}_\inp$) error mechanism.
If we were to only look at the output state fidelity, as shown in \Cref{fig:h4_errors}c, we would conclude that the PC circuit outperforms the 2-H4 circuit.
However, if we consider that we will likely use multiplexed fusions to build resource states, we should also consider the fusion heraldable properties.
\Cref{fig:h4_errors}d shows that using 2-H4 seeds gives a higher 4GHZ fidelity.
Calculation of the errors of a state after a fusion, conditional on fusion success, can be handled using the framework of this paper by simply considering the SSGs and fusion circuit as a single, larger SSG.

As discussed in \Cref{app:det_pattern_error_dependence}, when we look at the error response coefficients for the 2-H4 SSG without averaging over $\bm{p}$, we see that the error properties of this circuit have a strong dependence on the particular pattern which is measured.
This potentially motivates treating the $(3,2)$ and $(2,3)$ outcomes as failures, even if in the ideal case they are successful, so that the average error properties of the circuit improve.

\subsection{4-3DFT 4GHZ Generator}\label{sec:4-3dft}
Another new SSG we introduce in this work is the 4-3DFT 4GHZ state generator, shown in \Cref{fig:ssg_circuits}(f).
This SSG uses the same circuit as the $d=4$ qudit Bell state generator from \cite{bhatti2025heralding}, taking ${n_\inp}=12$ photons into a circuit formed from 4 copies of a 3DFT interferometer, where one output mode from each copy is routed to a $H^{\otimes 2}$ interferometer followed by $m_\detc=4$ detectors, but detecting 8 photons instead of 10.
There are 84 different successful 8-photon detector outcomes that can be sorted into 9 equivalence classes within which the heralding patterns are the same up to mode relabeling.
The success probability is $\Pr(\text{success};\Id) = 10/729$ with qubits defined over fixed output modes, or $15/729$ if adaptive reconfiguration of the output modes is allowed.
This is $>$75\% higher than the $\Pr(\text{success};\Id) = 1/128$ generator of \cite{gimeno2016towards} and, to our knowledge, is the highest success probability of any fully ballistic 4GHZ generator.

After applying the detector loss operator we project onto a state which contains only $(M, V, V, V)$ error level terms, up to qubit permutations, meaning that detector loss in this SSG is fully fusion heraldable.
Upon applying a detector noise operator we project into a state which contains both $(M,V,\ideal,\ideal)$ and $(V,V,M,\ideal)$ type terms, up to permutations.
The $(M, V, \ideal, \ideal)$ terms mean that detector noise is only fusion heralded to degree $k=2$, and so implies that we would prefer to use this SSG in contexts where at least 3 out of the 4 qubits are fused and multiplexed within the RSG scheme.

We note that this circuit can be adapted to produce states which are locally equivalent to 4-line graph states by replacing the $H^{\otimes 2}$ interferometer with one implementing the transfer matrix
\begin{align}
    U = \frac{1}{2} \begin{pmatrix}
        1 & i & -1 & i\\
        1 & i & 1 & -i\\
        -1 & i & i & -1\\
        1 & -i & i & -1
    \end{pmatrix} .
\end{align}

\section{Considerations for realization}\label{sec:realization}

\begin{figure}
    \centering 
    \includegraphics[width=\columnwidth]{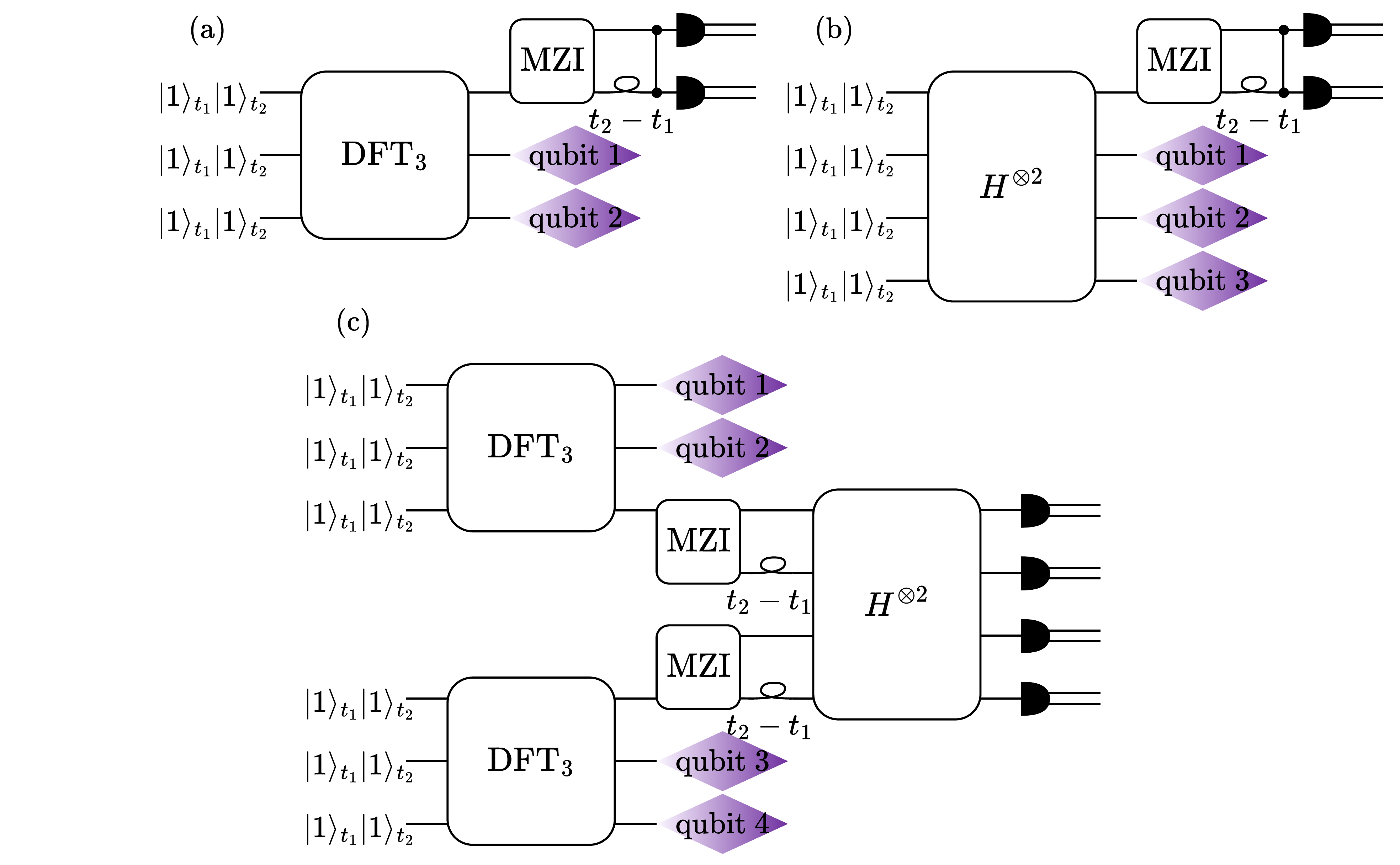}
    \caption{
    Native time-bin circuits which realize the 2-3DFT Bell state generator (a), the 2-H4 3GHZ state generator (b) and the 4-3DFT 4GHZ state generator (c).
    All these circuits allow for time-bin encoded qubits to be generated without performing switching or delays on the photons which form the seed state. 
    The MZI should be set to implement either an $\I_2$ or $X$ unitary transformation, such that photons from time $t_1$ are routed to a delay.
    The delay should be set such that the $t_1$ and $t_2$ photons become aligned in time and can interfere. 
    \label{fig:native_timebin}}
\end{figure}

One particularly beneficial aspect of the increased resilience to loss of the seed states described in this paper concerns their implementation in a real-world photonic FBQC scheme.
In order to implement a large-scale photonic quantum computer, it is vital to allow photons to travel between chips in optical fiber delays that both increase the connectivity between and within modules and allow classical feedforward logic to be implemented.
Although optical fiber already has very low propagation losses and there has been promising progress in chip to fiber coupling~\cite{alexander2024manufacturable}, the level of phase stabilization needed to ensure no relative phase drift between the two rails of a spatially-encoded qubit is challenging. 
This challenge can be avoided by propagating qubits between chips in a dual-rail time-bin encoding, where we would expect the phase noise in the fiber to drift much more slowly than the time separation in the time-bin encoding, so that any phase variations impart only a global phase.

One approach to generating time-bin qubits is to generate them in a path encoding, then convert to a time-bin encoding through the use of delays and switches.
These components add extra loss to the photon worldlines.
To mitigate this, in some cases we can use variants of the seed state generation circuits where the photons in the time-bin encoded output seed state do not need to propagate through any delays or switches.
Instead, we can apply the time-to-space conversion operations to the detected photons, as shown in \Cref{fig:native_timebin}.
This adds loss on the detected modes of the SSG but, for circuits with intrinsic removal this only changes the heralding probability to first order, while for circuits with fusion heraldable errors it creates correlated leakage that can be rejected by subsequent fusion and multiplexing stages.

\begin{figure}
    \centering 
    \includegraphics[width=\columnwidth]{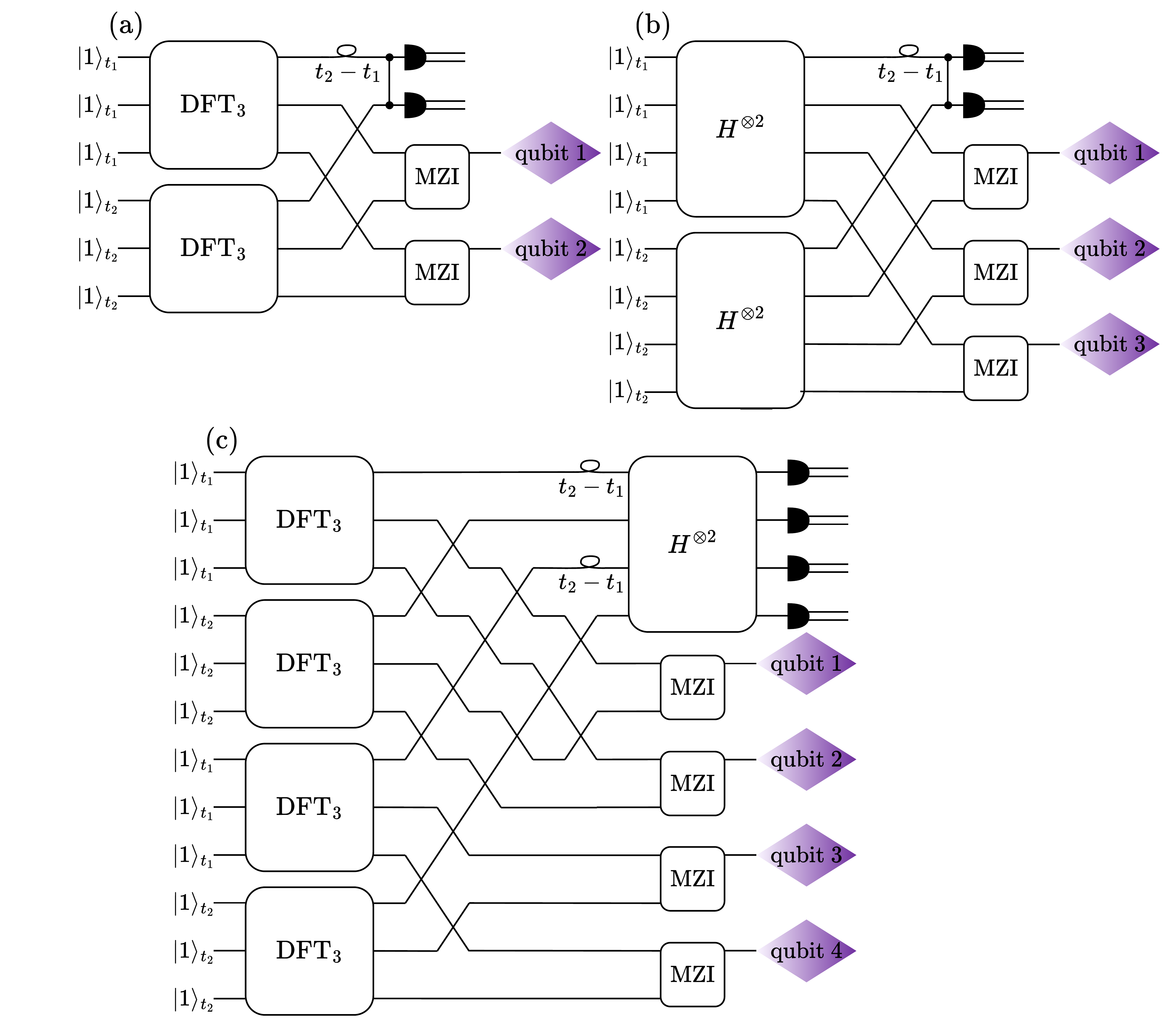}
    \caption{
    Circuits which emit time-bin qubits where the timescale at which we produce subsequent rounds of single photons is decoupled from the timescales of the time-bin qubits. 
    Shown for the 2-3DFT (a), 2-H4 (b) and 4-3DFT (c) SSGs.
     \label{fig:half_native_timebin}}
\end{figure}

It should be noted that these ``native time-bin'' circuits couple the timescale at which photons are arriving at the SSG with the time-bin separation of the time-bin qubits. 
However, we can construct additional variants of the SSGs which allow us to decouple these timescales by putting the switch on the output modes and delays on the detected modes.
The circuits for this are shown in \Cref{fig:half_native_timebin}.

The ability to mitigate the impact of loss on the detected modes of the SSG has further benefits when considering other aspects of the realization of these systems.
In particular, system considerations that can introduce loss such as finite detector fanout in a pseudo-photon number detector setup and connections between room-temperature and cryogenic modules on these modes become less detrimental.
The benefits of these SSGs can therefore enable additional flexibility in system design.

\section{Conclusion}

The results in this paper show that there are rich dynamics for how seed state generation schemes respond to errors.
There is no catch-all metric that fully captures the important error properties of an SSG, particularly in the context of a larger resource state generation or FBQC scheme.
This means that, despite much of the previous literature focusing on resource requirements such as ideal case success probability and number of input photons required, or metrics like output state fidelity, there are other important metrics to optimize for when searching for and assessing new SSG circuits.
The example given in \Cref{fig:h4_errors} highlights the insufficiency of the output state fidelity as a measure of SSG quality, particularly in the context of building larger states.

The framework we have introduced in this paper allows us to describe the state of a photonic architecture without the need for costly Fock state simulations by tracking a distribution of tuples which represent the probabilities that the SSG output state would be projected into a collection of qubit subspaces.
We are able to characterize each SSG circuit with these probabilities and approximate them as sums over leading order responses to particularly important error channels.
Under this framework, we can see that in certain ways, the circuits introduced here are optimal due to many-photon interference which produces error response coefficients of zero for some non-trivial errors.

In fact, by combining an SSG like the S6 SSG, which intrinsically removes both detector loss and detector noise to first order, with single qubit Cliffords and fully destructive $n$-way fusions, it is possible to create any dual-rail stabilizer state where any first order detector error does not affect the quality of the output state.
To see this, we first note that fully destructive measurements will fail under a first order detector loss or detector noise error. 
Next, we note that we can construct a totally destructive measurement which projects onto any stabilizer state by combining single qubit gates with type-I and type-II fusions~\cite{browne2005resource,pankovich2024flexible}.
When we combine Bell state generation with these destructive measurements, a first order error will either lead to state generation failure or fusion failure.
Although, for preparing large states, this first order approximation will break down more rapidly.

When choosing an SSG scheme for a given architecture, it is important to be able to perform simulations which can track errors and resource requirements across a large, fault tolerant system. 
Such simulations allow us to assess the trade-offs between input resources, success probabilities, and error rates of the SSGs in the architecture, enabling us to make informed architecture design choices.
This will also help inform an understanding of the necessary component performance and resource requirements to build a fault-tolerant fusion based quantum computer.
While this paper focuses specifically on seed state generation, we would like to emphasize that the same modeling framework is applicable to the state consumption, i.e. fusion measurements, and to larger resource state generation as well.
We hope that the framework and discussion in this paper will help others to guide their exploration of photonic quantum computing architectures.

A natural open question is whether there are further SSG schemes of interest for dual-rail qubit seed state generation which have similar or better error performance than those described in this paper.
There are some interesting areas to explore beyond this which we would also like to highlight.

One interesting direction comes from SSGs which produce states in other encodings. 
A notable example is multi-rail qudit encodings where there are SSGs published~\cite{paesani2021scheme, bhatti2025heralding} which have promising error response properties.
Other fixed energy qubit encodings beyond the single photon dual-rail qubit~\cite{chuang1997bosonic, wasilewski2007protecting} may also be of interest, as they would maintain some of the phase stability advantages discussed in \Cref{sec:realization}.

We are also interested to know more about bounds on error response coefficients. 
As we have shown in this paper, some responses can be zero, such as for detector loss and detector noise.
Conversely, for the class of SSG we consider here, we know some cannot ever be zero, such as for input loss and input multiphotons.
Going beyond the SSGs considered here, for example combining existing SSGs with constructions for photon indistinguishability distillation~\cite{birchall2018fundamental, sparrow2018quantum,marshall2022distillation, saied2024general, somhorst2024photon,faurby2024purifying,somhorst2026below}, allows for distinguishability induced Paulis to become arbitrarily low.
Understanding bounds on Pauli responses and success probability that generalize these combined constructions would also be of interest.

There are also SSG schemes which use intermediate stages of feedforward and adaptivity to improve success probability~\cite{bartolucci2021creation}.
We are interested in the tradeoff between the additional errors introduced by realistic implementations of these adaptive schemes and the resource reductions that they offer.

Similarly, we know of more examples of Bell state generators with intrinsic removal of both loss and noise detector errors than of SSGs which produce larger states.
These larger generators often rely on the fusion heraldability of errors for improved performance.
We are interested in examples of SSGs for creating 3 qubit states or larger which cannot be decomposed into Bell state generators plus fusions and which have intrinsic removal properties.

\section*{Acknowledgements}
We would like to thank the PsiQuantum architecture team for motivating discussions. 
We thank Reece Shaw and Joel Wallman for valuable feedback on the manuscript. 
Most of these results were derived a few years ago, and so predate LLM assisted research. 
We acknowledge the use of LLMs to assist with proofreading.

\bibliography{bib.bib}

\begin{thebibliography}{57}%
\makeatletter
\providecommand \@ifxundefined [1]{%
 \@ifx{#1\undefined}
}%
\providecommand \@ifnum [1]{%
 \ifnum #1\expandafter \@firstoftwo
 \else \expandafter \@secondoftwo
 \fi
}%
\providecommand \@ifx [1]{%
 \ifx #1\expandafter \@firstoftwo
 \else \expandafter \@secondoftwo
 \fi
}%
\providecommand \natexlab [1]{#1}%
\providecommand \enquote  [1]{``#1''}%
\providecommand \bibnamefont  [1]{#1}%
\providecommand \bibfnamefont [1]{#1}%
\providecommand \citenamefont [1]{#1}%
\providecommand \href@noop [0]{\@secondoftwo}%
\providecommand \href [0]{\begingroup \@sanitize@url \@href}%
\providecommand \@href[1]{\@@startlink{#1}\@@href}%
\providecommand \@@href[1]{\endgroup#1\@@endlink}%
\providecommand \@sanitize@url [0]{\catcode `\\12\catcode `\$12\catcode `\&12\catcode `\#12\catcode `\^12\catcode `\_12\catcode `\%12\relax}%
\providecommand \@@startlink[1]{}%
\providecommand \@@endlink[0]{}%
\providecommand \url  [0]{\begingroup\@sanitize@url \@url }%
\providecommand \@url [1]{\endgroup\@href {#1}{\urlprefix }}%
\providecommand \urlprefix  [0]{URL }%
\providecommand \Eprint [0]{\href }%
\providecommand \doibase [0]{https://doi.org/}%
\providecommand \selectlanguage [0]{\@gobble}%
\providecommand \bibinfo  [0]{\@secondoftwo}%
\providecommand \bibfield  [0]{\@secondoftwo}%
\providecommand \translation [1]{[#1]}%
\providecommand \BibitemOpen [0]{}%
\providecommand \bibitemStop [0]{}%
\providecommand \bibitemNoStop [0]{.\EOS\space}%
\providecommand \EOS [0]{\spacefactor3000\relax}%
\providecommand \BibitemShut  [1]{\csname bibitem#1\endcsname}%
\let\auto@bib@innerbib\@empty
%</preamble>
\bibitem [{\citenamefont {Bartolucci}\ \emph {et~al.}(2023)\citenamefont {Bartolucci}, \citenamefont {Birchall}, \citenamefont {Bombin}, \citenamefont {Cable}, \citenamefont {Dawson}, \citenamefont {Gimeno-Segovia}, \citenamefont {Johnston}, \citenamefont {Kieling}, \citenamefont {Nickerson}, \citenamefont {Pant} \emph {et~al.}}]{bartolucci2023fusion}%
  \BibitemOpen
  \bibfield  {author} {\bibinfo {author} {\bibfnamefont {S.}~\bibnamefont {Bartolucci}}, \bibinfo {author} {\bibfnamefont {P.}~\bibnamefont {Birchall}}, \bibinfo {author} {\bibfnamefont {H.}~\bibnamefont {Bombin}}, \bibinfo {author} {\bibfnamefont {H.}~\bibnamefont {Cable}}, \bibinfo {author} {\bibfnamefont {C.}~\bibnamefont {Dawson}}, \bibinfo {author} {\bibfnamefont {M.}~\bibnamefont {Gimeno-Segovia}}, \bibinfo {author} {\bibfnamefont {E.}~\bibnamefont {Johnston}}, \bibinfo {author} {\bibfnamefont {K.}~\bibnamefont {Kieling}}, \bibinfo {author} {\bibfnamefont {N.}~\bibnamefont {Nickerson}}, \bibinfo {author} {\bibfnamefont {M.}~\bibnamefont {Pant}}, \emph {et~al.},\ }\bibfield  {title} {\bibinfo {title} {Fusion-based quantum computation},\ }\href@noop {} {\bibfield  {journal} {\bibinfo  {journal} {Nature Communications}\ }\textbf {\bibinfo {volume} {14}},\ \bibinfo {pages} {912} (\bibinfo {year} {2023})}\BibitemShut {NoStop}%
\bibitem [{\citenamefont {Chuang}\ and\ \citenamefont {Yamamoto}(1995)}]{chuang1995simple}%
  \BibitemOpen
  \bibfield  {author} {\bibinfo {author} {\bibfnamefont {I.~L.}\ \bibnamefont {Chuang}}\ and\ \bibinfo {author} {\bibfnamefont {Y.}~\bibnamefont {Yamamoto}},\ }\bibfield  {title} {\bibinfo {title} {Simple quantum computer},\ }\href@noop {} {\bibfield  {journal} {\bibinfo  {journal} {Physical Review A}\ }\textbf {\bibinfo {volume} {52}},\ \bibinfo {pages} {3489} (\bibinfo {year} {1995})}\BibitemShut {NoStop}%
\bibitem [{\citenamefont {Birchall}(2018)}]{birchall2018fundamental}%
  \BibitemOpen
  \bibfield  {author} {\bibinfo {author} {\bibfnamefont {P.}~\bibnamefont {Birchall}},\ }\emph {\bibinfo {title} {Fundamental Advantages and Practicalities of Quantum-Photonic Metrology and Computing}},\ \href@noop {} {Ph.D. thesis},\ \bibinfo  {school} {University of Bristol} (\bibinfo {year} {2018})\BibitemShut {NoStop}%
\bibitem [{\citenamefont {Sparrow}(2018)}]{sparrow2018quantum}%
  \BibitemOpen
  \bibfield  {author} {\bibinfo {author} {\bibfnamefont {C.}~\bibnamefont {Sparrow}},\ }\emph {\bibinfo {title} {Quantum interference in universal linear optical devices for quantum computation and simulation}},\ \href@noop {} {Ph.D. thesis},\ \bibinfo  {school} {Imperial College London} (\bibinfo {year} {2018})\BibitemShut {NoStop}%
\bibitem [{\citenamefont {Marshall}(2022)}]{marshall2022distillation}%
  \BibitemOpen
  \bibfield  {author} {\bibinfo {author} {\bibfnamefont {J.}~\bibnamefont {Marshall}},\ }\bibfield  {title} {\bibinfo {title} {Distillation of indistinguishable photons},\ }\href@noop {} {\bibfield  {journal} {\bibinfo  {journal} {{Physical Review Letters}}\ }\textbf {\bibinfo {volume} {129}},\ \bibinfo {pages} {213601} (\bibinfo {year} {2022})}\BibitemShut {NoStop}%
\bibitem [{\citenamefont {Saied}\ \emph {et~al.}(2024)\citenamefont {Saied}, \citenamefont {Marshall}, \citenamefont {Anand},\ and\ \citenamefont {Rieffel}}]{saied2024general}%
  \BibitemOpen
  \bibfield  {author} {\bibinfo {author} {\bibfnamefont {J.}~\bibnamefont {Saied}}, \bibinfo {author} {\bibfnamefont {J.}~\bibnamefont {Marshall}}, \bibinfo {author} {\bibfnamefont {N.}~\bibnamefont {Anand}},\ and\ \bibinfo {author} {\bibfnamefont {E.~G.}\ \bibnamefont {Rieffel}},\ }\bibfield  {title} {\bibinfo {title} {General protocols for the efficient distillation of indistinguishable photons},\ }\href@noop {} {\bibfield  {journal} {\bibinfo  {journal} {arXiv preprint arXiv:2404.14217}\ } (\bibinfo {year} {2024})}\BibitemShut {NoStop}%
\bibitem [{\citenamefont {Somhorst}\ \emph {et~al.}(2024)\citenamefont {Somhorst}, \citenamefont {Sau{\"e}r}, \citenamefont {van~den Hoven},\ and\ \citenamefont {Renema}}]{somhorst2024photon}%
  \BibitemOpen
  \bibfield  {author} {\bibinfo {author} {\bibfnamefont {F.}~\bibnamefont {Somhorst}}, \bibinfo {author} {\bibfnamefont {B.~K.}\ \bibnamefont {Sau{\"e}r}}, \bibinfo {author} {\bibfnamefont {S.}~\bibnamefont {van~den Hoven}},\ and\ \bibinfo {author} {\bibfnamefont {J.~J.}\ \bibnamefont {Renema}},\ }\bibfield  {title} {\bibinfo {title} {Photon distillation schemes with reduced resource costs based on multiphoton {F}ourier interference},\ }\href@noop {} {\bibfield  {journal} {\bibinfo  {journal} {arXiv preprint arXiv:2404.14262}\ } (\bibinfo {year} {2024})}\BibitemShut {NoStop}%
\bibitem [{\citenamefont {Faurby}\ \emph {et~al.}(2024)\citenamefont {Faurby}, \citenamefont {Carosini}, \citenamefont {Cao}, \citenamefont {Sund}, \citenamefont {Hansen}, \citenamefont {Giorgino}, \citenamefont {Villadsen}, \citenamefont {Van Den~Hoven}, \citenamefont {Lodahl}, \citenamefont {Paesani} \emph {et~al.}}]{faurby2024purifying}%
  \BibitemOpen
  \bibfield  {author} {\bibinfo {author} {\bibfnamefont {C.~F.}\ \bibnamefont {Faurby}}, \bibinfo {author} {\bibfnamefont {L.}~\bibnamefont {Carosini}}, \bibinfo {author} {\bibfnamefont {H.}~\bibnamefont {Cao}}, \bibinfo {author} {\bibfnamefont {P.~I.}\ \bibnamefont {Sund}}, \bibinfo {author} {\bibfnamefont {L.~M.}\ \bibnamefont {Hansen}}, \bibinfo {author} {\bibfnamefont {F.}~\bibnamefont {Giorgino}}, \bibinfo {author} {\bibfnamefont {A.~B.}\ \bibnamefont {Villadsen}}, \bibinfo {author} {\bibfnamefont {S.~N.}\ \bibnamefont {Van Den~Hoven}}, \bibinfo {author} {\bibfnamefont {P.}~\bibnamefont {Lodahl}}, \bibinfo {author} {\bibfnamefont {S.}~\bibnamefont {Paesani}}, \emph {et~al.},\ }\bibfield  {title} {\bibinfo {title} {Purifying photon indistinguishability through quantum interference},\ }\href@noop {} {\bibfield  {journal} {\bibinfo  {journal} {{Physical Review Letters}}\ }\textbf {\bibinfo {volume} {133}},\ \bibinfo {pages} {033604} (\bibinfo {year} {2024})}\BibitemShut {NoStop}%
\bibitem [{\citenamefont {Somhorst}\ \emph {et~al.}(2026)\citenamefont {Somhorst}, \citenamefont {Saied}, \citenamefont {Kannan}, \citenamefont {Kassenberg}, \citenamefont {Marshall}, \citenamefont {de~Goede}, \citenamefont {Snijders}, \citenamefont {Stremoukhov}, \citenamefont {Lukianenko}, \citenamefont {Venderbosch} \emph {et~al.}}]{somhorst2026below}%
  \BibitemOpen
  \bibfield  {author} {\bibinfo {author} {\bibfnamefont {F.}~\bibnamefont {Somhorst}}, \bibinfo {author} {\bibfnamefont {J.}~\bibnamefont {Saied}}, \bibinfo {author} {\bibfnamefont {N.}~\bibnamefont {Kannan}}, \bibinfo {author} {\bibfnamefont {B.}~\bibnamefont {Kassenberg}}, \bibinfo {author} {\bibfnamefont {J.}~\bibnamefont {Marshall}}, \bibinfo {author} {\bibfnamefont {M.}~\bibnamefont {de~Goede}}, \bibinfo {author} {\bibfnamefont {H.}~\bibnamefont {Snijders}}, \bibinfo {author} {\bibfnamefont {P.}~\bibnamefont {Stremoukhov}}, \bibinfo {author} {\bibfnamefont {A.}~\bibnamefont {Lukianenko}}, \bibinfo {author} {\bibfnamefont {P.}~\bibnamefont {Venderbosch}}, \emph {et~al.},\ }\bibfield  {title} {\bibinfo {title} {Below-threshold error reduction in single photons through photon distillation},\ }\href@noop {} {\bibfield  {journal} {\bibinfo  {journal} {arXiv preprint arXiv:2601.05947}\ } (\bibinfo {year} {2026})}\BibitemShut {NoStop}%
\bibitem [{\citenamefont {Bennett}\ \emph {et~al.}(1996)\citenamefont {Bennett}, \citenamefont {Brassard}, \citenamefont {Popescu}, \citenamefont {Schumacher}, \citenamefont {Smolin},\ and\ \citenamefont {Wootters}}]{bennett1996purification}%
  \BibitemOpen
  \bibfield  {author} {\bibinfo {author} {\bibfnamefont {C.~H.}\ \bibnamefont {Bennett}}, \bibinfo {author} {\bibfnamefont {G.}~\bibnamefont {Brassard}}, \bibinfo {author} {\bibfnamefont {S.}~\bibnamefont {Popescu}}, \bibinfo {author} {\bibfnamefont {B.}~\bibnamefont {Schumacher}}, \bibinfo {author} {\bibfnamefont {J.~A.}\ \bibnamefont {Smolin}},\ and\ \bibinfo {author} {\bibfnamefont {W.~K.}\ \bibnamefont {Wootters}},\ }\bibfield  {title} {\bibinfo {title} {Purification of noisy entanglement and faithful teleportation via noisy channels},\ }\href@noop {} {\bibfield  {journal} {\bibinfo  {journal} {{Physical Review Letters}}\ }\textbf {\bibinfo {volume} {76}},\ \bibinfo {pages} {722} (\bibinfo {year} {1996})}\BibitemShut {NoStop}%
\bibitem [{\citenamefont {Pan}\ \emph {et~al.}(2001)\citenamefont {Pan}, \citenamefont {Simon}, \citenamefont {Brukner},\ and\ \citenamefont {Zeilinger}}]{pan2001entanglement}%
  \BibitemOpen
  \bibfield  {author} {\bibinfo {author} {\bibfnamefont {J.-W.}\ \bibnamefont {Pan}}, \bibinfo {author} {\bibfnamefont {C.}~\bibnamefont {Simon}}, \bibinfo {author} {\bibfnamefont {{\v{C}}.}~\bibnamefont {Brukner}},\ and\ \bibinfo {author} {\bibfnamefont {A.}~\bibnamefont {Zeilinger}},\ }\bibfield  {title} {\bibinfo {title} {Entanglement purification for quantum communication},\ }\href@noop {} {\bibfield  {journal} {\bibinfo  {journal} {Nature}\ }\textbf {\bibinfo {volume} {410}},\ \bibinfo {pages} {1067} (\bibinfo {year} {2001})}\BibitemShut {NoStop}%
\bibitem [{\citenamefont {Pan}\ \emph {et~al.}(2003)\citenamefont {Pan}, \citenamefont {Gasparoni}, \citenamefont {Ursin}, \citenamefont {Weihs},\ and\ \citenamefont {Zeilinger}}]{pan2003experimental}%
  \BibitemOpen
  \bibfield  {author} {\bibinfo {author} {\bibfnamefont {J.-W.}\ \bibnamefont {Pan}}, \bibinfo {author} {\bibfnamefont {S.}~\bibnamefont {Gasparoni}}, \bibinfo {author} {\bibfnamefont {R.}~\bibnamefont {Ursin}}, \bibinfo {author} {\bibfnamefont {G.}~\bibnamefont {Weihs}},\ and\ \bibinfo {author} {\bibfnamefont {A.}~\bibnamefont {Zeilinger}},\ }\bibfield  {title} {\bibinfo {title} {Experimental entanglement purification of arbitrary unknown states},\ }\href@noop {} {\bibfield  {journal} {\bibinfo  {journal} {Nature}\ }\textbf {\bibinfo {volume} {423}},\ \bibinfo {pages} {417} (\bibinfo {year} {2003})}\BibitemShut {NoStop}%
\bibitem [{\citenamefont {L{\"o}bl}\ \emph {et~al.}(2026)\citenamefont {L{\"o}bl}, \citenamefont {Capatos}, \citenamefont {Chan}, \citenamefont {Lodahl}, \citenamefont {S{\o}rensen},\ and\ \citenamefont {Paesani}}]{lobl2026purification}%
  \BibitemOpen
  \bibfield  {author} {\bibinfo {author} {\bibfnamefont {M.~C.}\ \bibnamefont {L{\"o}bl}}, \bibinfo {author} {\bibfnamefont {A.~A.}\ \bibnamefont {Capatos}}, \bibinfo {author} {\bibfnamefont {M.~L.}\ \bibnamefont {Chan}}, \bibinfo {author} {\bibfnamefont {P.}~\bibnamefont {Lodahl}}, \bibinfo {author} {\bibfnamefont {A.~S.}\ \bibnamefont {S{\o}rensen}},\ and\ \bibinfo {author} {\bibfnamefont {S.}~\bibnamefont {Paesani}},\ }\bibfield  {title} {\bibinfo {title} {Purification of photonic graph states},\ }\href@noop {} {\bibfield  {journal} {\bibinfo  {journal} {arXiv preprint arXiv:2609.01710}\ } (\bibinfo {year} {2026})}\BibitemShut {NoStop}%
\bibitem [{\citenamefont {Browne}\ and\ \citenamefont {Rudolph}(2005)}]{browne2005resource}%
  \BibitemOpen
  \bibfield  {author} {\bibinfo {author} {\bibfnamefont {D.~E.}\ \bibnamefont {Browne}}\ and\ \bibinfo {author} {\bibfnamefont {T.}~\bibnamefont {Rudolph}},\ }\bibfield  {title} {\bibinfo {title} {Resource-efficient linear optical quantum computation},\ }\href@noop {} {\bibfield  {journal} {\bibinfo  {journal} {{Physical Review Letters}}\ }\textbf {\bibinfo {volume} {95}},\ \bibinfo {pages} {010501} (\bibinfo {year} {2005})}\BibitemShut {NoStop}%
\bibitem [{\citenamefont {Zou}\ \emph {et~al.}(2005)\citenamefont {Zou}, \citenamefont {Pahlke},\ and\ \citenamefont {Mathis}}]{zou2005generation}%
  \BibitemOpen
  \bibfield  {author} {\bibinfo {author} {\bibfnamefont {X.}~\bibnamefont {Zou}}, \bibinfo {author} {\bibfnamefont {K.}~\bibnamefont {Pahlke}},\ and\ \bibinfo {author} {\bibfnamefont {W.}~\bibnamefont {Mathis}},\ }\bibfield  {title} {\bibinfo {title} {Generation of a multi-photon {G}reenberger--{H}orne--{Z}eilinger state with linear optical elements and photon detectors},\ }\href@noop {} {\bibfield  {journal} {\bibinfo  {journal} {Journal of Optics B: Quantum and Semiclassical Optics}\ }\textbf {\bibinfo {volume} {7}},\ \bibinfo {pages} {119} (\bibinfo {year} {2005})}\BibitemShut {NoStop}%
\bibitem [{\citenamefont {Zhang}\ \emph {et~al.}(2008)\citenamefont {Zhang}, \citenamefont {Bao}, \citenamefont {Lu}, \citenamefont {Zhou}, \citenamefont {Yang}, \citenamefont {Rudolph},\ and\ \citenamefont {Pan}}]{zhang2008demonstration}%
  \BibitemOpen
  \bibfield  {author} {\bibinfo {author} {\bibfnamefont {Q.}~\bibnamefont {Zhang}}, \bibinfo {author} {\bibfnamefont {X.-H.}\ \bibnamefont {Bao}}, \bibinfo {author} {\bibfnamefont {C.-Y.}\ \bibnamefont {Lu}}, \bibinfo {author} {\bibfnamefont {X.-Q.}\ \bibnamefont {Zhou}}, \bibinfo {author} {\bibfnamefont {T.}~\bibnamefont {Yang}}, \bibinfo {author} {\bibfnamefont {T.}~\bibnamefont {Rudolph}},\ and\ \bibinfo {author} {\bibfnamefont {J.-W.}\ \bibnamefont {Pan}},\ }\bibfield  {title} {\bibinfo {title} {Demonstration of a scheme for the generation of “event-ready” entangled photon pairs from a single-photon source},\ }\href@noop {} {\bibfield  {journal} {\bibinfo  {journal} {Physical Review A}\ }\textbf {\bibinfo {volume} {77}},\ \bibinfo {pages} {062316} (\bibinfo {year} {2008})}\BibitemShut {NoStop}%
\bibitem [{\citenamefont {Varnava}\ \emph {et~al.}(2008)\citenamefont {Varnava}, \citenamefont {Browne},\ and\ \citenamefont {Rudolph}}]{varnava2008good}%
  \BibitemOpen
  \bibfield  {author} {\bibinfo {author} {\bibfnamefont {M.}~\bibnamefont {Varnava}}, \bibinfo {author} {\bibfnamefont {D.~E.}\ \bibnamefont {Browne}},\ and\ \bibinfo {author} {\bibfnamefont {T.}~\bibnamefont {Rudolph}},\ }\bibfield  {title} {\bibinfo {title} {How good must single photon sources and detectors be for efficient linear optical quantum computation?},\ }\href@noop {} {\bibfield  {journal} {\bibinfo  {journal} {{Physical Review Letters}}\ }\textbf {\bibinfo {volume} {100}},\ \bibinfo {pages} {060502} (\bibinfo {year} {2008})}\BibitemShut {NoStop}%
\bibitem [{\citenamefont {Gimeno-Segovia}(2016)}]{gimeno2016towards}%
  \BibitemOpen
  \bibfield  {author} {\bibinfo {author} {\bibfnamefont {M.}~\bibnamefont {Gimeno-Segovia}},\ }\emph {\bibinfo {title} {Towards practical linear optical quantum computing}},\ \href@noop {} {Ph.D. thesis},\ \bibinfo  {school} {Imperial College London} (\bibinfo {year} {2016})\BibitemShut {NoStop}%
\bibitem [{\citenamefont {Gubarev}\ \emph {et~al.}(2020)\citenamefont {Gubarev}, \citenamefont {Dyakonov}, \citenamefont {Saygin}, \citenamefont {Struchalin}, \citenamefont {Straupe},\ and\ \citenamefont {Kulik}}]{gubarev2020improved}%
  \BibitemOpen
  \bibfield  {author} {\bibinfo {author} {\bibfnamefont {F.}~\bibnamefont {Gubarev}}, \bibinfo {author} {\bibfnamefont {I.}~\bibnamefont {Dyakonov}}, \bibinfo {author} {\bibfnamefont {M.~Y.}\ \bibnamefont {Saygin}}, \bibinfo {author} {\bibfnamefont {G.}~\bibnamefont {Struchalin}}, \bibinfo {author} {\bibfnamefont {S.}~\bibnamefont {Straupe}},\ and\ \bibinfo {author} {\bibfnamefont {S.}~\bibnamefont {Kulik}},\ }\bibfield  {title} {\bibinfo {title} {Improved heralded schemes to generate entangled states from single photons},\ }\href@noop {} {\bibfield  {journal} {\bibinfo  {journal} {Physical Review A}\ }\textbf {\bibinfo {volume} {102}},\ \bibinfo {pages} {012604} (\bibinfo {year} {2020})}\BibitemShut {NoStop}%
\bibitem [{\citenamefont {Fldzhyan}\ \emph {et~al.}(2021)\citenamefont {Fldzhyan}, \citenamefont {Saygin},\ and\ \citenamefont {Kulik}}]{fldzhyan2021compact}%
  \BibitemOpen
  \bibfield  {author} {\bibinfo {author} {\bibfnamefont {S.~A.}\ \bibnamefont {Fldzhyan}}, \bibinfo {author} {\bibfnamefont {M.~Y.}\ \bibnamefont {Saygin}},\ and\ \bibinfo {author} {\bibfnamefont {S.~P.}\ \bibnamefont {Kulik}},\ }\bibfield  {title} {\bibinfo {title} {Compact linear optical scheme for {B}ell state generation},\ }\href@noop {} {\bibfield  {journal} {\bibinfo  {journal} {Physical Review Research}\ }\textbf {\bibinfo {volume} {3}},\ \bibinfo {pages} {043031} (\bibinfo {year} {2021})}\BibitemShut {NoStop}%
\bibitem [{\citenamefont {Paesani}\ \emph {et~al.}(2021)\citenamefont {Paesani}, \citenamefont {Bulmer}, \citenamefont {Jones}, \citenamefont {Santagati},\ and\ \citenamefont {Laing}}]{paesani2021scheme}%
  \BibitemOpen
  \bibfield  {author} {\bibinfo {author} {\bibfnamefont {S.}~\bibnamefont {Paesani}}, \bibinfo {author} {\bibfnamefont {J.~F.}\ \bibnamefont {Bulmer}}, \bibinfo {author} {\bibfnamefont {A.~E.}\ \bibnamefont {Jones}}, \bibinfo {author} {\bibfnamefont {R.}~\bibnamefont {Santagati}},\ and\ \bibinfo {author} {\bibfnamefont {A.}~\bibnamefont {Laing}},\ }\bibfield  {title} {\bibinfo {title} {Scheme for universal high-dimensional quantum computation with linear optics},\ }\href@noop {} {\bibfield  {journal} {\bibinfo  {journal} {{Physical Review Letters}}\ }\textbf {\bibinfo {volume} {126}},\ \bibinfo {pages} {230504} (\bibinfo {year} {2021})}\BibitemShut {NoStop}%
\bibitem [{\citenamefont {Ruiz-Gonzalez}\ \emph {et~al.}(2023)\citenamefont {Ruiz-Gonzalez}, \citenamefont {Arlt}, \citenamefont {Petermann}, \citenamefont {Sayyad}, \citenamefont {Jaouni}, \citenamefont {Karimi}, \citenamefont {Tischler}, \citenamefont {Gu},\ and\ \citenamefont {Krenn}}]{ruiz2023digital}%
  \BibitemOpen
  \bibfield  {author} {\bibinfo {author} {\bibfnamefont {C.}~\bibnamefont {Ruiz-Gonzalez}}, \bibinfo {author} {\bibfnamefont {S.}~\bibnamefont {Arlt}}, \bibinfo {author} {\bibfnamefont {J.}~\bibnamefont {Petermann}}, \bibinfo {author} {\bibfnamefont {S.}~\bibnamefont {Sayyad}}, \bibinfo {author} {\bibfnamefont {T.}~\bibnamefont {Jaouni}}, \bibinfo {author} {\bibfnamefont {E.}~\bibnamefont {Karimi}}, \bibinfo {author} {\bibfnamefont {N.}~\bibnamefont {Tischler}}, \bibinfo {author} {\bibfnamefont {X.}~\bibnamefont {Gu}},\ and\ \bibinfo {author} {\bibfnamefont {M.}~\bibnamefont {Krenn}},\ }\bibfield  {title} {\bibinfo {title} {Digital discovery of 100 diverse quantum experiments with {PyTheus}},\ }\href@noop {} {\bibfield  {journal} {\bibinfo  {journal} {Quantum}\ }\textbf {\bibinfo {volume} {7}},\ \bibinfo {pages} {1204} (\bibinfo {year} {2023})}\BibitemShut {NoStop}%
\bibitem [{\citenamefont {Bhatti}\ and\ \citenamefont {Barz}(2025)}]{bhatti2025heralding}%
  \BibitemOpen
  \bibfield  {author} {\bibinfo {author} {\bibfnamefont {D.}~\bibnamefont {Bhatti}}\ and\ \bibinfo {author} {\bibfnamefont {S.}~\bibnamefont {Barz}},\ }\bibfield  {title} {\bibinfo {title} {Heralding higher-dimensional {B}ell and {G}reenberger--{H}orne--{Z}eilinger states using multiport splitters},\ }\href@noop {} {\bibfield  {journal} {\bibinfo  {journal} {New Journal of Physics}\ }\textbf {\bibinfo {volume} {27}},\ \bibinfo {pages} {033006} (\bibinfo {year} {2025})}\BibitemShut {NoStop}%
\bibitem [{\citenamefont {Rohde}\ and\ \citenamefont {Ralph}(2005)}]{rohde2005frequency}%
  \BibitemOpen
  \bibfield  {author} {\bibinfo {author} {\bibfnamefont {P.~P.}\ \bibnamefont {Rohde}}\ and\ \bibinfo {author} {\bibfnamefont {T.~C.}\ \bibnamefont {Ralph}},\ }\bibfield  {title} {\bibinfo {title} {Frequency and temporal effects in linear optical quantum computing},\ }\href@noop {} {\bibfield  {journal} {\bibinfo  {journal} {Physical Review A}\ }\textbf {\bibinfo {volume} {71}},\ \bibinfo {pages} {032320} (\bibinfo {year} {2005})}\BibitemShut {NoStop}%
\bibitem [{\citenamefont {Rohde}\ and\ \citenamefont {Ralph}(2006)}]{rohde2006error}%
  \BibitemOpen
  \bibfield  {author} {\bibinfo {author} {\bibfnamefont {P.~P.}\ \bibnamefont {Rohde}}\ and\ \bibinfo {author} {\bibfnamefont {T.~C.}\ \bibnamefont {Ralph}},\ }\bibfield  {title} {\bibinfo {title} {Error models for mode mismatch in linear optics quantum computing},\ }\href@noop {} {\bibfield  {journal} {\bibinfo  {journal} {Physical Review A}\ }\textbf {\bibinfo {volume} {73}},\ \bibinfo {pages} {062312} (\bibinfo {year} {2006})}\BibitemShut {NoStop}%
\bibitem [{\citenamefont {Barbieri}\ \emph {et~al.}(2009)\citenamefont {Barbieri}, \citenamefont {Weinhold}, \citenamefont {Lanyon}, \citenamefont {Gilchrist}, \citenamefont {Resch}, \citenamefont {Almeida},\ and\ \citenamefont {White}}]{barbieri2009parametric}%
  \BibitemOpen
  \bibfield  {author} {\bibinfo {author} {\bibfnamefont {M.}~\bibnamefont {Barbieri}}, \bibinfo {author} {\bibfnamefont {T.}~\bibnamefont {Weinhold}}, \bibinfo {author} {\bibfnamefont {B.}~\bibnamefont {Lanyon}}, \bibinfo {author} {\bibfnamefont {A.}~\bibnamefont {Gilchrist}}, \bibinfo {author} {\bibfnamefont {K.}~\bibnamefont {Resch}}, \bibinfo {author} {\bibfnamefont {M.}~\bibnamefont {Almeida}},\ and\ \bibinfo {author} {\bibfnamefont {A.}~\bibnamefont {White}},\ }\bibfield  {title} {\bibinfo {title} {Parametric downconversion and optical quantum gates: two's company, four's a crowd},\ }\href@noop {} {\bibfield  {journal} {\bibinfo  {journal} {Journal of Modern Optics}\ }\textbf {\bibinfo {volume} {56}},\ \bibinfo {pages} {209} (\bibinfo {year} {2009})}\BibitemShut {NoStop}%
\bibitem [{\citenamefont {Takeoka}\ \emph {et~al.}(2015)\citenamefont {Takeoka}, \citenamefont {Jin},\ and\ \citenamefont {Sasaki}}]{takeoka2015full}%
  \BibitemOpen
  \bibfield  {author} {\bibinfo {author} {\bibfnamefont {M.}~\bibnamefont {Takeoka}}, \bibinfo {author} {\bibfnamefont {R.-B.}\ \bibnamefont {Jin}},\ and\ \bibinfo {author} {\bibfnamefont {M.}~\bibnamefont {Sasaki}},\ }\bibfield  {title} {\bibinfo {title} {Full analysis of multi-photon pair effects in spontaneous parametric down conversion based photonic quantum information processing},\ }\href@noop {} {\bibfield  {journal} {\bibinfo  {journal} {New Journal of Physics}\ }\textbf {\bibinfo {volume} {17}},\ \bibinfo {pages} {043030} (\bibinfo {year} {2015})}\BibitemShut {NoStop}%
\bibitem [{\citenamefont {Thomas}\ \emph {et~al.}(2021)\citenamefont {Thomas}, \citenamefont {McCutcheon},\ and\ \citenamefont {McCutcheon}}]{thomas2021general}%
  \BibitemOpen
  \bibfield  {author} {\bibinfo {author} {\bibfnamefont {O.~F.}\ \bibnamefont {Thomas}}, \bibinfo {author} {\bibfnamefont {W.}~\bibnamefont {McCutcheon}},\ and\ \bibinfo {author} {\bibfnamefont {D.~P.}\ \bibnamefont {McCutcheon}},\ }\bibfield  {title} {\bibinfo {title} {A general framework for multimode {G}aussian quantum optics and photo-detection: Application to {H}ong--{O}u--{M}andel interference with filtered heralded single photon sources},\ }\href@noop {} {\bibfield  {journal} {\bibinfo  {journal} {APL Photonics}\ }\textbf {\bibinfo {volume} {6}} (\bibinfo {year} {2021})}\BibitemShut {NoStop}%
\bibitem [{\citenamefont {Shaw}\ \emph {et~al.}(2023)\citenamefont {Shaw}, \citenamefont {Jones}, \citenamefont {Yard},\ and\ \citenamefont {Laing}}]{shaw2023errors}%
  \BibitemOpen
  \bibfield  {author} {\bibinfo {author} {\bibfnamefont {R.~D.}\ \bibnamefont {Shaw}}, \bibinfo {author} {\bibfnamefont {A.~E.}\ \bibnamefont {Jones}}, \bibinfo {author} {\bibfnamefont {P.}~\bibnamefont {Yard}},\ and\ \bibinfo {author} {\bibfnamefont {A.}~\bibnamefont {Laing}},\ }\bibfield  {title} {\bibinfo {title} {Errors in heralded circuits for linear optical entanglement generation},\ }\href@noop {} {\bibfield  {journal} {\bibinfo  {journal} {arXiv preprint arXiv:2305.08452}\ } (\bibinfo {year} {2023})}\BibitemShut {NoStop}%
\bibitem [{\citenamefont {Wiesner}\ \emph {et~al.}(2024)\citenamefont {Wiesner}, \citenamefont {Chrzanowski}, \citenamefont {Pieplow}, \citenamefont {Schr{\"o}der}, \citenamefont {Pappa},\ and\ \citenamefont {Wolters}}]{wiesner2024influence}%
  \BibitemOpen
  \bibfield  {author} {\bibinfo {author} {\bibfnamefont {F.}~\bibnamefont {Wiesner}}, \bibinfo {author} {\bibfnamefont {H.~M.}\ \bibnamefont {Chrzanowski}}, \bibinfo {author} {\bibfnamefont {G.}~\bibnamefont {Pieplow}}, \bibinfo {author} {\bibfnamefont {T.}~\bibnamefont {Schr{\"o}der}}, \bibinfo {author} {\bibfnamefont {A.}~\bibnamefont {Pappa}},\ and\ \bibinfo {author} {\bibfnamefont {J.}~\bibnamefont {Wolters}},\ }\bibfield  {title} {\bibinfo {title} {The influence of experimental imperfections on photonic {GHZ} state generation},\ }\href@noop {} {\bibfield  {journal} {\bibinfo  {journal} {New Journal of Physics}\ }\textbf {\bibinfo {volume} {26}},\ \bibinfo {pages} {113021} (\bibinfo {year} {2024})}\BibitemShut {NoStop}%
\bibitem [{\citenamefont {Steinmetz}\ \emph {et~al.}(2024)\citenamefont {Steinmetz}, \citenamefont {Ostmann}, \citenamefont {Neville}, \citenamefont {Pankovich},\ and\ \citenamefont {Sohbi}}]{steinmetz2024simulating}%
  \BibitemOpen
  \bibfield  {author} {\bibinfo {author} {\bibfnamefont {J.}~\bibnamefont {Steinmetz}}, \bibinfo {author} {\bibfnamefont {M.}~\bibnamefont {Ostmann}}, \bibinfo {author} {\bibfnamefont {A.}~\bibnamefont {Neville}}, \bibinfo {author} {\bibfnamefont {B.}~\bibnamefont {Pankovich}},\ and\ \bibinfo {author} {\bibfnamefont {A.}~\bibnamefont {Sohbi}},\ }\bibfield  {title} {\bibinfo {title} {Simulating imperfect quantum optical circuits using unsymmetrized bases},\ }\href@noop {} {\bibfield  {journal} {\bibinfo  {journal} {arXiv preprint arXiv:2412.13330}\ } (\bibinfo {year} {2024})}\BibitemShut {NoStop}%
\bibitem [{\citenamefont {Randles}\ \emph {et~al.}(2026)\citenamefont {Randles}, \citenamefont {Mu{\~n}oz-Arias},\ and\ \citenamefont {Sarovar}}]{randles2026limits}%
  \BibitemOpen
  \bibfield  {author} {\bibinfo {author} {\bibfnamefont {K.~J.}\ \bibnamefont {Randles}}, \bibinfo {author} {\bibfnamefont {M.~H.}\ \bibnamefont {Mu{\~n}oz-Arias}},\ and\ \bibinfo {author} {\bibfnamefont {M.}~\bibnamefont {Sarovar}},\ }\bibfield  {title} {\bibinfo {title} {Limits of heralded photonic bell-state generation in the presence of loss},\ }\href@noop {} {\bibfield  {journal} {\bibinfo  {journal} {arXiv preprint arXiv:2608.01549}\ } (\bibinfo {year} {2026})}\BibitemShut {NoStop}%
\bibitem [{\citenamefont {Pittman}\ \emph {et~al.}(2002)\citenamefont {Pittman}, \citenamefont {Jacobs},\ and\ \citenamefont {Franson}}]{pittman2002single}%
  \BibitemOpen
  \bibfield  {author} {\bibinfo {author} {\bibfnamefont {T.}~\bibnamefont {Pittman}}, \bibinfo {author} {\bibfnamefont {B.}~\bibnamefont {Jacobs}},\ and\ \bibinfo {author} {\bibfnamefont {J.}~\bibnamefont {Franson}},\ }\bibfield  {title} {\bibinfo {title} {Single photons on pseudodemand from stored parametric down-conversion},\ }\href@noop {} {\bibfield  {journal} {\bibinfo  {journal} {Physical Review A}\ }\textbf {\bibinfo {volume} {66}},\ \bibinfo {pages} {042303} (\bibinfo {year} {2002})}\BibitemShut {NoStop}%
\bibitem [{\citenamefont {Migdall}\ \emph {et~al.}(2002)\citenamefont {Migdall}, \citenamefont {Branning},\ and\ \citenamefont {Castelletto}}]{migdall2002tailoring}%
  \BibitemOpen
  \bibfield  {author} {\bibinfo {author} {\bibfnamefont {A.}~\bibnamefont {Migdall}}, \bibinfo {author} {\bibfnamefont {D.}~\bibnamefont {Branning}},\ and\ \bibinfo {author} {\bibfnamefont {S.}~\bibnamefont {Castelletto}},\ }\bibfield  {title} {\bibinfo {title} {Tailoring single-photon and multiphoton probabilities of a single-photon on-demand source},\ }\href@noop {} {\bibfield  {journal} {\bibinfo  {journal} {Physical Review A}\ }\textbf {\bibinfo {volume} {66}},\ \bibinfo {pages} {053805} (\bibinfo {year} {2002})}\BibitemShut {NoStop}%
\bibitem [{\citenamefont {Bartolucci}\ \emph {et~al.}(2021{\natexlab{a}})\citenamefont {Bartolucci}, \citenamefont {Birchall}, \citenamefont {Bonneau}, \citenamefont {Cable}, \citenamefont {Gimeno-Segovia}, \citenamefont {Kieling}, \citenamefont {Nickerson}, \citenamefont {Rudolph},\ and\ \citenamefont {Sparrow}}]{bartolucci2021switch}%
  \BibitemOpen
  \bibfield  {author} {\bibinfo {author} {\bibfnamefont {S.}~\bibnamefont {Bartolucci}}, \bibinfo {author} {\bibfnamefont {P.}~\bibnamefont {Birchall}}, \bibinfo {author} {\bibfnamefont {D.}~\bibnamefont {Bonneau}}, \bibinfo {author} {\bibfnamefont {H.}~\bibnamefont {Cable}}, \bibinfo {author} {\bibfnamefont {M.}~\bibnamefont {Gimeno-Segovia}}, \bibinfo {author} {\bibfnamefont {K.}~\bibnamefont {Kieling}}, \bibinfo {author} {\bibfnamefont {N.}~\bibnamefont {Nickerson}}, \bibinfo {author} {\bibfnamefont {T.}~\bibnamefont {Rudolph}},\ and\ \bibinfo {author} {\bibfnamefont {C.}~\bibnamefont {Sparrow}},\ }\bibfield  {title} {\bibinfo {title} {Switch networks for photonic fusion-based quantum computing},\ }\href@noop {} {\bibfield  {journal} {\bibinfo  {journal} {arXiv preprint arXiv:2109.13760}\ } (\bibinfo {year} {2021}{\natexlab{a}})}\BibitemShut {NoStop}%
\bibitem [{\citenamefont {Karimi}\ \emph {et~al.}(2026)\citenamefont {Karimi}, \citenamefont {Mister}, \citenamefont {Simon},\ and\ \citenamefont {Wein}}]{karimi2026characterization}%
  \BibitemOpen
  \bibfield  {author} {\bibinfo {author} {\bibfnamefont {M.}~\bibnamefont {Karimi}}, \bibinfo {author} {\bibfnamefont {S.}~\bibnamefont {Mister}}, \bibinfo {author} {\bibfnamefont {C.}~\bibnamefont {Simon}},\ and\ \bibinfo {author} {\bibfnamefont {S.~C.}\ \bibnamefont {Wein}},\ }\bibfield  {title} {\bibinfo {title} {Characterization of errors in photon-heralded quantum operations between non-interacting quantum emitters},\ }\href@noop {} {\bibfield  {journal} {\bibinfo  {journal} {arXiv preprint arXiv:2606.04312}\ } (\bibinfo {year} {2026})}\BibitemShut {NoStop}%
\bibitem [{\citenamefont {Loudon}(1973)}]{loudon2000quantum}%
  \BibitemOpen
  \bibfield  {author} {\bibinfo {author} {\bibfnamefont {R.}~\bibnamefont {Loudon}},\ }\href@noop {} {\emph {\bibinfo {title} {The quantum theory of light}}},\ \bibinfo {edition} {3rd}\ ed.\ (\bibinfo  {publisher} {Clarendon Press, Oxford},\ \bibinfo {year} {1973})\BibitemShut {NoStop}%
\bibitem [{\citenamefont {Knill}(2005)}]{knill2005quantum}%
  \BibitemOpen
  \bibfield  {author} {\bibinfo {author} {\bibfnamefont {E.}~\bibnamefont {Knill}},\ }\bibfield  {title} {\bibinfo {title} {Quantum computing with realistically noisy devices},\ }\href@noop {} {\bibfield  {journal} {\bibinfo  {journal} {Nature}\ }\textbf {\bibinfo {volume} {434}},\ \bibinfo {pages} {39} (\bibinfo {year} {2005})}\BibitemShut {NoStop}%
\bibitem [{\citenamefont {Somhorst}\ and\ \citenamefont {Renema}(2026)}]{somhorst2026mitigating}%
  \BibitemOpen
  \bibfield  {author} {\bibinfo {author} {\bibfnamefont {F.}~\bibnamefont {Somhorst}}\ and\ \bibinfo {author} {\bibfnamefont {J.~J.}\ \bibnamefont {Renema}},\ }\bibfield  {title} {\bibinfo {title} {Mitigating quantum operation infidelity through engineering the distribution of photon losses},\ }\href@noop {} {\bibfield  {journal} {\bibinfo  {journal} {Optica Quantum}\ }\textbf {\bibinfo {volume} {4}},\ \bibinfo {pages} {130} (\bibinfo {year} {2026})}\BibitemShut {NoStop}%
\bibitem [{\citenamefont {Alexander}\ \emph {et~al.}(2025)\citenamefont {Alexander}, \citenamefont {Bahgat}, \citenamefont {Benyamini}, \citenamefont {Black}, \citenamefont {Bonneau}, \citenamefont {Burgos}, \citenamefont {Burridge}, \citenamefont {Campbell}, \citenamefont {Catalano}, \citenamefont {Ceballos} \emph {et~al.}}]{alexander2024manufacturable}%
  \BibitemOpen
  \bibfield  {author} {\bibinfo {author} {\bibfnamefont {K.}~\bibnamefont {Alexander}}, \bibinfo {author} {\bibfnamefont {A.}~\bibnamefont {Bahgat}}, \bibinfo {author} {\bibfnamefont {A.}~\bibnamefont {Benyamini}}, \bibinfo {author} {\bibfnamefont {D.}~\bibnamefont {Black}}, \bibinfo {author} {\bibfnamefont {D.}~\bibnamefont {Bonneau}}, \bibinfo {author} {\bibfnamefont {S.}~\bibnamefont {Burgos}}, \bibinfo {author} {\bibfnamefont {B.}~\bibnamefont {Burridge}}, \bibinfo {author} {\bibfnamefont {G.}~\bibnamefont {Campbell}}, \bibinfo {author} {\bibfnamefont {G.}~\bibnamefont {Catalano}}, \bibinfo {author} {\bibfnamefont {A.}~\bibnamefont {Ceballos}}, \emph {et~al.},\ }\bibfield  {title} {\bibinfo {title} {A manufacturable platform for photonic quantum computing},\ }\href@noop {} {\bibfield  {journal} {\bibinfo  {journal} {Nature}\ } (\bibinfo {year} {2025})}\BibitemShut {NoStop}%
\bibitem [{\citenamefont {Geller}\ and\ \citenamefont {Knill}(2026)}]{geller2026measuring}%
  \BibitemOpen
  \bibfield  {author} {\bibinfo {author} {\bibfnamefont {S.}~\bibnamefont {Geller}}\ and\ \bibinfo {author} {\bibfnamefont {E.}~\bibnamefont {Knill}},\ }\bibfield  {title} {\bibinfo {title} {Measuring multiparticle indistinguishability with the generalized bunching probability},\ }\href@noop {} {\bibfield  {journal} {\bibinfo  {journal} {Physical Review A}\ }\textbf {\bibinfo {volume} {113}},\ \bibinfo {pages} {042606} (\bibinfo {year} {2026})}\BibitemShut {NoStop}%
\bibitem [{\citenamefont {Berry}\ and\ \citenamefont {Lvovsky}(2011)}]{berry2011preservation}%
  \BibitemOpen
  \bibfield  {author} {\bibinfo {author} {\bibfnamefont {D.~W.}\ \bibnamefont {Berry}}\ and\ \bibinfo {author} {\bibfnamefont {A.}~\bibnamefont {Lvovsky}},\ }\bibfield  {title} {\bibinfo {title} {Preservation of loss in linear-optical processing},\ }\href@noop {} {\bibfield  {journal} {\bibinfo  {journal} {Physical Review A—Atomic, Molecular, and Optical Physics}\ }\textbf {\bibinfo {volume} {84}},\ \bibinfo {pages} {042304} (\bibinfo {year} {2011})}\BibitemShut {NoStop}%
\bibitem [{\citenamefont {Dittel}\ \emph {et~al.}(2018{\natexlab{a}})\citenamefont {Dittel}, \citenamefont {Dufour}, \citenamefont {Walschaers}, \citenamefont {Weihs}, \citenamefont {Buchleitner},\ and\ \citenamefont {Keil}}]{dittel2018totally}%
  \BibitemOpen
  \bibfield  {author} {\bibinfo {author} {\bibfnamefont {C.}~\bibnamefont {Dittel}}, \bibinfo {author} {\bibfnamefont {G.}~\bibnamefont {Dufour}}, \bibinfo {author} {\bibfnamefont {M.}~\bibnamefont {Walschaers}}, \bibinfo {author} {\bibfnamefont {G.}~\bibnamefont {Weihs}}, \bibinfo {author} {\bibfnamefont {A.}~\bibnamefont {Buchleitner}},\ and\ \bibinfo {author} {\bibfnamefont {R.}~\bibnamefont {Keil}},\ }\bibfield  {title} {\bibinfo {title} {Totally destructive many-particle interference},\ }\href@noop {} {\bibfield  {journal} {\bibinfo  {journal} {{Physical Review Letters}}\ }\textbf {\bibinfo {volume} {120}},\ \bibinfo {pages} {240404} (\bibinfo {year} {2018}{\natexlab{a}})}\BibitemShut {NoStop}%
\bibitem [{\citenamefont {Dittel}\ \emph {et~al.}(2018{\natexlab{b}})\citenamefont {Dittel}, \citenamefont {Dufour}, \citenamefont {Walschaers}, \citenamefont {Weihs}, \citenamefont {Buchleitner},\ and\ \citenamefont {Keil}}]{dittel2018totally2}%
  \BibitemOpen
  \bibfield  {author} {\bibinfo {author} {\bibfnamefont {C.}~\bibnamefont {Dittel}}, \bibinfo {author} {\bibfnamefont {G.}~\bibnamefont {Dufour}}, \bibinfo {author} {\bibfnamefont {M.}~\bibnamefont {Walschaers}}, \bibinfo {author} {\bibfnamefont {G.}~\bibnamefont {Weihs}}, \bibinfo {author} {\bibfnamefont {A.}~\bibnamefont {Buchleitner}},\ and\ \bibinfo {author} {\bibfnamefont {R.}~\bibnamefont {Keil}},\ }\bibfield  {title} {\bibinfo {title} {Totally destructive interference for permutation-symmetric many-particle states},\ }\href@noop {} {\bibfield  {journal} {\bibinfo  {journal} {Physical Review A}\ }\textbf {\bibinfo {volume} {97}},\ \bibinfo {pages} {062116} (\bibinfo {year} {2018}{\natexlab{b}})}\BibitemShut {NoStop}%
\bibitem [{\citenamefont {Bartolucci}\ \emph {et~al.}(2021{\natexlab{b}})\citenamefont {Bartolucci}, \citenamefont {Birchall}, \citenamefont {Gimeno-Segovia}, \citenamefont {Johnston}, \citenamefont {Kieling}, \citenamefont {Pant}, \citenamefont {Rudolph}, \citenamefont {Smith}, \citenamefont {Sparrow},\ and\ \citenamefont {Vidrighin}}]{bartolucci2021creation}%
  \BibitemOpen
  \bibfield  {author} {\bibinfo {author} {\bibfnamefont {S.}~\bibnamefont {Bartolucci}}, \bibinfo {author} {\bibfnamefont {P.~M.}\ \bibnamefont {Birchall}}, \bibinfo {author} {\bibfnamefont {M.}~\bibnamefont {Gimeno-Segovia}}, \bibinfo {author} {\bibfnamefont {E.}~\bibnamefont {Johnston}}, \bibinfo {author} {\bibfnamefont {K.}~\bibnamefont {Kieling}}, \bibinfo {author} {\bibfnamefont {M.}~\bibnamefont {Pant}}, \bibinfo {author} {\bibfnamefont {T.}~\bibnamefont {Rudolph}}, \bibinfo {author} {\bibfnamefont {J.}~\bibnamefont {Smith}}, \bibinfo {author} {\bibfnamefont {C.}~\bibnamefont {Sparrow}},\ and\ \bibinfo {author} {\bibfnamefont {M.~D.}\ \bibnamefont {Vidrighin}},\ }\bibfield  {title} {\bibinfo {title} {Creation of entangled photonic states using linear optics},\ }\href@noop {} {\bibfield  {journal} {\bibinfo  {journal} {arXiv preprint arXiv:2106.13825}\ } (\bibinfo {year} {2021}{\natexlab{b}})}\BibitemShut {NoStop}%
\bibitem [{\citenamefont {Tichy}\ \emph {et~al.}(2010)\citenamefont {Tichy}, \citenamefont {Tiersch}, \citenamefont {de~Melo}, \citenamefont {Mintert},\ and\ \citenamefont {Buchleitner}}]{tichy2010zero}%
  \BibitemOpen
  \bibfield  {author} {\bibinfo {author} {\bibfnamefont {M.~C.}\ \bibnamefont {Tichy}}, \bibinfo {author} {\bibfnamefont {M.}~\bibnamefont {Tiersch}}, \bibinfo {author} {\bibfnamefont {F.}~\bibnamefont {de~Melo}}, \bibinfo {author} {\bibfnamefont {F.}~\bibnamefont {Mintert}},\ and\ \bibinfo {author} {\bibfnamefont {A.}~\bibnamefont {Buchleitner}},\ }\bibfield  {title} {\bibinfo {title} {Zero-transmission law for multiport beam splitters},\ }\href@noop {} {\bibfield  {journal} {\bibinfo  {journal} {{Physical Review Letters}}\ }\textbf {\bibinfo {volume} {104}},\ \bibinfo {pages} {220405} (\bibinfo {year} {2010})}\BibitemShut {NoStop}%
\bibitem [{\citenamefont {Bruzda}\ \emph {et~al.}(2006)\citenamefont {Bruzda}, \citenamefont {Tadej},\ and\ \citenamefont {{\.Z}yczkowski}}]{bruzda2006complex}%
  \BibitemOpen
  \bibfield  {author} {\bibinfo {author} {\bibfnamefont {W.}~\bibnamefont {Bruzda}}, \bibinfo {author} {\bibfnamefont {W.}~\bibnamefont {Tadej}},\ and\ \bibinfo {author} {\bibfnamefont {K.}~\bibnamefont {{\.Z}yczkowski}},\ }\bibfield  {title} {\bibinfo {title} {Complex hadamard matrices - a catalog (since 2006)},\ }\href@noop {} {\bibfield  {journal} {\bibinfo  {journal} {\url{https://chaos.if.uj.edu.pl/~karol/hadamard/}}\ } (\bibinfo {year} {2006})}\BibitemShut {NoStop}%
\bibitem [{\citenamefont {Pankovich}\ \emph {et~al.}(2024)\citenamefont {Pankovich}, \citenamefont {Neville}, \citenamefont {Kan}, \citenamefont {Omkar}, \citenamefont {Wan},\ and\ \citenamefont {Br{\'a}dler}}]{pankovich2024flexible}%
  \BibitemOpen
  \bibfield  {author} {\bibinfo {author} {\bibfnamefont {B.}~\bibnamefont {Pankovich}}, \bibinfo {author} {\bibfnamefont {A.}~\bibnamefont {Neville}}, \bibinfo {author} {\bibfnamefont {A.}~\bibnamefont {Kan}}, \bibinfo {author} {\bibfnamefont {S.}~\bibnamefont {Omkar}}, \bibinfo {author} {\bibfnamefont {K.~H.}\ \bibnamefont {Wan}},\ and\ \bibinfo {author} {\bibfnamefont {K.}~\bibnamefont {Br{\'a}dler}},\ }\bibfield  {title} {\bibinfo {title} {Flexible entangled-state generation in linear optics},\ }\href@noop {} {\bibfield  {journal} {\bibinfo  {journal} {Physical Review A}\ }\textbf {\bibinfo {volume} {110}},\ \bibinfo {pages} {032402} (\bibinfo {year} {2024})}\BibitemShut {NoStop}%
\bibitem [{\citenamefont {Chuang}\ \emph {et~al.}(1997)\citenamefont {Chuang}, \citenamefont {Leung},\ and\ \citenamefont {Yamamoto}}]{chuang1997bosonic}%
  \BibitemOpen
  \bibfield  {author} {\bibinfo {author} {\bibfnamefont {I.~L.}\ \bibnamefont {Chuang}}, \bibinfo {author} {\bibfnamefont {D.~W.}\ \bibnamefont {Leung}},\ and\ \bibinfo {author} {\bibfnamefont {Y.}~\bibnamefont {Yamamoto}},\ }\bibfield  {title} {\bibinfo {title} {Bosonic quantum codes for amplitude damping},\ }\href@noop {} {\bibfield  {journal} {\bibinfo  {journal} {Physical Review A}\ }\textbf {\bibinfo {volume} {56}},\ \bibinfo {pages} {1114} (\bibinfo {year} {1997})}\BibitemShut {NoStop}%
\bibitem [{\citenamefont {Wasilewski}\ and\ \citenamefont {Banaszek}(2007)}]{wasilewski2007protecting}%
  \BibitemOpen
  \bibfield  {author} {\bibinfo {author} {\bibfnamefont {W.}~\bibnamefont {Wasilewski}}\ and\ \bibinfo {author} {\bibfnamefont {K.}~\bibnamefont {Banaszek}},\ }\bibfield  {title} {\bibinfo {title} {Protecting an optical qubit against photon loss},\ }\href@noop {} {\bibfield  {journal} {\bibinfo  {journal} {Physical Review A}\ }\textbf {\bibinfo {volume} {75}},\ \bibinfo {pages} {042316} (\bibinfo {year} {2007})}\BibitemShut {NoStop}%
\bibitem [{\citenamefont {Adamson}\ \emph {et~al.}(2008)\citenamefont {Adamson}, \citenamefont {Turner}, \citenamefont {Mitchell},\ and\ \citenamefont {Steinberg}}]{adamson2008detecting}%
  \BibitemOpen
  \bibfield  {author} {\bibinfo {author} {\bibfnamefont {R.}~\bibnamefont {Adamson}}, \bibinfo {author} {\bibfnamefont {P.}~\bibnamefont {Turner}}, \bibinfo {author} {\bibfnamefont {M.}~\bibnamefont {Mitchell}},\ and\ \bibinfo {author} {\bibfnamefont {A.}~\bibnamefont {Steinberg}},\ }\bibfield  {title} {\bibinfo {title} {Detecting hidden differences via permutation symmetries},\ }\href@noop {} {\bibfield  {journal} {\bibinfo  {journal} {Physical Review A—Atomic, Molecular, and Optical Physics}\ }\textbf {\bibinfo {volume} {78}},\ \bibinfo {pages} {033832} (\bibinfo {year} {2008})}\BibitemShut {NoStop}%
\bibitem [{\citenamefont {Czab\'{a}n}\ \emph {et~al.}(2024)\citenamefont {Czab\'{a}n}, \citenamefont {Kolarovszki}, \citenamefont {Kar\'{a}csony},\ and\ \citenamefont {Zimbor\'{a}s}}]{czaban2024suppressing}%
  \BibitemOpen
  \bibfield  {author} {\bibinfo {author} {\bibfnamefont {C.}~\bibnamefont {Czab\'{a}n}}, \bibinfo {author} {\bibfnamefont {Z.}~\bibnamefont {Kolarovszki}}, \bibinfo {author} {\bibfnamefont {M.}~\bibnamefont {Kar\'{a}csony}},\ and\ \bibinfo {author} {\bibfnamefont {Z.}~\bibnamefont {Zimbor\'{a}s}},\ }\bibfield  {title} {\bibinfo {title} {Suppressing photon detection errors in nondeterministic state preparation},\ }in\ \href {https://doi.org/10.1145/3665870.3665873} {\emph {\bibinfo {booktitle} {Proceedings of Recent Advances in Quantum Computing and Technology}}},\ \bibinfo {series and number} {ReAQCT '24}\ (\bibinfo  {publisher} {Association for Computing Machinery},\ \bibinfo {address} {New York, NY, USA},\ \bibinfo {year} {2024})\ p.\ \bibinfo {pages} {7–11}\BibitemShut {NoStop}%
\bibitem [{\citenamefont {Aralov}\ \emph {et~al.}(2026)\citenamefont {Aralov}, \citenamefont {Gillet}, \citenamefont {Nguyen}, \citenamefont {Cosentino}, \citenamefont {Walschaers},\ and\ \citenamefont {Frigerio}}]{aralov2026photon}%
  \BibitemOpen
  \bibfield  {author} {\bibinfo {author} {\bibfnamefont {A.}~\bibnamefont {Aralov}}, \bibinfo {author} {\bibfnamefont {{\'E}.}~\bibnamefont {Gillet}}, \bibinfo {author} {\bibfnamefont {V.}~\bibnamefont {Nguyen}}, \bibinfo {author} {\bibfnamefont {A.}~\bibnamefont {Cosentino}}, \bibinfo {author} {\bibfnamefont {M.}~\bibnamefont {Walschaers}},\ and\ \bibinfo {author} {\bibfnamefont {M.}~\bibnamefont {Frigerio}},\ }\bibfield  {title} {\bibinfo {title} {Photon catalysis for general multimode multi-photon quantum state preparation},\ }\href@noop {} {\bibfield  {journal} {\bibinfo  {journal} {PRX Quantum}\ }\textbf {\bibinfo {volume} {7}},\ \bibinfo {pages} {020323} (\bibinfo {year} {2026})}\BibitemShut {NoStop}%
\bibitem [{\citenamefont {Chin}\ \emph {et~al.}(2024)\citenamefont {Chin}, \citenamefont {Ryu},\ and\ \citenamefont {Kim}}]{chin2024exponentially}%
  \BibitemOpen
  \bibfield  {author} {\bibinfo {author} {\bibfnamefont {S.}~\bibnamefont {Chin}}, \bibinfo {author} {\bibfnamefont {J.}~\bibnamefont {Ryu}},\ and\ \bibinfo {author} {\bibfnamefont {Y.-S.}\ \bibnamefont {Kim}},\ }\bibfield  {title} {\bibinfo {title} {Exponentially enhanced scheme for the heralded qudit {G}reenberger-{H}orne-{Z}eilinger state in linear optics},\ }\href@noop {} {\bibfield  {journal} {\bibinfo  {journal} {{Physical Review Letters}}\ }\textbf {\bibinfo {volume} {133}},\ \bibinfo {pages} {253601} (\bibinfo {year} {2024})}\BibitemShut {NoStop}%
\bibitem [{\citenamefont {Hurwitz}(1897)}]{hurwitz1897erzeugung}%
  \BibitemOpen
  \bibfield  {author} {\bibinfo {author} {\bibfnamefont {A.}~\bibnamefont {Hurwitz}},\ }\bibfield  {title} {\bibinfo {title} {{\"U}ber die erzeugung der invarianten durch integration},\ }\href@noop {} {\bibfield  {journal} {\bibinfo  {journal} {Nachrichten von der Gesellschaft der Wissenschaften zu G{\"o}ttingen, Mathematisch-Physikalische Klasse}\ }\textbf {\bibinfo {volume} {1897}} (\bibinfo {year} {1897})}\BibitemShut {NoStop}%
\bibitem [{\citenamefont {Diaconis}\ and\ \citenamefont {Forrester}(2017)}]{diaconis2017hurwitz}%
  \BibitemOpen
  \bibfield  {author} {\bibinfo {author} {\bibfnamefont {P.}~\bibnamefont {Diaconis}}\ and\ \bibinfo {author} {\bibfnamefont {P.~J.}\ \bibnamefont {Forrester}},\ }\bibfield  {title} {\bibinfo {title} {Hurwitz and the origins of random matrix theory in mathematics},\ }\href@noop {} {\bibfield  {journal} {\bibinfo  {journal} {Random Matrices: Theory and Applications}\ }\textbf {\bibinfo {volume} {6}},\ \bibinfo {pages} {1730001} (\bibinfo {year} {2017})}\BibitemShut {NoStop}%
\bibitem [{\citenamefont {Reck}\ \emph {et~al.}(1994)\citenamefont {Reck}, \citenamefont {Zeilinger}, \citenamefont {Bernstein},\ and\ \citenamefont {Bertani}}]{reck1994experimental}%
  \BibitemOpen
  \bibfield  {author} {\bibinfo {author} {\bibfnamefont {M.}~\bibnamefont {Reck}}, \bibinfo {author} {\bibfnamefont {A.}~\bibnamefont {Zeilinger}}, \bibinfo {author} {\bibfnamefont {H.~J.}\ \bibnamefont {Bernstein}},\ and\ \bibinfo {author} {\bibfnamefont {P.}~\bibnamefont {Bertani}},\ }\bibfield  {title} {\bibinfo {title} {Experimental realization of any discrete unitary operator},\ }\href@noop {} {\bibfield  {journal} {\bibinfo  {journal} {{Physical Review Letters}}\ }\textbf {\bibinfo {volume} {73}},\ \bibinfo {pages} {58} (\bibinfo {year} {1994})}\BibitemShut {NoStop}%
\end{thebibliography}%
\appendix

\begin{table*}[t]
    \centering
    \begin{tabular}{l | l }
    \textbf{Symbol} & \textbf{Meaning} \\ \hline
    $n_\inp$ & Number of input single photons \\
    $m$ & Number of modes \\
    $n$ & Number of output photons / qubits \\
    $a^\dagger$ & Bosonic creation operator \\ 
    $U$ & Transfer matrix of the interferometer  \\
    $\mathcal{U}$ & Fock space unitary generated by $U$ (eq. \ref{eq:mathcalU}) \\ 
    \hline 
    $n_\detc$ & number of detected photons \\
    $m_\detc$ & number of modes which go to detectors \\
    $m_\out$ & number of modes in the output state \\
    $\bm{p}$ & Successful detection pattern \\ 
 \hline 
    $\ket{\psi_{\inp}}$ & Non-vacuum part of the ideal input state (eq. \ref{eq:psi_inp}) \\ 
    $\ket{\psi_{\inter}}$ & Ideal state after the interferometer (eq. \ref{eq:psi_int}) \\
    $\ket{\psi_{\out}^{(\bm{p})}}$ & Ideal output state (eq. \ref{eq:psi_out}) \\ 
    $\Pi^{(\bm{p})}_\Id$ & Ideal measurement operator (eq. \ref{eq:DI}) \\ 
    $\succ$ & Success via any possible successful detection pattern \\ \hline 
    $\rho_\inp$ & Non-vacuum part of the noisy input state (eq. \ref{eq:rho_inp})\\ 
    $\rho_\inter$ & Noisy intermediate state (eq. \ref{eq:rho_int}) \\ 
    $\rho_\out^{(\bm{p})}$ & Noisy output state (eq. \ref{eq:rho_h_p}) \\
    $\Pi^{(\bm{p})}$ & Noisy measurement operator (eq. \ref{eq:noisy_measurement})\\ \hline 
    $\Pr$ & Probability \\
    $Q_i$ & Output subspace qubit label \\ 
    $\bm{Q}$ & Output subspace label tuple \\ 
    $\Id$ & Ideal operation \\ 
    $\text{E}$ & Error mechanism \\
    $\mathcal{E}_\inp$ & Set of input error mechanisms \\
    $\mathcal{E}_\detc$ & Set of detection error mechanisms \\
    $\mathcal{E}$ & Set of all error mechanisms \\ \hline 
    $\Lambda_{\text{E},i}$ & Channel for error mechanism $\text{E}$ on mode $i$ \\
    $\text{L}$ & Loss error mechanism \\
    $\text{A}$ & Coherent photon addition error mechanism \\
    $\tilde{\text{A}}$ & Incoherent photon addition error mechanism \\
    $\text{D}$ & Distinguishability error mechanism \\ \hline
    $\Hil$ & Hilbert space \\ 
    $\Hil_\text{dr}$ & Dual-rail qubit subspace \\
    $\Hil_\sph$ & Qubit subspace with a single photon \\ 
    $\ideal$ & Ideal subspace \\
    $V$ & Vacuum subspace \\
    $M$ & Multiphoton subspace \\ 
    $B$ & Bad photon subspace \\  
    $P$ & Pauli subspace $(X,Y,Z)$ \\
    $BP$ & Bad photon + Pauli subspace $(BX, BY, BZ)$\\  \hline 
    $\nu_\text{E}$ & Scaling coefficient for the first-order contribution of error mechanism $\text{E}$ \\
    $c_{\text{E} \to \bm{Q}}^{(\bm{p})}$ & Response coefficient relating error $\text{E}$ to subspace tuple $\bm{Q}$ (eq. \ref{error_coeff})\\ 
    $c_{\text{E} \to \bm{Q}}$ & Averaged response coefficient relating error $\text{E}$ to subspace tuple $\bm{Q}$ (eq. \ref{avg_error_coeff})
    \end{tabular}
    \caption{
        Summary of symbols used in this work and their meanings.
        See the main text for more explicit definitions. 
        \label{tab:notation}
    }
\end{table*}

\begin{figure}
    \centering 
    \includegraphics[width=1\columnwidth]{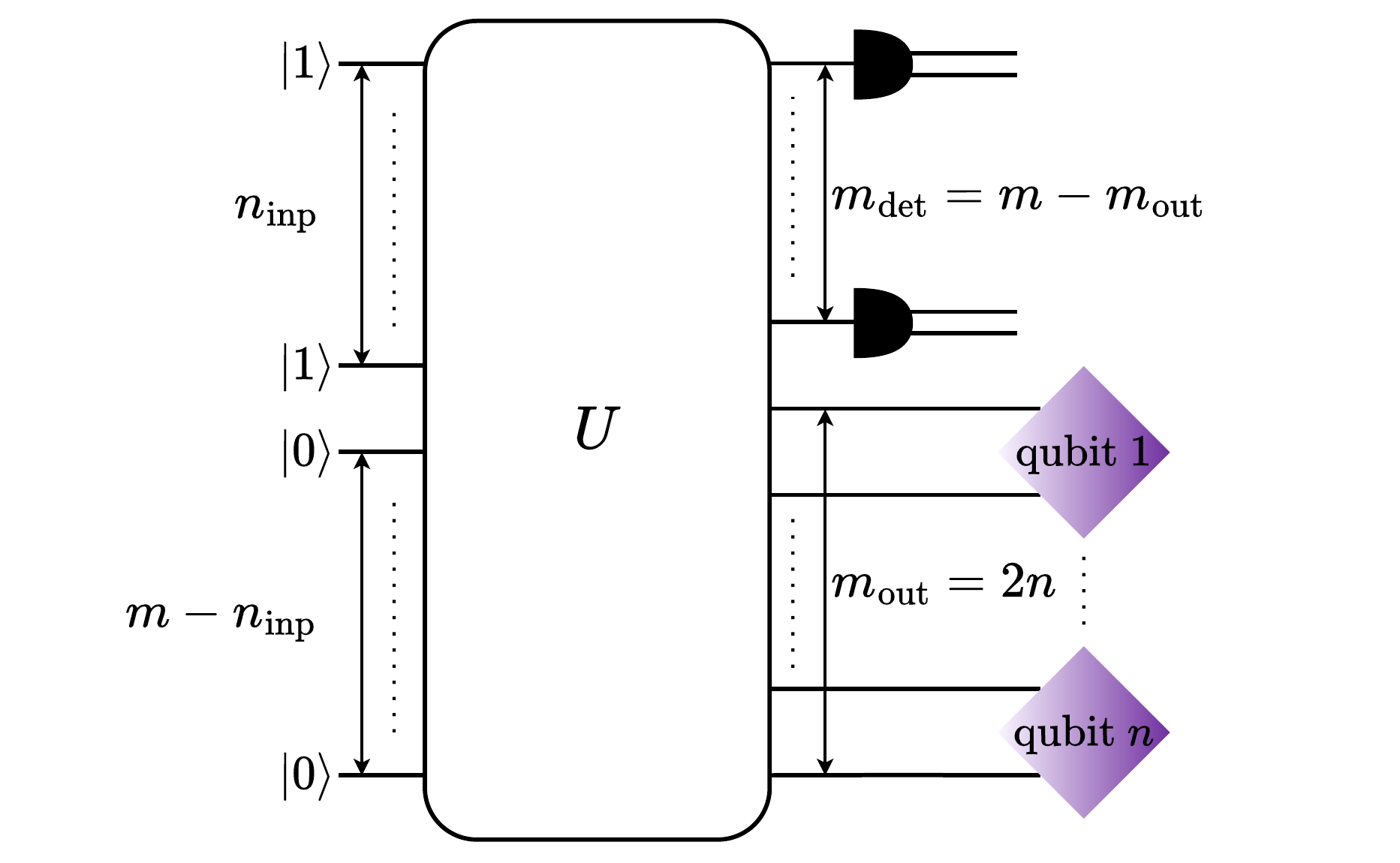}
    \caption{
        Schematic for a general ballistic seed-state generation circuit, with $m$ modes, $n_\inp$ single photons, $m_\detc$ photon number resolving detectors and $\mout$ output modes which support $n=\mout/2$ dual-rail qubits.
        We use $\ket{0},\ket{1}$ to denote the input Fock states but, for simplicity, later on we will just use the yellow circles to show modes which are to have single photon inputs.
    \label{fig:general_ssg}}
\end{figure}

\section{Worked Example \label[appendix]{app:example}}
This appendix section walks through the method of calculating response coefficients outlined in the main text of this paper.
\Cref{tab:notation} contains a reference summary of the notation used in the main body of the paper and \Cref{fig:general_ssg} illustrates some of the basic concepts.

We consider the 5DFT Bell state generator from \Cref{sec:5dft}.
It is an $m = 5, n = 2$ seed state generator with $n_\inp = 5$ single photon inputs, one in each input mode, that is,
\begin{equation}
    \ket{\psi_\inp} = \ket{11111}.
\end{equation}
After propagating through the 5DFT interferometer, the input state $\ket{\psi_\inp}$ is transformed into the intermediate state
\begin{equation}
    \begin{split}
        \ket{\psi_\inter} &= \sqrt{\frac{296}{625}}\ket{0}\ket{\xi_0} + \sqrt{\frac{165}{625}}\ket{1}\ket{\xi_1} + \sqrt{\frac{80}{625}}\ket{2}\ket{\xi_2}\\
        &+ \sqrt{\frac{60}{625}}\sqrt{\frac{1}{2}}\ket{3}\bigg(\ket{1001} + \ket{0110}\bigg)\\
        &+ \sqrt{\frac{24}{625}}\ket{5}\ket{\xi_5}
    \end{split}
\end{equation}
where the states $\ket{\xi_i}, i \in \{0, 1, 2, 5\}$ are the states present in the $m_\out = 4$ output modes when the state in the $m_\detc = 1$ detected mode is $\ket{i}$.
As we see in the equation above, this SSG heralds the creation of a Bell state upon the detection of $n_\detc = 3$ photons in the detector, that is, it has a single successful detection pattern $\bm{p} = (3)$.
Thus, we can write
\begin{align}
    \begin{split}
    \tr_{[1]}\left(\ketbra{\psi_\inter}{\psi_\inter}\Pi_\Id^{((3))}\right) & = \Pr(\succ ; \Id) \ketbra{\psi^{((3))}_\out}{\psi^{((3))}_\out} \\
    & = \frac{12}{125}\ketbra{\Psi^+}{\Psi^+}
    \end{split}
\end{align}
where $\ket{\Psi^+}$ is the Bell state stabilized by $\langle +XX, -ZZ\rangle$.

We now derive the coefficients which show the response of this SSG to multiphotons at the input, i.e.\ the case where $\text{E} = \text{A}_\inp$.
From \Cref{eq:decomp_input} and \Cref{app:multi_channel} we have
\begin{align}
    \rho_\inp & \approx \left( 1 - 5 \beta_{\text{A}_\inp}\right) \rho_{\inp, \Id} + 5 \beta_{\text{A}_\inp} \rho_{\inp, \text{A}_\inp}  \\
    \rho_{\inp, \Id} & = \ketbra{\psi_\inp}{\psi_\inp} \\
    \begin{split}
    \rho_{\inp, \text{A}_\inp}  & = \frac{1}{5}\Big(\ketbra{21111}{21111}\\
    &\qquad  + \ketbra{12111}{12111}\\
    &\qquad + \ketbra{11211}{11211}\\
    &\qquad + \ketbra{11121}{11121}\\
    &\qquad + \ketbra{11112}{11112}\Big). \end{split}
\end{align}

Since we are only concerned here with an input error, the measurement operator of interest remains
\begin{equation}
\Pi_\Id^{((3))} = \ketbra{3}{3}
\end{equation}
and we can restrict our focus to the portion of the intermediate state where the state in the detected mode is in the state $\ketbra{3}{3}$.
Evolving $\rho_{\inp,\text{A}_\inp}$ to $\rho_{\inter, \text{A}_\inp}$, acting with the projector $\Pi_\Id^{((3))}$, tracing over the detected mode, and renormalizing, we get

\begin{widetext}
    \begin{equation}
        \begin{split}
            \rho^{((3))}_{\out, \text{A}_\inp} = & \bigg[\frac{1}{10}\big(\ket{1200} + \ket{0102} + \ket{2010} + \ket{0021}\big) \big(\bra{1200} + \bra{0102} + \bra{2010} + \bra{0021}\big)\\
            & + \frac{1}{20}\big(\ket{0111} + \sqrt{2}\ket{1002}\big)\big(\bra{0111} + \sqrt{2}\bra{1002}\big)\\
            & + \frac{1}{20}\big(\ket{1101} + \sqrt{2}\ket{0210}\big)\big(\bra{1101} + \sqrt{2}\bra{0210}\big)\\
            & + \frac{1}{20}\big(\ket{1110} + \sqrt{2}\ket{2001}\big)\big(\bra{1110} + \sqrt{2}\bra{2001}\big)\\
            & + \frac{1}{20}\big(\ket{1011} + \sqrt{2}\ket{0120}\big)\big(\bra{1011} + \sqrt{2}\bra{0120}\big)\bigg]\\
        \end{split}
    \end{equation}
\end{widetext}

This state contains terms in four non-ideal subspaces which are labeled by the tuples $(M, I), (I, M), (M, V)$, and $(V, M)$.
Because we do not have any Pauli subspaces here, we can find the subspace projectors by taking tensor product of the single qubit subspace projectors (truncated at 3 photons per qubit)
\begin{align}
    M \rightarrow &\ketbra{11}{11} + \ketbra{20}{20} + \ketbra{02}{02}\\
    &+ \ketbra{21}{21} + \ketbra{12}{12} + \ketbra{30}{30} + \ketbra{03}{03}\\
    I \rightarrow &\ketbra{10}{10} + \ketbra{01}{01}\\
    V \rightarrow &\ketbra{00}{00}
\end{align}
which gives
\begin{equation}
    \begin{split}
    \Pr\left(\bm{Q}=(M, I)\ |\ \bm{p} = ((3))\ ;\ \text{A}_\inp\right) &= \frac25 \\
    \Pr\left(\bm{Q}=(I, M)\ |\ \bm{p} = ((3))\ ;\ \text{A}_\inp\right) & = \frac25 \\
    \end{split}
\end{equation}
\begin{equation}
    \begin{split}
    \Pr\left(\bm{Q}=(M, V)\ |\ \bm{p} = ((3))\ ;\ \text{A}_\inp\right) &= \frac{1}{10} \\
    \Pr\left(\bm{Q}=(V, M)\ |\ \bm{p} = ((3))\ ;\ \text{A}_\inp\right) &= \frac{1}{10} .\\
    \end{split}
\end{equation}
For example, the $(M, V)$ probability comes from the term $\ketbra{1200}{1200}$ and $(I, M)$ comes from the terms above containing 0102, 0111, 1002, 1011.

Since there is only one successful detection pattern, we can write
\begin{equation}
c_{\text{A}_\inp\rightarrow \bf{Q}} = c^{(\bf{p})}_{\text{A}_\inp\rightarrow \bf{Q}}
\end{equation}
For this error mechanism and detector pattern, the relative heralding factor in \Cref{error_coeff} is equal to 1,
\begin{equation}
    \frac{\Pr((3);\text{A}_\inp)}{\Pr((3);\Id)} = 1,
\end{equation}
since $\Pr((3);\text{A}_\inp)=\Pr((3);\Id)=12/125$.
Therefore, multiplying the above probabilities by ${\nu_{\text{A}_\inp} = n_\inp=5}$ gives us
\begin{equation}
    \begin{split}
        c_{\text{A}_\inp\rightarrow (M, I)} &= c_{\text{A}_\inp\rightarrow (I, M)} = 2\\
        c_{\text{A}_\inp\rightarrow (M, V)} &= c_{\text{A}_\inp\rightarrow (V, M)} = \frac{1}{2} .
    \end{split}
\end{equation}

\section{Error mechanisms}\label[appendix]{app:error_channels}
Here we provide a more detailed description of the channels that we use in this work to describe the error mechanisms that we consider and their first-order effects on the SSG's heralded states.

\begin{table}[t]
    \centering
    \begin{tabular}{c|c|c}
        \textbf{Error mechanism}, $\text{E}$ & \textbf{Location} & $\nu_\text{E}$ \\ \hline
        $\text{L}_\inp$ &  &  \\
        $\text{A}_\inp$ & input modes & $n_\inp$ \\
        $\text{D}_\inp$ &  &  \\ \hline
        $\text{L}_\detc$ &  & $n_\detc + 1$ \\
        $\text{A}_\detc$ & detected modes & $(m_\detc+n_\detc-1)/2$ \\
        $\tilde{\text{A}}_\detc$ &  & $m_\detc$
    \end{tabular}
    \caption{
        Summary of the first-order scaling coefficients $\nu_\text{E}$ used in \Cref{eq:decomp_POVM,eq:decomp_output}.
        Here $n_\inp$ is the number of single-photon inputs, $m_\detc$ is the number of detector modes, and $n_\detc=\sum_i p_i$ is the number of photons in the successful detector pattern $\bm{p}$.
    }\label{tab:nu_values}
\end{table}

First, we will look at how we can apply a channel's Kraus operators to the measurement operator rather than the state's density operator.
We observe that if we decompose the channel on the detected modes into Kraus operators
\begin{align}
    \left[ \bigotimes_{j=1}^{m_\detc} \bigcirc_{\text{E} \in \mathcal{E}_\detc} \Lambda_{\text{E}, j} \right] \left( \rho_\inter \right) = \sum_i K_i \rho_\inter K_i^\dagger , 
\end{align}
then we can rewrite our definition of $\rho_\out^{(\bm{p})}$ from \Cref{eq:rho_h_p} like 
\begin{align}
\begin{split}
    \rho_\out^{(\bm{p})} & = \frac{1}{\Pr(\bm{p})} \tr_{[1,m_\detc]} \left( \sum_i K_i \rho_\inter K_i^\dagger \ketbra{\bm{p}}{\bm{p}} \right)  \\
    & = \frac{1}{\Pr(\bm{p})} \tr_{[1,m_\detc]} \left(  \rho_\inter \sum_i K_i^\dagger \ketbra{\bm{p}}{\bm{p}}  K_i \right) \\
     & = \frac{1}{\Pr(\bm{p})} \tr_{[1,m_\detc]} \left( \rho_\inter \cdot \Pi^{(\bm{p})} \right) 
\end{split}\\ 
    \Pi^{(\bm{p})} & = \sum_i K_i^\dagger \ketbra{\bm{p}}{\bm{p}} K_i . \label{eq:kraus_povm} 
\end{align}
When error probabilities are low, we can construct this sum to first order as being a sum over different error mechanisms.
This establishes the noisy-measurement representation used in \Cref{eq:decomp_POVM}.

With this, we will take a closer look at the first order expansion of $\rho_\out^{(\bm{p})}$ by noticing that we can write it such that only the error mechanisms acting on the input will impact $\rho_\inp$ and only the error mechanisms acting on the detected modes will impact $\Pi^{(\bm{p})}$.
Expanding to first order in $\beta_\text{E}$ for $\beta_\text{E} \ll 1$ we find
\begin{align}
    \begin{split} \label{eq:rho_h_lim}
     \rho_\out^{(\bm{p})}  & =\   \tr_{[1,m_\detc]} \Biggl(\frac{1}{\Pr(\bm{p})}\rho_{\inter}\cdot \Pi^{(\bm{p})} \Biggr) \\
      = &  \tr_{[1,m_\detc]}  \Biggl[ \frac{1}{\Pr(\bm{p})}\Biggl(\beta_{\Id} \cdot \rho_{\inter, \Id}  \cdot \Pi^{(\bm{p})}_\Id \\
     & +\sum_{\text{E} \in \mathcal{E}_\inp} \nu_\text{E} \beta_\text{E}  \cdot\mathcal{U} (\rho_{\inp,\text{E}} \otimes \ketbra{0}{0}^{\otimes(m-n_\inp)}) \mathcal{U}^\dagger \cdot \Pi^{(\bm{p})}_\Id  \\
    & + \sum_{\text{E} \in \mathcal{E}_\detc} \nu_\text{E} \beta_\text{E}  \cdot\rho_{\inter, \Id} \cdot \Pi^{(\bm{p})}_\text{E} \Biggr) \Biggr] \\
    &
         \quad + \bigO\left(\sum_{\text{E}_1, \text{E}_2 \in \mathcal{E}} \beta_{\text{E}_1} \beta_{\text{E}_2}\right).
    \end{split}
\end{align}
This is the explicit channel-level expansion underlying the output-state decomposition in \Cref{eq:decomp_output}.
Recall that we use $\nu_\text{E}$ to denote the scaling coefficients used in \Cref{eq:decomp_POVM,eq:decomp_output}, with values given in \Cref{tab:nu_values}.

Collecting contributions to the ideal term from the input error mechanisms, we can expand $\beta_{\Id}$ as 
\begin{align}
    \beta_{\Id} = \beta_{\detc, \Id} - n_\inp \sum_{\text{E} \in \mathcal{E}_\inp}  \beta_\text{E}
\end{align}
and in the following derivations, we derive that our detector error mechanisms give us
\begin{align}
    \beta_{\detc, \Id} = 1 - n_\detc \beta_{\text{L}_\detc} - \frac{n_\detc + m_\detc}{2} \beta_{\text{A}_\detc} - m_\detc \beta_{\tilde{A}_\detc} .
\end{align}
We can see how \Cref{eq:rho_h_lim} can be rewritten as a sum over different states each corresponding to the different error mechanism channels applied.
In this appendix, we will derive expressions for $\rho_{\inp,\text{E}}$, $\Pi^{(\bm{p})}_\text{E}$ and $\nu_\text{E}$ for the different error mechanisms we use in this work.

\subsection{Loss}
We consider linear photon loss either occurring at the input of the SSG interferometer or just prior to detection.
The photon loss channel on mode $i$, parameterized by the single photon loss probability, $\beta_{\text{L}}$, can be written like 
\begin{multline}
    \Lambda_{\text{L},i}(\rho) = \\ \tr_{i'}\left( \text{BS}(\beta_{\text{L}})_{i,i'} \left( \rho \otimes \ketbra{0}{0}_{i'} \right) \text{BS}(\beta_{\text{L}})_{i,i'}^\dagger  \right)
\end{multline}
where $\text{BS}(r)_{i,j}$ is the beam splitter unitary operator, $a^\dagger_i \to \sqrt{t} a^\dagger_i + \sqrt{r} a^\dagger_j$, $a^\dagger_j \to \sqrt{r} a^\dagger_i - \sqrt{t} a^\dagger_j$ for some real, positive $r \leq 1$ and $t = {1 - r}$. 

We can write the channel $\Lambda_{\text{L},i}$ using Kraus operators as:
\begin{align}
    \Lambda_{\text{L},i}(\rho) & = \sum_{l=0}^\infty K^{(\text{L})}_{l,i} \rho K^{(\text{L})\dagger}_{l,i} \\
    K^{(\text{L})}_{l,i} & = \sum_{q=l}^\infty \sqrt{ \binom{q}{l} \beta_{\text{L}}^l (1 - \beta_{\text{L}})^{q-l}} \ketbra{q-l}{q}_i .
\end{align}
Later we will further specify $\text{L}$ as $\text{L}_\inp$, $\text{L}_\detc$ or $\text{L}_\out$ depending on whether we are looking at loss channels on the input, detected, or output modes of the SSG.

If we expand and look at the first order terms in this expression, we can approximately write this channel as 
\begin{align}
    \Lambda_{\text{L},i}(\rho) & \approx \tilde{K}^{(\text{L})}_{0,i} \rho \tilde{K}^{(\text{L})\dagger}_{0,i} + \tilde{K}^{(\text{L})}_{1,i} \rho \tilde{K}^{(\text{L})\dagger}_{1,i}\\
    & \text{where} \nonumber \\
    \tilde{K}_{0,i}^{(\text{L})} & =  \sum_{q=0}^\infty \sqrt{1 - q\beta_{\text{L}}} \ketbra{q}{q}_i \\
    \tilde{K}_{1,i}^{(\text{L})} & = \sum_{q=1}^\infty \sqrt{q \beta_{\text{L}}} \ketbra{q-1}{q}_i.
\end{align}

We can see how this channel leads us to a map which takes the ideal input state $\ketbra{1}{1} \to \ketbra{0}{0}$ with probability $\beta_{\text{L}}$.

As we look toward applying a loss channel across $m$ modes, we now consider the first order terms for this channel.
We will use $\qb$ to denote an $m$-length tuple of integers $(q_1, q_2, \dots, q_m)$ describing the Fock state in each mode, and we will use $\onei$ to denote the tuple with $0$s in all entries except with a $1$ as the $i$th entry: \begin{equation}
\onei = (0_1, \dots 0_{i-1}, 1_i, 0_{i+1}, \dots 0_m).
\end{equation}
Let $|\qb|$ denote the total photon number, $\sum_{i=1}^m q_i$.
Here, we can consider the following Kraus operators
\begin{align}
    \left[ \bigotimes_{i=1}^m \Lambda_{\text{L},i}\right](\rho) & \approx \tilde{K}^{(\text{L})}_{0} \rho \tilde{K}^{(\text{L})\dagger}_{0} + \sum_{i=1}^{m} \tilde{K}^{(\text{L})}_{1,i} \rho \tilde{K}^{(\text{L})\dagger}_{1,i} \\
    & \text{where} \nonumber \\
    \tilde{K}_{0}^{(\text{L})} & =  \sum_{\substack{\qb \\ |\qb|\geq 0}}^\infty \sqrt{1 - |\qb|\beta_{\text{L}}} \ketbra{\qb}{\qb} \\
    \tilde{K}_{1,i}^{(\text{L})} & =  \sum_{\substack{\qb \\ |\qb|\geq 1}}^\infty \sqrt{q_i\beta_{\text{L}}}  \ketbra{\qb-\onei}{\qb} .
\end{align}

Applying these to the ideal input state, we can find an expression for $\rho_{\inp, \text{L}_\inp}$
\begin{align}
    \rho_{\inp, \text{L}_\inp} = \frac{1}{n_\inp} \sum_{i=1}^{n_\inp} \bigotimes_{j=1}^{n_\inp} \ketbra{1-\delta_{ij}}{1 - \delta_{ij}} 
\end{align}
which implies $\nu_{\text{L}_\inp} = n_\inp$.
This is the loss-error component, i.e. $\text{E} = \text{L}_\inp$, appearing in the general input-state decomposition \Cref{eq:decomp_input}.

We next will derive the noisy measurement operator, using $\pb$ as the tuple for the photon number basis measurement outcome and $n_\detc$ for the total detected photon number $\sum_{i=1}^{m_\detc} p_i$.
It is helpful to compute expressions for $\bra{\pb} \tilde{K}_0^{(\text{L})}$ and $\bra{\pb} \tilde{K}_{1,i}^{(\text{L})}$.
\begin{align}
    \bra{\pb} \tilde{K}_0^{(\text{L})} & = \sqrt{1 - n_\detc\beta_{\text{L}}} \bra{\pb} \\
    \bra{\pb} \tilde{K}_{1,i}^{(\text{L})} & = \sqrt{(p_i+1)\beta_{\text{L}}} \bra{\pb+\onei} .
\end{align}

This is because we can compute the probability via a POVM element as outlined in \Cref{eq:kraus_povm}.
With this, we write the expression for the POVM element for detector loss on the pattern $\pb$ to first order in detector loss probability, which we denote $\beta_{\text{L}_\detc}$, as 
\begin{align}
    \begin{split}
    \Pi^{(\pb)} \approx &\ (1 - n_\detc \beta_{\text{L}_\detc}) \Pi^{(\pb)}_\Id \\
    & + (n_\detc+1) \beta_{\text{L}_\detc} \Pi^{(\pb)}_{\text{L}_\detc} \end{split}\\
    & \text{where} \nonumber \\
    \begin{split}
    \Pi^{(\pb)}_{\text{L}_\detc} = & \frac{1}{(n_\detc+1) \beta_{\text{L}_\detc}} \sum_i  \tilde{K}_{1,i}^{(\text{L})\dagger} \ket{\pb} \!  \bra{\pb} \tilde{K}_{1,i}^{(\text{L})} \\
    \approx &\ \frac{1}{n_\detc+1}\sum_{i=1}^{m_\detc} (p_i+1) \ketbra{\pb + \onei}{\pb + \onei}
    \end{split}
\end{align}
which is the $\text{E} = \text{L}_\detc$ term in \Cref{eq:decomp_POVM}.
This implies that $\nu_{\text{L}_\detc} = n_\detc+1$.

The set of elements $\Pi^{(\pb)}_{\text{L}_\detc}$ does not itself constitute a full POVM, as it does not contain a term for the vacuum outcome. 
To complete the POVM, we could also include the vacuum outcome, denoted \(\perp\), from which no photon can be lost.
This would give the additional POVM element
\begin{align}
    \Pi^{(\perp)}_{\text{L}_\detc} = \ketbra{\bm{0}}{\bm{0}}
\end{align}
such that 
\begin{align}
    \Pi^{(\perp)}_{\text{L}_\detc} + \sum_{\pb}  \Pi^{(\pb)}_{\text{L}_\detc} = \I .
\end{align}

\subsection{Coherent Photon addition} \label{app:multi_channel}

We start by modeling this process through the application of a beam splitter between some mode $i$ and an ancilla mode $i'$ which contains a single photon. 
\begin{align}
    \rho \to \text{BS}(r)_{i,i'} \left( \rho \otimes \ketbra{1}{1}_{i'} \right) \text{BS}(r)_{i,i'}^\dagger 
\end{align}

A closer look at the case where we input some Fock state $\ket{q}_i$ into this channel and then end up with vacuum in the ancilla mode $i'$ shows
\begin{align}
    \ket{q}_i \to & \bra{0}_{i'} \text{BS}(r)_{i,i'} \ket{q}_{i} \ket{1}_{i'} \\
     & = \bra{0}_{i'} \text{BS}(r)_{i,i'} \frac{(a^\dagger_i)^q a^\dagger_{i'}}{\sqrt{q!}} \ket{0}_{i} \ket{0}_{i'} \\
     & = \frac{ \sqrt{r t^{q}} (a^\dagger_i)^{q+1}}{\sqrt{q!}} \ket{0}_i \\ 
     & = \sqrt{r t^{q}} \sqrt{q+1} \ket{q+1}_i
\end{align}

This process will occur with probability of $rt^{q} (q+1)$.
Similarly, we can consider the scenario where we herald that the photon has not been added to the state e.g.\ 
\begin{align}
    \ket{q}_i \to & \bra{1}_{i'} \text{BS}(r)_{i,i'} \ket{q}_{i} \ket{1}_{i'} \\
    & = \sqrt{t^{q-1}}\left(t - q r \right) \ket{q}_i
\end{align}
which occurs with probability $t^{q-1}\left(t - q r \right)^2$.

As we approach $r \to 0$, the probability of $\geq 2$ photons going into the ancillary mode vanishes.
In this regime, we also see that addition of the ancilla photon to mode $i$ occurs with probability \begin{equation}
\sim  (q+1)r
\end{equation} and the absence of photon addition occurs with probability 
\begin{equation}
\sim 1 - (3q + 1)r .
\end{equation}
With this, we get the first order expression for the coherent photon addition state transformation 
\begin{align}
    \ket{q}_i \to\ \approx  \sqrt{q+1} \sqrt{r} \ket{q+1}_i 
\end{align}
and for the absence of photon addition 
\begin{align}
    \ket{q}_i \to\ \approx \sqrt{1 - (3q + 1)r}\ket{q}_i .
\end{align}

Another process which occurs to first order in $r$ is that a photon from mode $i$ can couple into mode $i'$. 
Because we want to decouple the effects of photon addition with the effects of loss, we will postselect such that these events are removed.
For example we could postselect on projecting onto $\ketbra{0}{0}_{i'} + \ketbra{1}{1}_{i'}$.
Using this postselection, we are left with first-order-approximate Kraus operators 
\begin{align}
    \rho \to \ \approx \tilde{K}_{0}^{(\text{A})} \rho \tilde{K}_{0}^{(\text{A})\dagger} + \sum_{i=1}^m \tilde{K}_{1,i}^{(\text{A})} \rho \tilde{K}_{1,i}^{(\text{A})\dagger}
\end{align}
where
\begin{align}
        \tilde{K}_{0}^{(\text{A})} & = \sum_{\qb} \sqrt{1-(3|\qb| + m)r} \ketbra{\qb}{\qb} \\
    \tilde{K}_{1,i}^{(\text{A})} & = \sqrt{r} \sum_{\qb} \sqrt{q_i+1} \ketbra{\qb+\onei}{\qb} .
\end{align}
To ensure that the state remains unit trace after the application of this postselected error mechanism, we will renormalize the state. 
We use $\Lambda_{\text{A}}$ to denote this postselected, renormalized operation. 
Working through this postselection and renormalization, we can see the action of this operation on a Fock state
\begin{align}
    \begin{split}
	    \Lambda_{\text{A}}(\ketbra{\fockidx}{\fockidx}) \approx &\  \mathcal{N} \bigl( [1-(3\fockidx+1)r] \ketbra{\fockidx}{\fockidx}  \\ & \quad + (\fockidx+1)r \ketbra{\fockidx+1}{\fockidx+1} \bigr)
	     \\
	    \approx &\ [1 - (\fockidx + 1)r] \ketbra{\fockidx}{\fockidx} \\
	    & \quad + (\fockidx + 1)r\ketbra{\fockidx+1}{\fockidx+1} \\ 
		    \approx &\ [1 - \frac12 \beta_\text{A}(\fockidx + 1)] \ketbra{\fockidx}{\fockidx} \\ & \quad + \frac12 \beta_\text{A}(\fockidx + 1)\ketbra{\fockidx+1}{\fockidx+1}  \end{split}
\end{align}
	where to first order in $r$ we have the normalization factor $\mathcal{N} \approx (1-2\fockidx r)^{-1} \approx 1 + 2\fockidx r$.
We have also set $\beta_\text{A} = 2r$ as the probability which parameterizes the channel, this gives the channel map for the single photon input state $\ketbra{1}{1} \to \ketbra{2}{2}$ as occurring with probability $\beta_\text{A}$ to first order. 

Note that because the postselection is not trace preserving and the renormalization is not linear, this process is not a CPTP map.
However, in the case where we have a fixed photon number input state, we end up with the same mapping as the conventional CPTP multiphoton addition channel as could be derived e.g.\ by setting $a^\dagger$ as a Lindblad jump operator.
We have chosen to derive the channel in this way here to help set up our results in \Cref{app:multiphoton_commuting}.

To first order, the full CPTP channel can be written with Kraus operators
\begin{align}
    K_{0}^{(\text{A})} & = \sum_{\qb} \sqrt{1-(|\qb| + m_\detc)r}\, \ketbra{\qb}{\qb} \\
    K_{1,i}^{(\text{A})} & = \sqrt{r} \sum_{\qb} \sqrt{q_i+1}\, \ketbra{\qb+\onei}{\qb},
\end{align}
which satisfy $K_{0}^{(\text{A})\dagger}K_{0}^{(\text{A})} + \sum_{i=1}^{m_\detc} K_{1,i}^{(\text{A})\dagger}K_{1,i}^{(\text{A})} = \I$ up to $\mathcal{O}(r^2)$.

We also can write the coherent photon addition term of the input state, i.e. the $\text{E} = \text{A}_\inp$ term in \Cref{eq:decomp_input}:
\begin{align}
    \rho_{\inp, \text{A}_\inp} = \frac{1}{n_\inp} \sum_{i=1}^{n_\inp} \bigotimes_{j=1}^{n_\inp} \ketbra{1+\delta_{ij}}{1 + \delta_{ij}} 
\end{align}
which shows that $\nu_{\text{A}_\inp} = n_\inp$.

Next we move on to the measurement operator found by considering this error mechanism on the detected modes.
As we did for loss, it is helpful to compute $\bra{\pb}K_{0}^{(\text{A})}$ and $\bra{\pb}K_{1,i}^{(\text{A})}$.
\begin{align}
    \bra{\pb}K_{0}^{(\text{A})} & = \sqrt{1-(n_\detc + m_\detc)r} \bra{\pb} \\
    \bra{\pb}K_{1,i}^{(\text{A})} & = \sqrt{r p_i}\bra{\pb-\onei}
\end{align}
This helps us to derive the noisy POVM element which captures the first order impact of detector photon addition errors.
We substitute back in the channel probability parameter $\beta_{\text{A}_\detc} = 2r$.
\begin{align}
    \begin{split}
    \Pi^{(\pb)} \approx &\  \left(1 - \frac{m_\detc+n_\detc}{2}\beta_{\text{A}_\detc}\right) \Pi^{(\pb)}_\Id \\
    & \quad + \frac{m_\detc+n_\detc-1}{2}\beta_{\text{A}_\detc} \Pi^{(\pb)}_{\text{A}_\detc}
    \end{split}\\
    & \text{where} \nonumber \\
    \Pi^{(\pb)}_{\text{A}_\detc} & = \frac{1}{m_\detc+n_\detc-1} \sum_{\substack{i=1 \\ p_i >0 }}^{m_\detc} p_i \ketbra{\pb-\onei}{\pb-\onei}
\end{align}
Therefore we have $\nu_{\text{A}_\detc} = \frac12(m_\detc+n_\detc-1)$.

When $m_\detc=1, n_\detc = 0$, we define the POVM element
\begin{align}
    \Pi^{((0))}_{\text{A}_\detc} = 0 .
\end{align}

\subsection{Incoherent Photon Addition}
For detector noise such as dark counts, we consider a phenomenological model in which the addition channel is independent of the number of photons in the mode.
We call this error channel \textit{incoherent photon addition} and denote it $\Lambda_{\tilde{\text{A}}}$ to highlight the comparison with $\Lambda_{\text{A}}$ in \Cref{app:multi_channel} which does have a dependence on the photon number.

The channel is parameterized by the probability of adding a single noise photon $\beta_{\tilde{\text{A}}}$.
We can consider the $m$-mode Kraus operator for this channel for small $\beta_{\tilde{\text{A}}}$ to first order.
\begin{align}
    \rho \to & \approx \tilde{K}^{(\tilde{\text{A}})}_{0} \rho \tilde{K}^{(\tilde{\text{A}})\dagger}_{0} + \sum_{i=1}^m \tilde{K}^{(\tilde{\text{A}})}_{1,i} \rho \tilde{K}^{(\tilde{\text{A}})\dagger}_{1,i}  \\
    \tilde{K}^{(\tilde{\text{A}})}_{0} & = \sqrt{1-m\beta_{\tilde{\text{A}}}} \sum_{\qb} \ketbra{\qb}{\qb} \\
    \tilde{K}^{(\tilde{\text{A}})}_{1,i} & = \sqrt{\beta_{\tilde{\text{A}}}} \sum_{\qb}  \ketbra{\qb+\onei}{\qb} 
\end{align}

When we consider this noise channel prior to detection, it is again helpful to compute expressions for $\bra{\pb} \tilde{K}^{(\tilde{\text{A}})}_{0}$ and $\bra{\pb} \tilde{K}^{(\tilde{\text{A}})}_{1,i}$.
\begin{align}
    \bra{\pb} \tilde{K}^{(\tilde{\text{A}})}_{0} = & \sqrt{1-m\beta_{\tilde{\text{A}}}} \bra{\pb} \\
    \bra{\pb} \tilde{K}^{(\tilde{\text{A}})}_{1,i} = & \sqrt{\beta_{\tilde{\text{A}}}} \bra{\pb-\onei} 
\end{align}
This allows us to write a POVM element for detection under incoherent noise to first order in the incoherent noise probability, which we write as $\beta_{\tilde{\text{A}}}$.
\begin{align}
    \Pi^{(\pb)} \approx &\ (1 - m_\detc \beta_{\tilde{\text{A}}}) \Pi^{(\pb)}_\Id + m_\detc \beta_{\tilde{\text{A}}} \Pi^{(\pb)}_{\tilde{\text{A}}_\detc} \\ 
    & \text{where} \nonumber \\ 
    \Pi^{(\pb)}_{\tilde{\text{A}}_\detc} = &\ \frac{1}{m_\detc} \sum_{\substack{i=1 \\ p_i >0 }}^{m_\detc} \ketbra{\pb-\onei}{\pb-\onei}
\end{align}
This implies $\nu_{\tilde{\text{A}}_\detc} = m_\detc$.

It can also be shown that the POVM elements $\Pi^{(\pb)}$ are equivalent to the ones obtained for a ``log-tree'' photon-number resolving detector (PNRD) in which light from a single mode is evenly split among $F$ bucket detectors, each with a dark count probability $p_d$ in the limit
$$
    F\gg 1,\quad p_d\ll 1, \quad\beta_{\tilde{\text{A}}}= Fp_d\ll 1.
$$

\subsection{Distinguishability}
Because we do not consider commuting the distinguishability channel through the linear optical interferometer, we can afford to define it only via its action on the input state.
Also note that, as shown in \Cref{fig:channels}, we choose the distinguishability channel $\Lambda_\text{D}$ to act first, i.e.\ directly onto the state $\ketbra{1}{1}_g \ketbra{0}{0}_b$.

So defining 
\begin{align}
    \begin{split}
    \Lambda_{\text{D}}(\ket{1}_g\ket{0}_b \! \bra{1}_g \bra{0}_b) = \ & (1 - \beta_\text{D})\ket{1}_g\ket{0}_b \! \bra{1}_g \bra{0}_b \\
    &  + \beta_\text{D} \ket{0}_g\ket{1}_b \! \bra{0}_g \bra{1}_b
    \end{split}
\end{align}
is sufficient for our needs in this paper.
For these first order errors, this is equivalent to the \textit{orthogonal bad bit} (OBB) model~\cite{sparrow2018quantum,birchall2018fundamental}.
More motivation for treating distinct spectral components incoherently is given in \Cref{app:obb_motivation}.

Using this channel, we can write the $\text{E} = \text{D}_\inp$ term of \Cref{eq:decomp_input}:
\begin{align}
    \rho_{\inp, \text{D}_\inp} = \frac{1}{n_\inp}\sum_{i=1}^{n_\inp} \bigotimes_{j=1}^{n_\inp} \ketbra{1 - \delta_{ij}}{1-\delta_{ij}}_g \otimes \ketbra{\delta_{ij}}{\delta_{ij}}_b
\end{align}
which gives $\nu_{\text{D}_\inp} = n_\inp$.

With this, we have all the necessary expressions to expand $\rho_\out^{(\bm{p})}$ to first order in all error mechanisms considered in this work.

\section{Output state subspaces}\label[appendix]{app:subspaces}
In this section, we give a detailed description for how the Hilbert space of a dual-rail stabilizer state can be decomposed into subspaces which each correspond to different forms of errors on the state.
We will explicitly define these subspaces and write down projectors onto each subspace.
As a starting point, recall our definition of the dual-rail qubit's Hilbert space as $\mathcal{H}_\text{dr} = \mathcal{H}_\text{rail}^{\otimes 2}$ with $\mathcal{H}_\text{rail}$ defined in \Cref{eq:H_rail}.

\subsection{Leakage subspaces}

We will write Fock basis vectors in the space $\Hil_\text{dr}$ like 
\begin{equation}
 \ket{i,j}_\good \ket{k,l}_\bad = \frac{(a_{1\good}^\dagger)^{i} (a_{2\good}^\dagger)^{j} (a_{1\bad}^\dagger)^{k} (a_{2\bad}^\dagger)^{l}}{\sqrt{i! j! k! l!}} \ket{0,0}_\good \ket{0,0}_\bad
\end{equation}
where $(i,j)$ are the Fock occupancies of the 2 ``good'' modes, labeled $1\good$, $2\good$ and $(k,l)$ are the occupancies of the 2 ``bad'' modes, labeled $1\bad$, $2\bad$.

This lets us write projectors onto each subspace $\Hil_\text{dr} = \Hil_V \oplus \Hil_\sph \oplus \Hil_M$ as 
\begin{align}
    \Pi_V & = \ket{0,0}_\good \! \ket{0,0}_\bad \! \bra{0,0}_\good \! \bra{0,0}_\bad \\
    \Pi_{\sph} & = \sum_{i+j+k+l=1} \ket{i,j}_\good \! \ket{k,l}_\bad \! \bra{i,j}_\good \! \bra{k,l}_\bad \\
    \Pi_M & =  \sum_{i+j+k+l>1} \ket{i,j}_\good \! \ket{k,l}_\bad \! \bra{i,j}_\good \! \bra{k,l} _\bad .
\end{align}

We can define labels for each qubit which show whether they have leaked and which single photon or leakage subspace they are in. 
These labels are stored in a tuple $\bm{Q}_l = (Q_{l,1}, Q_{l,2}, 
\dots Q_{l,n})$ where $Q_{l,i} \in \{V, \sph, M\}$ corresponding to vacuum, single photon and multiphoton subspaces respectively. 
We call these the leakage labels.
We also can define a projector which corresponds to measuring the qubits in the subspaces given by $\bm{Q}_l$.
\begin{align}
    \Pi_{\bm{Q}_l} & = \bigotimes_{i=1}^n \Pi_{Q_{l,i}} 
\end{align}

\subsection{Single photon subspaces}

Next, we look at resolving the $\Hil_\sph$ subspace. 
To do this, we first look at the subspace which is labeled by $\bm{Q}_l$ and write down the subnormalized projected state corresponding to the qubits which are in $\Hil_\sph$.
We say that $n_\sph$ is the number of unleaked qubits and $n_L$ the number of leaked qubits, so $n_\sph + n_L = n$. 
The positions of the leaked qubits are given by the set $\bm{L} = \left\{i \in [1,n] : Q_{l,i} \in \{ V, M \} \right\}$, so $|\bm{L}|=n_L$. 
\begin{align}
    \tilde{\rho}_{\out,\sph}^{(\bm{p}, \bm{Q}_l)} =
    \tr_{\bm{L}} \left( \Pi_{\bm{Q}_l} \cdot \rho_\out^{(\bm{p})}  \cdot \Pi_{\bm{Q}_l}  \right)
    \label{eq:rho_unleaked}
\end{align}
The $\tr_{\bm{L}}$ operation traces out the qubits indicated by $\bm{L}$. 
This is a positive, subnormalized operator on the space $\Hil_\sph^{\otimes n_\sph}$, whose trace is the probability of projecting onto the single photon and leakage subspaces indicated by $\bm{Q}_l$.

To start to decompose $\Hil_\sph^{\otimes n_\sph}$ we map the single photon subspace onto an isomorphic space which is formed by a tensor product between the photon's \textit{visible} information, which rail it is in, and its \textit{hidden} information, which for us is just whether it is in the good or bad mode
\begin{align}
    \Hil_\sph \cong \Hil_\visible \otimes \Hil_\hidden .
\end{align}
This information is ``hidden'' as we model the detectors as being insensitive to this degree of freedom.
We can outline this mapping explicitly
\begin{align}
    \begin{split}
    &  \ket{1,0}_\good \! \ket{0,0}_\bad \to \ket{0}_\visible \ket{\good}_{\hidden}  \\ 
    &  \ket{0,1}_\good \! \ket{0,0}_\bad \to \ket{1}_\visible \ket{\good}_{\hidden}  \\ 
    &  \ket{0,0}_\good \! \ket{1,0}_\bad \to \ket{0}_\visible \ket{\bad}_{\hidden}  \\ 
    &  \ket{0,0}_\good \! \ket{0,1}_\bad \to \ket{1}_\visible \ket{\bad}_{\hidden}  .
    \end{split}
\end{align}
This map defines how states in $\Hil_\sph$ are identified with states in $\Hil_\visible \otimes \Hil_\hidden$ so we choose to denote these vectors as equal e.g.\ $\ket{0,1}_\good \ket{0,0}_\bad = \ket{1}_\visible \ket{\good}_\hidden$.
To make it clear when this mapping has been applied, we will keep the $\visible$ and $\hidden$ labels.

To resolve the visible part of the single photon subspace, we partition  $\Hil_\visible^{\otimes n_\sph}$.
The ideal state after mapping into visible and hidden Hilbert spaces is written 
\begin{equation}
\ket{\psi_\out^{({\bm{p}})}} = \ket{\psi_\out^{({\bm{p}})}}_{\visible} \ket{\good}_{\hidden}^{\otimes n} .
\end{equation}
Recall that all the SSGs we consider make dual-rail stabilizer states in the ideal case, so we can write that 
\begin{align}
    \ket{\psi_\out^{({\bm{p}})}}_{\visible} = \ket{\Psi_{\mathcal{S}}}
\end{align}
i.e.\ it is a state which can be defined by being stabilized by the elements of the group $\mathcal{S}$.
\begin{align}
    \mathcal{S} & = \left\langle S_1, S_2, \dots S_n \right\rangle \\
    S_i & \ket{\Psi_\mathcal{S}}  = \ket{\Psi_\mathcal{S}} \ \forall \ S_i
\end{align}

Using the positions of the leaked qubits, $\bm{L}$, we can write down the ideal reduced state in the visible subspace of the unleaked qubits
\begin{align}
     \tr_{\bm{L}} \left( \ketbra{\Psi_{\mathcal{S}}}{\Psi_{\mathcal{S}}}\right) =
      \rho_{\mathcal{S}_{\bm{L}}} .
\end{align}
This reduced state can be defined as the maximally mixed state on the $+1$ eigenspace of the stabilizer group $\mathcal{S}_{\bm{L}}$ which is formed by only keeping the elements from $\mathcal{S}$ which act trivially on the leaked qubits.
\begin{align}
    \begin{split}
    \mathcal{S}_{\bm{L}} & = \big\langle \ S|_\sph \ : \ S|_{\bm{L}} = \I_{\bm{L}}  ,\ S \in \mathcal{S} \big\rangle \\
    S \rho_{\mathcal{S}_{\bm{L}}} & =  \rho_{\mathcal{S}_{\bm{L}}} \quad \forall \ S \in \mathcal{S}_{\bm{L}} 
    \end{split}
\end{align}
where $\sph$ labels the qubits in the unleaked subspace and $S|_\sph$, $S|_{\bm{L}}$ denote the parts of the Pauli operator $S$ which act on the states given by $\sph$ or $\bm{L}$.
So $\mathcal{S}_{\bm{L}}$ can be used to define a subspace of $H_{\visible}^{\otimes n_\sph}$.

We want to identify and label the other subspaces (or cosets) of the $2^{n_\sph}$-dimensional $H_{\visible}^{\otimes n_\sph}$. 
To do this, we define the $n_\sph$-length Pauli string
\begin{align}
    \bm{P} = (P_1, P_2, \dots, P_{n_\sph}), \quad P_i \in \{\I, X, Y, Z\}
\end{align}
which we use as a label and also as an $n_\sph$-qubit operator $\bm{P} = \bigotimes_{i=1}^{n_\sph} P_i$.
The subspaces generated by $\bm{P} \mathcal{S}_{\bm{L}} \bm{P}^\dagger$ form an orbit which fills the space $H_{\visible}^{\otimes n_\sph}$.
Each subspace may be generated by multiple different $\bm{P}$s.
To label a subspace, we first find all the minimal weight (number of non-identity components) Pauli strings which generate that subspace.
We then evaluate the Pauli-$Y$ weight of these strings, keeping only the lowest $Y$ weight of the minimal overall Pauli weight strings.
We pick one of these strings which acts as the label of the space through some tie-breaker strategy.
All such labels are added to the set $\mathcal{P}_{\bm{L}}$.

To allow us to ensure later that all symmetries of the errors on the state are reflected by the error probabilities we will calculate, for each chosen label $\bm{P} \in \mathcal{P}_{\bm{L}}$, we also store all the minimal weight and minimal $Y$ weight Pauli strings for that subspace in the set 
\begin{align} \label{eq:M_set}
    \begin{split}
   \mathcal{M}_{\bm{P}} & = \left\{\bm{P}^{(1)}, \bm{P}^{(2)}, \bm{P}^{(3)}, \dots \ \big| \ \bm{P}^{(i)} {\mathcal{S}_{\bm{L}}} \bm{P}^{(i)\dagger} = \bm{P}  { \mathcal{S}_{\bm{L}}} \bm{P}^\dagger \right\} \\
   &\ \forall \ \bm{P} \in \mathcal{P}_{\bm{L}} .
    \end{split}
\end{align}
We can define a function $\#_{\bm{L}}(\bm{P})$ which given a Pauli string $\bm{P}'$ returns the size of the set $\mathcal{M}_{\bm{P}}$ which it is a member of.
\begin{align}
    \#_{\bm{L}}(\bm{P}') = |\mathcal{M}_{\bm{P}}|\quad  \text{for}\ \bm{P}' \in \mathcal{M}_{\bm{P}}
\end{align}

This lets us decompose the $n_\sph$-qubit logical subspace as
\begin{align}
    \Hil_\visible^{\otimes n_\sph} & = \bigoplus_{\bm{P} \in \mathcal{P}_{\bm{L}}} \Hil_{\visible}^{(\bm{P})} \\ 
    \Hil_{\visible}^{(\bm{P})} & = \text{supp} \left( \bm{P} \rho_{\mathcal{S}_{\bm{L}}} \bm{P}^\dagger \right)
\end{align}
where $\Hil_{\visible}^{(\bm{P})}$ is the subspace we associate with $\bm{P}$.
Here we write this as the support of $\bm{P} \rho_{\mathcal{S}_{\bm{L}}} \bm{P}^\dagger$. 
The projector onto the $\Hil_\visible^{(\bm{P})}$ subspace can be written 
\begin{align}
    \Pi_\visible^{(\bm{P})} = \frac{1}{|\mathcal{S}_{\bm{L}}|} \sum_{S \in \bm{P}\mathcal{S}_{\bm{L}}\bm{P}^\dagger} S .
\end{align}

The SSGs we focus on in this paper produce GHZ states.
Therefore for $1 \leq n_\sph<n$, we have
\begin{align}
    \rho_{\mathcal{S}_{\bm{L}}} = \frac12 \left(\ket{0}^{\otimes n_\sph}\bra{0}^{\otimes n_\sph} + \ket{1}^{\otimes n_\sph}\bra{1}^{\otimes n_\sph} \right) .
\end{align}
This is the state formed by the maximally mixed state on the +1 eigenspace of the stabilizer group
\begin{align}
    \mathcal{S}_{\bm{L}} = \big\langle Z_i Z_{i+1} : i \in [1, n_\sph-1] \big\rangle 
\end{align}
and so we can define the projectors onto the logical subspaces as 
\begin{align}
    \Pi_\visible^{(\bm{P})} = \bm{P} \left( \ket{0}^{\otimes n_\sph}\bra{0}^{\otimes n_\sph} + \ket{1}^{\otimes n_\sph}\bra{1}^{\otimes n_\sph} \right) \bm{P}^\dagger
\end{align}
which corresponds to a 2-dimensional Hilbert space, meaning we need to define $2^{n_\sph-1}$ Pauli strings to decompose the logical part of the Hilbert space for the unleaked qubits.
For $n = n_\sph$, we have $\mathcal{S}$ as the stabilizers of the GHZ state and so each subspace is 1-dimensional and we need $2^n$ Pauli strings.

As an example, for an $n=3$ GHZ state with one leaked qubit, so $n_\sph=2$, $n_L=1$, $\bm{L} = \{1\}$, we have 
\begin{align}
    \rho_{\mathcal{S}_{\bm{L}}} & = \frac12 \left( \ketbra{0,0}{0,0} + \ketbra{1,1}{1,1}\right) \\
    \mathcal{S}_{\bm{L}} & = \left\langle  Z Z \right\rangle = \{\I \I, ZZ \} \\
    \Pi_\visible^{(\bm{P})} & = \bm{P} \frac{\I\I + ZZ}{2}\bm{P} \\
    & = \bm{P} \left( \ketbra{0,0}{0,0} + \ketbra{1,1}{1,1} \right) \bm{P}.
\end{align}
We will always have a subspace labeled by an all-identity Pauli string, which in this case is just $\bm{P}_\ideal = (\I, \I)$.
This is the only lowest-weight label so we have $\mathcal{M}_{\bm{P}_\ideal} = \{\bm{P}_\ideal\}$ and $\#_{\bm{L}}(\bm{P}_\ideal)=1$. 
In this example we have just one other subspace, which can be labeled by $\bm{P}_X = (X, \I)$.
So $\mathcal{P}_{\bm{L}} = \{\bm{P}_\ideal, \bm{P}_X\}$ labels all parts of the decomposition of the logical subspace. 
For $\bm{P}_X$, we have another minimal weight label option, $\bm{P}_X' = (\I, X)$ which maps onto the same state as $\bm{P}_X$, so $\mathcal{M}_{\bm{P}_X} = \{\bm{P}_X, \bm{P}_X'\}$.
This means that $\#_{\bm{L}}(\bm{P_X}) = \#_{\bm{L}}(\bm{P_X}') = 2.$

\subsection{Internal mode subspaces}

With the single photon subspace mapping, we define a projector onto the ``bad'', distinguishable subspace of the single photon subspace
\begin{align}
    \Pi_B^{(\hidden)} = \ketbrasub{\bad}{\bad}{\hidden} 
\end{align}
and the indistinguishable ``good'' subspace
\begin{align}
    \Pi_{\ideal}^{(\hidden)} = \ketbrasub{\good}{\good}{\hidden} .
\end{align}
If we have a tuple of qubit labels $\bm{Q}_\hidden = (Q_{\hidden,1}, Q_{\hidden,2}, \dots, Q_{\hidden,n_\sph})$ which stores labels $Q_{\hidden,i} \in \{\ideal, B\}$ corresponding to ideal, indistinguishable, ``good'' photons and distinguishable, ``bad'' mode photons, we can write the projector 
\begin{align}
    \Pi_{\bm{Q}_\hidden} = \bigotimes_{i=1}^{n_\sph} \Pi_{Q_{\hidden,i}}^{(\hidden)} .
\end{align}

When considering symmetrizing errors over the minimum weight labels, we sometimes also wish to extend the definitions in this section to include distinguishability errors when picking a minimum weight label.
This modifies \Cref{eq:M_set} so that it considers only minimum weight (and minimum $Y$) labels over both Pauli and distinguishability errors. 
For example for a Bell state, we may consider a label $\bm{Q} = (BX, I)$ to be lower weight than $\bm{Q} = (B, X)$.
We use this modification for the tables in \Cref{app:error_responses}.

\subsection{Full Hilbert space decomposition}
We are now equipped to write down a full decomposition of the Hilbert space that is isomorphic to the initial Hilbert space
\begin{multline}
    \Hil_\text{dr}^{\otimes n} \cong \\ \bigoplus_{\bm{L} \subseteq [1,n]} 
    \left(\Hil_V \oplus \Hil_M \right)^{\otimes n_L }
    \otimes \left( \bigoplus_{\bm{P} \in \mathcal{P}_{\bm{L}}} \Hil_\visible^{(\bm{P})} \right) \otimes \Hil_\hidden^{\otimes n_\sph}
\end{multline}
which is a direct sum of the possible placements of the leaked qubits. 
Each term of the direct sum contains the available vacuum and multiphoton spaces of the leaked qubits and the visible and hidden states of the non-leaked qubits.
For notational convenience, this decomposition is written with the leaked qubits coming first, followed by the unleaked qubits, and this should be mapped onto the actual ordering of leaked/unleaked qubits as given by $\bm{L}$ when mapping states into this decomposed space.

We can find probabilities associated with all combinations of vacuum, multiphoton, Pauli error and distinguishability on these qubits. 
To label this, each qubit is assigned a label $Q_i$ where now we have available the following options \begin{equation}
Q_i \in \{\ideal, V, M, B, X, Y, Z, BX, BY, BZ\} .
\end{equation}
We can calculate the probability associated with $\bm{Q} = (Q_1, Q_2, \dots Q_n)$ by extracting the leakage and distinguishability labels $\bm{Q}_l$, $\bm{Q}_\hidden$ and Pauli string $\bm{P}$ as shown in \Cref{tab:labels}, and compute
\begin{align}
    \Pr(\bm{Q} | \bm{p}) = \tr\left( \left[ \Pi_{\bm{P}}^{(\visible)} \otimes \Pi_{\bm{Q}_\hidden}\right] \cdot \tilde{\rho}_{\out,\sph}^{(\bm{p}, \bm{Q}_l)} \right) / \#_{\bm{L}}(\bm{P}) .
    \label{eq:Q_given_p}
\end{align}
So we first project into the appropriate single photon and leakage subspaces and trace out the leaked qubits, to give $\tilde{\rho}_{\out,\sph}^{(\bm{p}, \bm{Q}_l)}$ as defined in \Cref{eq:rho_unleaked}, then we project onto the appropriate logical and internal subspaces on the remaining unleaked qubits.

This procedure for decomposing the Hilbert space shows which $\bm{Q}$ are valid.
Note that we will allow the Pauli strings $\bm{P}$ which form $\bm{Q}$ to be from any of the sets of minimal weight labels $\mathcal{M}_{\bm{P}}$.

\subsection{Subspace probabilities conditional on error mechanisms}\label[appendix]{app:sub_probs}
If we let $\Pi_{\bm{Q}}$ be a POVM element which combines the action of $\Pi_{\bm{Q}_l}$, $\Pi_{\bm{Q}_\hidden}$, $\Pi_{\bm{P}}^{(\visible)}$ and $\#_{\bm{L}}(\bm{P})$ such that 
\begin{align}
    \Pr(\bm{Q} | \bm{p}) = \tr\left( \rho_\out^{(\bm{p})} \cdot \Pi_{\bm{Q}} \right)
\end{align}
then we can write the first order expression of $\Pr(\bm{Q}|\bm{p})$ for small $\beta_\text{E}$ as 
\begin{align}
    \begin{split}
    \Pr(\bm{Q} | & \bm{p})  = \tr\left( \rho_\out^{(\bm{p})} \cdot \Pi_{\bm{Q}} \right) \\
     = &\ \tr \Biggl[ \frac{1}{\Pr(\bm{p})} \Bigl( \beta_{\Id} \Pr(\bm{p};\Id)\rho_{\out,\Id}^{(\bm{p})} \\
     &\quad + \sum_{\text{E} \in \mathcal{E}} \nu_\text{E} \beta_{\text{E}} \Pr(\bm{p};\text{E})\rho_{\out,\text{E}}^{(\bm{p})} \Bigr) \cdot \Pi_{\bm{Q}} \Biggr] \\
    & \quad + \bigO\left( \sum_{\text{E}_1, \text{E}_2 \in \mathcal{E}} \beta_{\text{E}_1} \beta_{\text{E}_2}\right) .
    \end{split} 
\end{align}
As we discuss in \Cref{sec:response_coefficients}, we know that for $\bm{Q} \neq \bm{Q}_\ideal$, we have no contribution from $\rho_{\out, \Id}^{(\bm{p})}$.
For these non-ideal subspaces, we can also replace $\Pr(\bm{p})$ in the denominator by its zeroth-order value $\Pr(\bm{p};\Id)$, since corrections to this denominator only contribute at second order.
This lets us simplify the expression to become:
\begin{multline}
        \Pr(\bm{Q} | \bm{p}) = \sum_{\text{E} \in \mathcal{E}} \nu_\text{E} \beta_\text{E} \frac{\Pr(\bm{p};\text{E})}{\Pr(\bm{p};\Id)}\Pr(\bm{Q} | \bm{p}; \text{E})\\ + \bigO\left(\sum_{\text{E}_1, \text{E}_2 \in \mathcal{E}} \beta_{\text{E}_1} \beta_{\text{E}_2}\right)
\end{multline}
where 
\begin{align}
    \Pr(\bm{Q} | \bm{p}; \text{E}) = \tr\Bigl( \rho_{\out,\text{E}}^{(\bm{p})} \cdot \Pi_{\bm{Q}} \Bigr) .
\end{align}

\begin{table}
\begin{tabular}{l|lll}
$Q_i$    & $Q_{l,i}$ & $Q_{\hidden,i}$ & $P_i$ \\ \hline
$\ideal$ & $\sph$    & $\ideal$           & $\I$               \\
$V$      &  $V$      &         -   &              -      \\
$M$      &  $M$      &         -   &              -      \\
$B$      &  $\sph$   &     $B$       &         $\I$           \\
$X$      &  $\sph$   &     $\ideal$       &            $X$        \\
$BX$     &  $\sph$   &   $B$              &           $X$         \\
$Y$      &  $\sph$   &     $\ideal$       &           $Y$       \\
$BY$     &  $\sph$   &   $B$              &           $Y$       \\
$Z$      &  $\sph$   &     $\ideal$       &            $Z$        \\
$BZ$     &  $\sph$   &   $B$              &             $Z$      
\end{tabular}
\caption{Abbreviated labels for the qubit subspaces $Q_i$ which combine the leakage labels $Q_{l,i}$, the internal mode subspace labels $Q_{\hidden,i}$ and the logical subspaces, defined by Pauli strings, $P_i$, that act on the ideal logical state. 
Because $\bm{Q}_\hidden, P$ only label qubits in the $\Hil_\sph$ subspace, they are undefined for $Q_i = V$ or $M$, which we indicate with a dash ( - ). \label{tab:labels}}
\end{table}

\section{Output subspace probabilities as an approximate state representation}\label[appendix]{app:error_levels}

Performing a full Fock state simulation of a resource state generation scheme containing multiple seed state generators and fusions can become intractable, especially if using local encodings means the resource states become large with tens of photons~\cite{bartolucci2023fusion}.
However, as we have described here, we can tractably compute the set of output subspace probabilities $\Pr(\bm{Q} | \bm{p})$ for the SSGs.
If we then use $\Pr(\bm{Q} | \bm{p})$ as an approximate representation of the state, this ignores information within the subspaces that are more than 1-dimensional. 
It also ignores any coherences between the subspaces.
In this section, we show that applying some random operations to a general quantum state of multiple qubits leaving a seed state generator serves to remove many of these inter-subspace coherences.

One example of a multidimensional subspace is the multiphoton space.
We note that first quantization methods~\cite{adamson2008detecting,shaw2023errors,steinmetz2024simulating} can allow a decomposition of the multiphoton space but we do not invoke these methods here. 
This structure of the multiphoton space will not change that the qubit is in a leakage state, and will only change the structure of the non-leakage errors if higher order combinations of multiphoton contamination, distinguishability and loss errors are considered, so is not relevant for the low-order analysis in this work.

\subsection{Random Optical Phase Application}

The first step we take to simplify the structure of the errors is to apply a random optical phase to each pair of rails comprising a qubit.
In Fock space, we can represent the density matrix elements of the state in this pair of rails as
\begin{equation}
\ket{\mathbf{n, m}}\bra{\mathbf{q, p}}
\end{equation}
where $\mathbf{n, m, p}$, and $\mathbf{q}$ are $k$-length tuples of integers specifying the occupancy of all spectral modes in the two rails. 

When a random phase channel is applied which acts identically on both rails, these matrix elements are transformed as
\begin{align}
\ket{\mathbf{n}, \mathbf{m}}\bra{\mathbf{q}, \mathbf{p}} \rightarrow &\int_0^{2\pi} \frac{d\theta}{2\pi} e^{i\theta (\lvert \mathbf{n} \rvert + \lvert \mathbf{m} \rvert)} \ket{\mathbf{n}, \mathbf{m}}\bra{\mathbf{q}, \mathbf{p}} e^{-i\theta (\lvert \mathbf{q} \rvert + \lvert \mathbf{p} \rvert)} \\
& = \delta_{\lvert \mathbf{n} \rvert + \lvert \mathbf{m} \rvert, \lvert \mathbf{q} \rvert + \lvert \mathbf{p} \rvert} \ket{\mathbf{n}, \mathbf{m}}\bra{\mathbf{q}, \mathbf{p}}
\end{align}
where $\delta_{x, y}$ is the Kronecker delta function.
As the above equation shows, this operation removes coherence between the parts of a state with different photon numbers in a pair of rails which comprise a qubit.
This means that if we assume that the dephasing channel gets applied to all states, any coherences between different subspaces which have different corresponding $\bm{Q}_{l}$ can be ignored.

The model is motivated both from architectures using optical fiber connections which do not require phase stability and due to multiplexing which randomizes over different optical paths with different optical phases (common to both rails of each qubit) for the heralded states produced in resource state generation.

\subsection{Random Pauli Application to Seed States}
The second operation we add to our model is a \textit{stabilizer twirl} on the visible (logical) degrees of freedom: we apply a uniformly random Pauli from the stabilizer group of the ideal seed state.
Let $\mathcal{S}=\langle S_1,\ldots,S_n\rangle$ be the stabilizer group generated by $n$ independent, commuting Pauli operators, so $|\mathcal{S}|=2^n$.
Equivalently, we may apply each generator $S_i$ independently with probability $1/2$.

For simplicity, we write the argument below for the case where no qubits are leaked.
For fixed positions of the leaked qubits $\bm{L}$, the same argument can be applied to the remaining visible degrees of freedom by replacing $\mathcal{S}$ with the reduced stabilizer group $\mathcal{S}_{\bm{L}}$ from \Cref{app:subspaces}.
In that case the twirl removes coherences between distinct $\mathcal{S}_{\bm{L}}$ eigenspaces, while any coherence within a degenerate eigenspace is part of the intra-subspace information that the probability representation does not track.

To see the effect of this within the single photon subspace, we separate out the visible and hidden degrees of freedom of the single photon and write the unleaked part of the SSG output state as
\begin{equation}
\rho = \sum_{\mathbf{q},\mathbf{r},w,v}\alpha_{\mathbf{q},\mathbf{r},w,v}\,
\ketbra{\psi_{\mathbf{q}}}{\psi_{\mathbf{r}}}_\visible\otimes \ketbra{\xi_w}{\xi_v}_\hidden,
\end{equation}
where $\ket{\xi_w}_\hidden$ labels hidden degrees of freedom and $\mathbf{q}$ is an $n$-bit label for orthonormal visible states $\ket{\psi_{\mathbf{q}}}_\visible$.
We will choose these $\ket{\psi_{\mathbf{q}}}_\visible$ to lie in distinct stabilizer-eigenspaces, i.e.\ for each generator $S_i$ we have
\begin{equation}
S_i\ket{\psi_{\mathbf{q}}}_\visible = (-1)^{q_i}\ket{\psi_{\mathbf{q}}}_\visible.
\end{equation}

Define the stabilizer-twirl channel acting on visible degrees of freedom as
\begin{equation}
\mathcal{T}_{\mathcal{S}}(\sigma)=\frac{1}{|\mathcal{S}|}\sum_{S\in\mathcal{S}} S\,\sigma\,S^\dagger .
\end{equation}
For each generator,
\begin{equation}
S_i\ketbra{\psi_{\mathbf{q}}}{\psi_{\mathbf{r}}}_\visible S_i^\dagger
=(-1)^{q_i+r_i}\ketbra{\psi_{\mathbf{q}}}{\psi_{\mathbf{r}}}_\visible .
\end{equation}
If $\mathbf{q}\neq\mathbf{r}$, then $q_i\neq r_i$ for at least one generator, so the average over the choices $\I$ and $S_i$ cancels the off-diagonal operator.
This yields
\begin{equation}
\mathcal{T}_{\mathcal{S}}\!\left(\ketbra{\psi_{\mathbf{q}}}{\psi_{\mathbf{r}}}_\visible\right)
 =
\delta_{\mathbf{q},\mathbf{r}}\,\ketbra{\psi_{\mathbf{q}}}{\psi_{\mathbf{r}}}_\visible.
\end{equation}
Our model assumes that any linear optical transformation acts trivially on the hidden degrees of freedom, so if we assume this Pauli twirling is implemented with linear optical single qubit gates, we obtain
\begin{align}
(\mathcal{T}_{\mathcal{S}}\otimes \mathcal{I})(\rho)
&=
\sum_{\mathbf{q},w,v}\alpha_{\mathbf{q},\mathbf{q},w,v}\,
\ketbra{\psi_{\mathbf{q}}}{\psi_{\mathbf{q}}}_\visible\otimes \ketbra{\xi_w}{\xi_v}_\hidden \\
&=
\sum_{\mathbf{q}}\ketbra{\psi_{\mathbf{q}}}{\psi_{\mathbf{q}}}_\visible\otimes \tilde{\kappa}_{\mathbf{q}},
\end{align}
where
\begin{equation}
\tilde{\kappa}_{\mathbf{q}}=\sum_{w,v}\alpha_{\mathbf{q},\mathbf{q},w,v}\ketbra{\xi_w}{\xi_v}_\hidden.
\end{equation}
Writing $\beta_{\mathbf{q}}=\tr(\tilde{\kappa}_{\mathbf{q}})$ and (when $\beta_{\mathbf{q}}>0$) $\kappa_{\mathbf{q}}=\tilde{\kappa}_{\mathbf{q}}/\beta_{\mathbf{q}}$, this has the classical--quantum form
\begin{equation}
(\mathcal{T}_{\mathcal{S}}\otimes \mathcal{I})(\rho)
=
\sum_{\mathbf{q}}\beta_{\mathbf{q}}\,
\ketbra{\psi_{\mathbf{q}}}{\psi_{\mathbf{q}}}_\visible\otimes \kappa_{\mathbf{q}}.
\end{equation}
Thus, stabilizer twirling removes coherences between distinct logical (visible) subspaces labeled by $\mathbf{q}$, leaving only classical correlations between the logical state and the hidden degrees of freedom.

\subsection{Ignoring Coherence Between States with Different Spectral Modes}\label{app:obb_motivation}

In our model for distinguishability, we treat the hidden spectral degree of freedom as an incoherent label.
Equivalently, in a basis $\{\ket{\xi_w}_\hidden\}$, we drop density-matrix elements connecting distinct spectral components,
\begin{equation}
    \bra{\xi_w}_{\hidden} \rho \ket{\xi_v}_{\hidden} = 0 \quad \text{for } w\neq v.
\end{equation}
Distinguishability between photons can be caused by spectral entanglement between the signal and idler photons at each photon source, by differences between the sources, or by different transformations on the spectral states of each photon.
In an architecture, photons which are just leaving a seed state generator have been through relatively few optical components and similar, if not the same, lengths of fiber and so transformations on their spectra are thought to be small contributors to distinguishability relative to effects related to the photon sources.

If all of the photon sources are the same, and spectral entanglement between the signal and idler photons is the only mechanism causing distinguishability, then there would not be any coherence between different spectral modes.
Each photon source would emit a single photon state that would be diagonal in some basis $\sum_i p_i \ketbra{\xi_i}{\xi_i}_\hidden$, due to the detector for the idler photon being insensitive to the hidden mode information, and thus no coherence between these spectral modes would exist.

If the sources are different due to random fabrication differences, then the process of multiplexing both at the single photon level and following the seed state generation would diminish the impact of the properties of any one individual source.
This motivates our assumption that coherences between different spectral modes can be ignored for the purposes of this work.
Although beyond the scope here, we also have seen that using more exact methods and realistic component performance metrics validates that this approximation can return accurate results.

This motivates the approximation that coherences between subspaces which contain different numbers of photons in the ``bad'' mode can be ignored.

\subsection{Seed state qubits do not reinterfere}\label{app:obb_motivation2}

Another coherence to consider is between subspaces with the same number of ``bad'' photons, for example $\bm{Q} = (\ideal, B)$ and $\bm{Q} = (B, \ideal)$ in a Bell state.
To first order, coherences between these subspaces will only impact measurement probabilities if qubits from the same SSG are later interfered with one another in a downstream optical circuit. So if we assume that qubits from SSGs will always be routed to independent destinations, for example, separate fusion circuits, these coherences will not impact measurement outcomes.

\section{Multiphoton Responses}\label[appendix]{app:multiphoton_commuting}

\subsection{A useful circuit identity}\label{app:multi_identity}
\begin{figure}
    \centering
    \includegraphics[width=1\columnwidth]{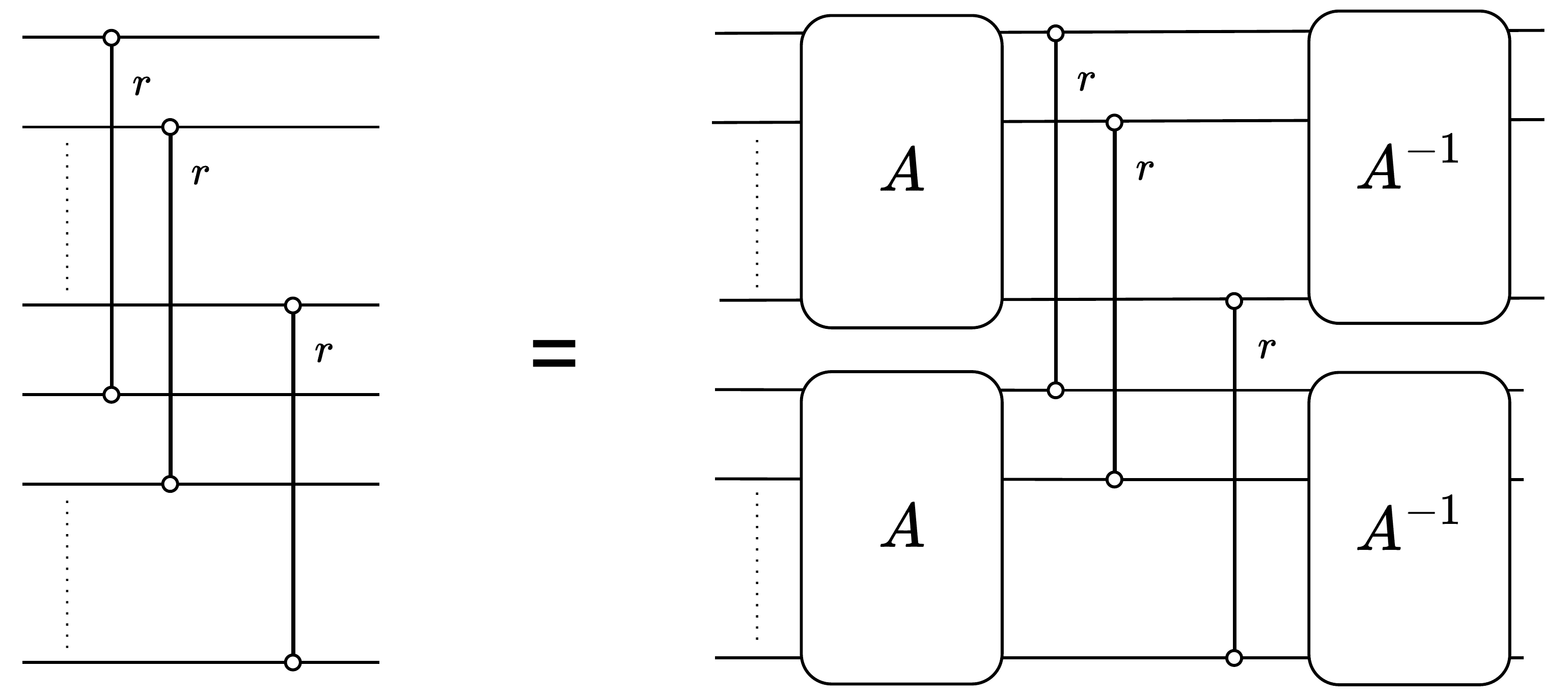}
    \caption{A circuit identity.
    All the beam splitters have the same unitary transformation, so the circuit on the left hand side can be written $U_\text{BS} \otimes \I_M$ where $\I_M$ is the $M\times M$ identity matrix.
    This is equal to the right hand side, $(A \oplus A) (U_\text{BS} \otimes \I_M) (A^{-1} \oplus A^{-1})$.
    \label{fig:matrix_id}}
\end{figure}

If we take the Kronecker product of an $m \times m$ identity matrix $\I_m$ with some invertible $n \times n$ matrix $U$, we get a block diagonal matrix formed of $m$ copies of $U$.
\begin{align}
    \I_m \otimes U = U^{\oplus m}
\end{align}
If we have some transformation which is formed by applying in parallel $n$ copies of some $m \times m$ matrix $V$, such that the overall transformation is given by $V \otimes \I_n$, we can see that this transformation commutes with any transformation of the form $\I_m \otimes U$.
\begin{align}
    (\I_m \otimes U) (V \otimes \I_n) = (V \otimes \I_n) (\I_m \otimes U) 
\end{align}
Combining these observations lets us write a useful circuit identity 
\begin{align}
    V \otimes \I_n \equiv U^{\oplus m} (V \otimes \I_n) [U^{-1}]^{\oplus m} 
\end{align}
which if we set $V$ to be a beam splitter unitary transformation $U_\text{BS}$, so $m=2$, and let $U$ be unitary, $A$, we get the circuit identity shown in \Cref{fig:matrix_id}
\begin{equation}
    U_\text{BS} \otimes \I_n \equiv (A \oplus A)(U_\text{BS} \otimes \I_n)(A^{-1} \oplus A^{-1}).
        \label{eq:matrix_id}
\end{equation}

\subsection{Commuting multiphoton contamination through linear optics}

We can use this circuit identity to help us consider the impact of multiphoton addition in an SSG where ideal occupied inputs each contain one photon, $n_\inp = m$.
Let $U_\text{SSG}$ be the unitary transformation for the SSG and 
\begin{equation}
    \rho_{\bm{a}} = \frac{1}{n_\inp} \sum_{i=1}^{n_\inp} \bigotimes_{j=1}^{n_\inp} \ketbra{\delta_{ij}}{\delta_{ij}}_j
\end{equation}
be the maximally mixed state of one photon in $n_\inp$ modes which occupies a set of ancilla modes $\bm{a}$.
This state is coupled to the SSG input modes with beamsplitters of reflectivity $r$.
To first order in $\beta_{\text{A}_\inp}$, the input state of the SSG is
\begin{align}
    \rho_{\inp} \approx  n_\inp \beta_{\text{A}_\inp} \rho_{\inp, \text{A}_\inp} + (1 - n_\inp \beta_{\text{A}_\inp})\rho_{\text{other}}
\end{align}
where $\rho_\text{other}$ contains terms for the ideal state and terms for the other error mechanisms.
Recall that
\begin{equation}
\rho_{\inp, \text{A}_\inp} = \frac{1}{n_\inp} \sum_{i=1}^{n_\inp} \bigotimes_{j=1}^{n_\inp} \ketbra{1+\delta_{ij}}{1+\delta_{ij}}_j .
\end{equation}

We can create a state with the same first order term in $\beta_{\text{A}_\inp}$ by coupling each mode of $\rho_{\bm{a}}$ to a mode of $\rho_{\inp, \Id}$, all with the same reflectivity $r=n_\inp \beta_{\text{A}_\inp}/2$, and conditioning on having $<2$ photons across all modes of $\bm{a}$.

\begin{figure*}
    \includegraphics[width=1\textwidth]{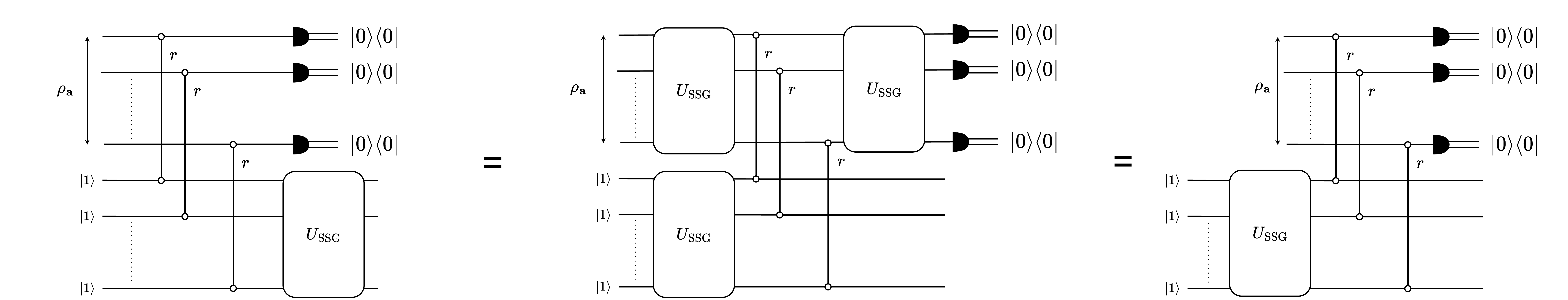}
    \caption{
    The commutation of multiphoton addition from SSG input to SSG output. 
    We start with a mixture of one photon spread over $n_\inp$ ancilla modes which mix with the single photon states at the input of the interferometer. 
    The vacuum projection on the ancilla modes ensures that the ancilla mode photon is added to the ideal input state.
    This circuit is then transformed using the circuit identity from \Cref{fig:matrix_id}.
    Finally, we can remove interferometers on the ancilla modes as they do not change the state or projection. 
    This results in a circuit where additional photons mix with the ideal state after the interferometer.
    \label{fig:multi_commute}}
\end{figure*}

The identity \Cref{eq:matrix_id} then tells us that we can commute $U_{\text{SSG}}$ across the beamsplitter array.
Noting that $\rho_{\mathbf{a}}$ and the vacuum state $\ket{0}_{\mathbf{a}}\bra{0}_{\mathbf{a}}$ are both invariant under the action of $U_{\text{SSG}}$, we can draw the circuit as on the right side of \Cref{fig:multi_commute}, showing that we have, in effect, commuted the photon addition channel which creates the multiphoton state at the SSG input, to the SSG output.

\subsection{Multiphoton contamination error responses}\label[appendix]{app:multiphoton_responses}

The result of the above section allows us to easily compute the probability of the
\begin{equation}
    \bm{Q}_M = (M, \ideal, \ldots, \ideal)
\end{equation}
subspace tuple when detector photon addition errors are protected against, either by being intrinsically removed or producing fusion heraldable errors.
In this case we know that, to first order, photon addition on the detected modes of an SSG will never produce a state in $\bm{Q}_M$.
Thus, all single qubit multiphoton errors on the output state will be the result of the photon addition channel coupling a photon into one of the output modes of the SSG.

The output state from an $m = n_\inp$ mode SSG in the computational basis will have one output mode per qubit which has 0 photons and one which has 1.
For each of the output modes, there is a $\frac{1}{n_\inp}$ probability that the ancilla photon is in the corresponding ancilla mode.
If the output mode contains vacuum, the ancilla photon will couple in with probability $r = \frac{n_\inp\psm}{2}$, while if the output mode already has a photon, to first order it will couple in the ancilla with probability $2r = n_\inp\psm$ due to HOM-type interference.
This gives a total multiphoton probability given success per qubit of
\begin{align}
    \Pr(\textbf{Q}_M | \text{success}) &= \frac{1}{n_\inp}\frac{3n_\inp}{2}\psm + \ldots\\
    & = \frac{3}{2}\psm + \ldots,
\end{align}
or equivalently,
\begin{equation}
    c_{\text{A}_\inp\rightarrow \textbf{Q}_M} = \frac{3}{2}.
\end{equation}

We now show how this coherent photon addition channel followed by a loss channel, both commuted through the interferometer, applied on the same qubit $i$ produces the state
\begin{multline}
    \rho_{\text{out}} = \frac{3}{4}\rho_{\text{out}, \Id}\\
     + \frac{1}{12}\big(X_i\rho_{\text{out}, \Id}X_i + Y_i\rho_{\text{out}, \Id}Y_i + Z_i\rho_{\text{out}, \Id}Z_i\big),
\end{multline}
where $X_i, Y_i, Z_i$ are the Pauli operators acting on qubit $i$.
We show the calculation assuming the SSG ideally produces the $\ket{\phi^+}$ Bell state but the same calculation can be performed for any stabilizer state and we arrive at the same error response coefficient.

The state of the $\ket{\phi^+}$ Bell state after application of the photon addition channel on the first qubit is
\begin{align}
    \begin{split}
    \rho_{\out,\text{A}_\out} = \frac{1}{2}\bigg[&\bigg(\sqrt{\frac{2}{3}}\ket{2010} + \sqrt{\frac{1}{3}}\ket{1101}\bigg) \\
    & \bigg(\sqrt{\frac{2}{3}}\bra{2010} + \sqrt{\frac{1}{3}}\bra{1101}\bigg)\\
    + & \bigg(\sqrt{\frac{1}{3}}\ket{1110} + \sqrt{\frac{2}{3}}\ket{0201}\bigg)\\
    & \bigg(\sqrt{\frac{1}{3}}\bra{1110} + \sqrt{\frac{2}{3}}\bra{0201}\bigg)\bigg].
    \end{split}
\end{align}

When a loss channel is applied to the same qubit, it produces the state
\begin{align}
    \begin{split}
    \rho_{\out, (\text{A}_\out,\text{L}_\out)} = \frac{1}{2}\bigg[\frac{5}{6}&\bigg(\sqrt{\frac{4}{5}}\ket{1010} + \sqrt{\frac{1}{5}}\ket{0101}\bigg)\\
    & \bigg(\sqrt{\frac{4}{5}}\bra{1010} + \sqrt{\frac{1}{5}}\bra{0101}\bigg)\\
    + \frac{5}{6}&\bigg(\sqrt{\frac{1}{5}}\ket{1010} + \sqrt{\frac{4}{5}}\ket{0101}\bigg)\\
    & \bigg(\sqrt{\frac{1}{5}}\bra{1010} + \sqrt{\frac{4}{5}}\bra{0101}\bigg)\\
    + \frac{1}{6}&\ketbra{1001}{1001} + \frac{1}{6}\ketbra{0110}{0110}\bigg] 
    \end{split}
\end{align}
which has $\frac{3}{4}$ overlap with $\ket{\phi^+}$ and $\frac{1}{12}$ overlap with each of the other Bell states.
If loss is applied to a different qubit, this results in leakage rather than Pauli error.

The probability of getting any particular Pauli error $P$ on the output state due \textit{only} to the components of the input multiphoton and input loss channels which commute to the output modes of the SSG is then
\begin{equation}
    \frac{1}{12} \cdot \frac{3}{2} \beta_{\text{A}_\inp} \cdot 2\beta_{\text{L}_\inp} = \frac{1}{4}\beta_{\text{A}_\inp}\beta_{\text{L}_\inp}
\end{equation}
where $\frac{3}{2}\beta_{\text{A}_\inp}$ is the probability that the multiphoton state is present and $2\beta_{\text{L}_\inp}$ is the probability that one of the two photons in that state is lost.
This gives the unavoidable lower bound of 
\begin{equation}
    c_{(\text{A}_\inp,\text{L}_\inp)\rightarrow \bm{Q}_P} \geq \frac{1}{4}
\end{equation}
seen in \Cref{eq:SM_SL_P}.

\section{Distinguishability Error Response Coefficients}\label[appendix]{app:dist}
In this section we show an approach for deducing the effects of distinguishable input photons on the outputs of SSGs.
We use this to calculate values of response coefficients relating input photon distinguishability to ``bad'' output qubit subspaces for many of the SSGs analyzed in this paper.

We begin by writing the intermediate state of an SSG as
\begin{equation}
    \ket{\psi_{\text{int}}} =
    \sum_{\pb\in S}\ket{\pb}\ket{\tilde{\Phi}_{\pb}}
    + \sum_{\pb'\notin S}\ket{\pb'}\ket{\tilde{\Phi}_{\pb'}}
\end{equation}
where
\begin{equation}
    \ket{\tilde{\Phi}_{\pb}} := \braket{\pb|\psi_{\text{int}}}
\end{equation}
is the unnormalized output state associated with detector pattern $\pb$, and $S$ is the set of successful detector patterns.
For $\pb\in S$, the corresponding normalized heralded output state is
\begin{equation}
    \ket{\Phi_{\pb}} =
    \frac{\ket{\tilde{\Phi}_{\pb}}}{\sqrt{\Pr(\pb;\Id)}}
    = \frac{\ket{\tilde{\Phi}_{\pb}}}{c(\pb)},
    \qquad
    c(\pb) := \sqrt{\Pr(\pb;\Id)} .
\end{equation}
Equivalently, $\ket{\tilde{\Phi}_{\pb}} = c(\pb)\ket{\Phi_{\pb}}$, with the phase of $\ket{\Phi_{\pb}}$ chosen so that $c(\pb)$ is real and non-negative.
In general, the normalized states $\ket{\Phi_{\pb}}$ are not orthogonal for different successful detection patterns $\pb$.

We model input photon distinguishability by saying that each single photon input of the SSG has a probability $\beta_\text{D}$ of being distinguishable from the others.
Distinguishability in single photon input $i$ transforms the input state as
\begin{equation}
    \ket{\psi_\inp} \rightarrow \ket{\psi_\inp^{(i)}} := a_i d_i^\dagger \ket{\psi_\inp},
\end{equation}
where $d_i^\dagger$ is the creation operator for a distinguishable photon in rail $i$. 

We assume no dispersion in the interferometers so both the good and distinguishable mode operators obey \Cref{eq:mathcalU}.
Therefore, we have
\begin{equation}
    \ket{\psi_\text{int}^{(i)}} := \mathcal{U}\ket{\psi_\inp^{(i)}} = \sum_{kj}U_{ki}^* U_{ji} a_k d_j^\dagger \ket{\psi_\text{int}}.
\end{equation}

\begin{figure}
    \centering 
    \includegraphics[width=1\columnwidth]{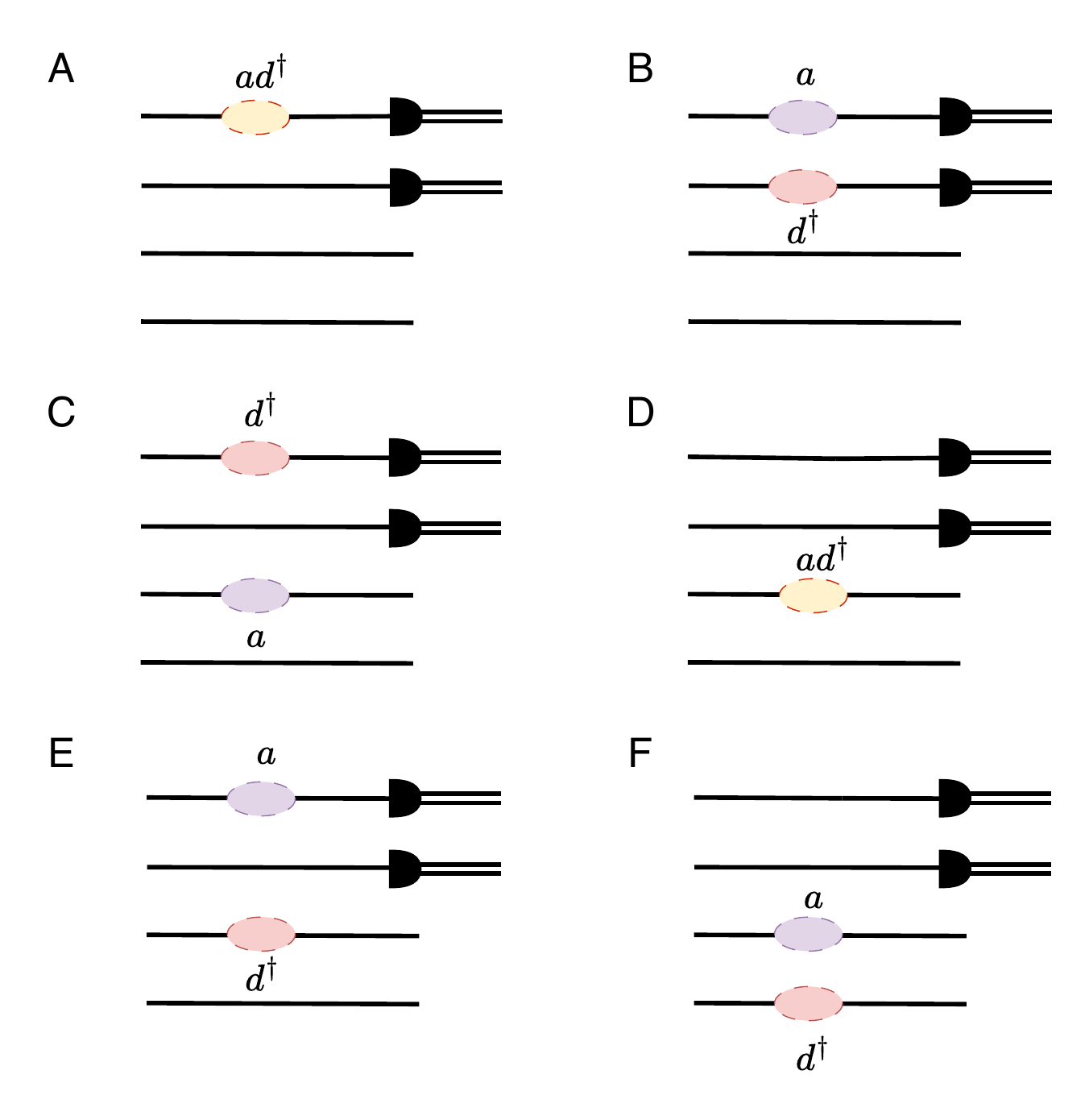}
    \caption{Categories of processes by which distinguishable photons can be created and ideal photons annihilated across the detected and output modes of an SSG. \label{fig:dist_process}}
\end{figure}

Fig. \ref{fig:dist_process} lays out the different categories of processes which contribute to $\ket{\psi_\text{int}^{(i)}}$.
The strategy of this appendix consists of calculating the contributions of particular processes A - F to some response coefficients and of showing that the others cannot contribute for interferometers with certain symmetries.

\subsection{The $(B, I, \ldots, I)$ Subspace}\label{sec:q_b}

To get the response corresponding to this subspace we must simply turn a good output qubit into a ``bad'' one.
Let that qubit live in the rails
\begin{equation}
    R_1 = \{r_1, r_2\}.
\end{equation}
Consider now the terms
\begin{equation}\label{eq:proc_d_terms}
    \ket{\psi_\text{int}^{(i)}} = \sum_{\pb\in S} c(\pb)\ket{\pb}\left(\sum_{k, j \in R_1} U_{ki}^* U_{ji} a_k d_j^\dagger \ket{\Phi_{\pb}}\right) + \ldots
\end{equation}
We begin by focusing on the simplest type of distinguishability process where a good photon is annihilated and a distinguishable one created in the same visible mode, i.e.\ process D from \Cref{fig:dist_process}.
If the target post-measurement state is an $n$-GHZ state $\ket{\phi^\pm_n}$, as it is for all of the SSGs in this paper, then the contribution from process D is
\begin{multline}\label{eq:proc_d_contribution}
        \sum_{k\in R_1} |U_{ki}|^2 a_k d_k^\dagger \ket{\phi^\pm_n} = \\
         \frac{1}{\sqrt{2}}\left(|U_{r_1 i}|^2\ket{1_d, 0, 1, 0 \dots} \pm |U_{r_2, i}|^2 \ket{0, 1_d, 0, 1 \dots}\right).
\end{multline}
Summing over the contributions from all the possible inputs, we see that the probability of detecting pattern $\pb$ and then measuring the tuple $\bm{Q}_B = (B, I, \ldots, I)$ is
\begin{equation}
     \sum_{i = 1}^{n_\inp}\left(\frac{|U_{r_1 i}|^2 + |U_{r_2 i}|^2}{2}\right)^2  |c(\pb)|^2 \beta_\text{D}
\end{equation}
or equivalently, the contribution to $\bm{Q}_B$ from process D is
\begin{equation}\label{eq:proc_d_bound}
    c^{(D)}_{\text{D}_\inp\rightarrow \bm{Q}_B} = \sum_{i = 1}^{n_\inp}\left(\frac{|U_{r_1 i}|^2 + |U_{r_2 i}|^2}{2}\right)^2
\end{equation}

It is clear that process F from \Cref{fig:dist_process} cannot contribute to $c_{\text{D}_\inp\rightarrow \bm{Q}_B}$ since the creation of a distinguishable photon in a different output mode to the annihilated good photon would cause either a Pauli error or leakage into a non-computational subspace.
Thus, we focus on the contributions from process E.

\subsubsection{Process E}

Consider the terms
\begin{equation}
    \ket{\psi_\text{int}^{(i)}} = \sum_{\pb\notin S}\sum_{k\in\mu_\text{det}} U^*_{ki}a_k \ket{\pb}\sum_{j\in R_1} U_{ji}d_j^\dagger\ket{\tilde{\Phi}_{\pb}} + \ldots
\end{equation}
which correspond to a distinguishable photon being created in an output mode, and a photon being annihilated in a detector pattern $\pb$ which does not herald success, leading to a pattern $\pb^*$ which does.
Ideally, when we herald $\pb^*$ we would obtain the unnormalized post-measurement state
\begin{equation}
    \ket{\tilde{\Phi}_{\pb^*}} = \braket{\pb^*|\psi_\text{int}} = \frac{1}{\sqrt{p_k}} \bra{\pb}a_k^\dagger\ket{\psi_\text{int}}.
\end{equation}
However, what we actually get is a state proportional to
\begin{equation}\label{eq:proc_e_bad_state}
    \sum_{j\in R_1} U_{ji}d_j^\dagger\ket{\tilde{\Phi}_{\pb}}.
\end{equation}

The value of $c_{\text{D}_\inp\rightarrow \bm{Q}_B}$ depends on the overlaps between \Cref{eq:proc_e_bad_state} and the states
\begin{equation}\label{eq:dist_basis}
    d_j^\dagger a_j \ket{\tilde{\Phi}_{\pb^*}}\quad \forall j\in R_1
\end{equation}
For a GHZ target, these two orthogonal branch states span the single-bad-photon sector in which the first output qubit has hidden label $B$ and no visible bit flip.
This sector decomposes into the $\bm{Q}_B$ and $\bm{Q}_{BZ}$ subspaces, corresponding to the appropriate even and odd GHZ superpositions.
Therefore, if the process E output state has zero overlap with both states in \Cref{eq:dist_basis}, it has zero overlap with $\bm{Q}_B$.
This is the case if
\begin{equation}
    Q_{kj}^{(E)}(\pb) := \braket{\psi_{\text{int}} | \pb}\bra{\pb}a_k^\dagger a_j\ket{\psi_{\text{int}}} = 0
\end{equation}
for any $k\in\mu_{\text{det}}$ and $j\in R_1$.

We now prove that $Q_{kj}^{(E)}(\pb) = 0$ for the SSGs we highlight in this paper due to symmetries of the interferometers, similar to refs.~\cite{saied2024general, dittel2018totally,dittel2018totally2}.
First, we rewrite the quantity as
\begin{equation}\label{eq:sub_u_dist}
    Q_{kj}^{(E)}(\pb) = \bra{\psi_\inp}\mathcal{U}^\dagger \ketbra{\pb}{\pb} a_k^\dagger a_j \mathcal{U}\ket{\psi_\inp}.
\end{equation}
Consider now a symmetry of the form
\begin{equation}\label{eq:symm}
    UPD_\inp = \left(D_{\text{det}}\oplus T_\out\right)U
\end{equation}
where $P$ is a permutation matrix, $D_\inp$, and $D_{\text{det}}$ are diagonal matrices acting on the input and detection modes respectively, and $T_\out$ is a unitary transfer matrix describing an interferometer acting on the output modes.
We use the operators $\mathcal{D}_\inp$, $\mathcal{P}$, $\mathcal{D}_\text{det}$, $\mathcal{T}_\out$ to describe the Fock space unitary generated by the linear optical transformations $D_\inp, P, D_\detc, T_\out$.
We also note that, for all the SSGs considered in \Cref{sec:example_circuits}, the single photon inputs obey
\begin{equation} \label{eq:inp_perm}
\ket{\psi_\inp} = \mathcal{P}\ket{\psi_\inp}
\end{equation}
and, furthermore, $\mathcal{D}_\inp$ has no bearing since it will only introduce a global phase to the state, so we set $D_\inp=\I$.

Using these observations from \Cref{eq:symm,eq:inp_perm} we rewrite \Cref{eq:sub_u_dist} as
\begin{equation}
    \begin{split}
        Q_{kj}^{(E)}(\pb) &= \bra{\psi_\inp}\mathcal{U}^\dagger \mathcal{D}_\detc^\dagger\ketbra{\pb}{\pb} a_k^\dagger \mathcal{D}_\detc \mathcal{T}^\dagger_\out a_j \mathcal{T}_\out \mathcal{U}\ket{\psi_\inp}\\
        & = e^{-i \alpha_k}\braket{\psi_{\text{int}}| \pb}\!\bra{\pb} a_k^\dagger \mathcal{T}^\dagger_\out a_j \mathcal{T}_\out\ket{\psi_{\text{int}}}
    \end{split}
\end{equation}
where in the last step we've used the fact that $a_k^\dagger \mathcal{D}_\detc = e^{-i \alpha_k} \mathcal{D}_\detc a_k^\dagger$ for the phase array $D_\detc = \text{diag}(e^{i\alpha_1}, e^{i\alpha_2}, \ldots)$.
$T_{jn}$ denotes the transfer matrix elements corresponding to the unitary operator $\mathcal{T}_\out$
\begin{equation}
    \mathcal{T}_\out^\dagger a_j \mathcal{T}_\out = \sum_{n\in\mu_\text{out}} T_{jn}a_n.
\end{equation}
so this gives us the set of linear equations:
\begin{equation}
    \begin{split}
        Q_{kj}^{(E)}(\pb) &= e^{-i \alpha_k}\sum_{n\in\mu_\text{out}}T_{jn}Q_{kn}^{(E)}(\pb)\\
        & \Rightarrow \sum_{n\in\mu_\text{out}}\left(\delta_{jn} - e^{-i\alpha_k}T_{jn}\right)q_n = 0
    \end{split}
    \label{eq:nonzero_qn}
\end{equation}
for a fixed $k$ and $\pb$ where $q_j := Q_{kj}^{(E)}$.

Recall that process E contributing to the probability of projecting the SSG output into the bad photon subspace requires a non-zero $q_n$.
\Cref{eq:nonzero_qn} means that a non-zero $q_n$ requires
\begin{equation}\label{eq:proc_e_test}
    \Xi_k := \det\left(I - e^{-i\alpha_k}T_\out\right) = 0.
\end{equation}
Conversely, if we can find some symmetry for which $\Xi_k \neq 0$ for all $k\in \mu_\text{det}$ then $Q_{kj}^{(E)} = 0$ and process E does not contribute to $c_{\text{D}_\inp\rightarrow\bm{Q}_B}$.

We also note here a property that can make it easier to find symmetries which satisfy the test in \Cref{eq:proc_e_test}.
Let $\left(D_{\text{det}}\oplus T_\out\right)$ be a symmetry of $U$ and consider the matrix
\begin{equation}
    U_1 = \left(\I_{\text{det}}\oplus T_1\right)U,
\end{equation}
where $T_1$ is any unitary applied to the output modes of the circuit.
Notice that
\begin{equation}
    \begin{split}
        U_1P &= \left(\I_{\text{det}}\oplus T_1\right)UP = \left(\I_{\text{det}}\oplus T_1\right)\left(D_{\text{det}}\oplus T_\out\right)U\\
        & = \left(D_{\text{det}}\oplus T_1T_\out T_1^{-1}\right)U_1.
    \end{split}
\end{equation}
This shows that the pair $D_{\text{det}}, T'_\out = T_1T_\out T_1^{-1}$ is a symmetry of $U_1$.
Substituting this into \Cref{eq:proc_e_test} we get
\begin{equation}
    \begin{split}
        \Xi_k &= \det\left(I - e^{-i\alpha_k}T'_\out\right) = |\det(T_1)|^2\det\left(I - e^{-i\alpha_k}T_\out\right)\\
        &= \det\left(I - e^{-i\alpha_k}T_\out\right).
    \end{split}
\end{equation}
Therefore, the extra operation $\left(\I_{\text{det}}\oplus T_1\right)$ doesn't affect \Cref{eq:proc_e_test}.

\subsubsection*{Symmetries for our SSGs}
For all of the following, let $\omega_{n} = e^{2\pi \ii/{n}}$.
For compactness, we list only the detector and output blocks, $D_\detc$ and $T_\out$, of each symmetry.
These were obtained by checking all possible permutations of the inputs.
In each case, we have verified by direct multiplication that there exist an input-mode permutation $P$ and diagonal input phase matrix $D_\inp$ such that \Cref{eq:symm} holds.

5DFT BSG:
\begin{equation}
D_\text{det} = \begin{pmatrix} 1 \end{pmatrix}, 
T_\out = \begin{pmatrix}
    \omega_5 & 0 & 0 & 0\\
    0 & \omega_5^2 & 0 & 0\\
    0 & 0 & \omega_5^3 & 0\\
    0 & 0 & 0 & \omega_5^4\\
\end{pmatrix}
\end{equation}
\begin{equation}
\Xi_1 = (1 - \omega_5)(1 - \omega_5^2)(1 - \omega_5^3)(1 - \omega_5^4)
\end{equation}
2-3DFT BSG:
\begin{equation}
D_\text{det} = \begin{pmatrix}
        1 & 0\\
        0 & 1\\
\end{pmatrix},
T_\out = \begin{pmatrix}
    \omega_3 & 0 & 0 & 0\\
    0 & \omega_3 & 0 & 0\\
    0 & 0 & \omega_3^2 & 0\\
    0 & 0 & 0 & \omega_3^2\\
\end{pmatrix}
\end{equation}
\begin{equation}
\Xi_1 = \Xi_2 = (1 - \omega_3)^2(1 - \omega_3^2)^2 = 9
\end{equation}
S6 BSG (for the $S_6$ unitary):
\begin{equation}
    D_\text{det} = \begin{pmatrix}
            1 & 0\\
            0 & \omega_3^2\\
    \end{pmatrix},
    T_\out = \frac{1}{2}\begin{pmatrix}
        \omega_3 & \omega_3 & -\omega_3^2 & \omega_3^2\\
        -\omega_3 & \omega_3 & \omega_3^2 & \omega_3^2\\
        1 & -1 & \omega_3 & \omega_3\\
        1 & 1 & \omega_3 & -\omega_3\\
    \end{pmatrix}
\end{equation}
\begin{equation}
\Xi_1 = 3-i\sqrt{3},\quad \Xi_2 = 3+i\sqrt{3}
\end{equation}
H8 BSG:
\begin{equation}
D_\text{det} = \begin{pmatrix}
    1 & 0 & 0 & 0\\
    0 & 1 & 0 & 0\\
    0 & 0 & 1 & 0\\
    0 & 0 & 0 & 1\\
\end{pmatrix},
T_\out = \begin{pmatrix}
    -1 & 0 & 0 & 0\\
    0 & -1 & 0 & 0\\
    0 & 0 & -1 & 0\\
    0 & 0 & 0 & -1\\
\end{pmatrix}
\end{equation}
\begin{equation}
\Xi_1 = \Xi_2 = \Xi_3 = \Xi_4 = 16
\end{equation}
2-H4 3GHZ Generator:
\begin{equation}
D_\text{det} = \begin{pmatrix}
    1 & 0\\
    0 & -1\\
\end{pmatrix},
T_\out = \begin{pmatrix}
    0 & -1 & 0 & 0 & 0 & 0\\
    1 & 0 & 0 & 0 & 0 & 0\\
    0 & 0 & 0 & 0 & 0 & 1\\
    0 & 0 & 1 & 0 & 0 & 0\\
    0 & 0 & 0 & -1 & 0 & 0\\
    0 & 0 & 0 & 0 & 1 & 0\\
\end{pmatrix}
\end{equation}
\begin{equation}
\Xi_1 = \Xi_2 = 4
\end{equation}
4-3DFT 4GHZ Generator:
\begin{equation}
D_\text{det} = \begin{pmatrix}
    1 & 0 & 0 & 0\\
    0 & 1 & 0 & 0\\
    0 & 0 & 1 & 0\\
    0 & 0 & 0 & 1\\
\end{pmatrix},
\end{equation}
\begin{equation}
T_\out = \begin{pmatrix}
    \omega_3 & 0 & 0 & 0 & 0 & 0 & 0 & 0\\
    0 & \omega_3 & 0 & 0 & 0 & 0 & 0 & 0\\
    0 & 0 & \omega_3^2 & 0 & 0 & 0 & 0 & 0\\
    0 & 0 & 0 & \omega_3^2 & 0 & 0 & 0 & 0\\
    0 & 0 & 0 & 0 & \omega_3 & 0 & 0 & 0\\
    0 & 0 & 0 & 0 & 0 & \omega_3 & 0 & 0\\
    0 & 0 & 0 & 0 & 0 & 0 & \omega_3^2 & 0\\
    0 & 0 & 0 & 0 & 0 & 0 & 0 & \omega_3^2\\
\end{pmatrix}
\end{equation}
\begin{equation}
\Xi_1 = \Xi_2 = \Xi_3 = \Xi_4 = 81
\end{equation}

Thus, for all of the SSGs in \Cref{sec:example_circuits}, process E, where a ``bad'' photon appears at an output mode and a ``good'' photon is annihilated at a detected mode, does not contribute to bad photon response coefficients considered here.

\subsubsection{Bounds}

We can bound the response coefficient $c_{\text{D}_\inp\rightarrow \bm{Q}_B}$ using the results of the previous sections.
In \Cref{eq:proc_d_bound}, we showed that the contribution from process D of \Cref{fig:dist_process} is
\begin{equation}\label{eq:b_response}
    c^{(D)}_{\text{D}_\inp\rightarrow \bm{Q}_B} = \sum_{i = 1}^{n_\inp}\left(\frac{|U_{r_1 i}|^2 + |U_{r_2 i}|^2}{2}\right)^2
\end{equation}
while the results presented up to this point of the section prove that the contributions from all other processes are 0.

To obtain a bound on this value we set
\begin{equation}
    x_i = \frac{|U_{r_1 i}|^2 + |U_{r_2 i}|^2}{2} \geq 0.
\end{equation}
Then,
\begin{equation}
    c_{\text{D}_\inp\rightarrow \bm{Q}_B} = (x_1^2 + \ldots + x_{n_\inp}^2).
\end{equation}
Using the power-mean inequality
\begin{equation}
    \left(\frac{x_1^2 + \ldots + x_{n_\inp}^2}{{n_\inp}}\right)^{\frac{1}{2}} \geq \frac{x_1 + \ldots + x_{n_\inp}}{{n_\inp}}
\end{equation}
we write
\begin{equation}
    c_{\text{D}_\inp\rightarrow \bm{Q}_B} \geq \frac{(x_1 + \ldots + x_{n_\inp})^2}{{n_\inp}} .
\end{equation}
For SSGs with no vacuum input modes, $m=n_\inp$, unitarity gives
\begin{equation}
    x_1 + \ldots + x_{n_\inp}
    = \frac{1}{2}\sum_{i=1}^{n_\inp}\left(|U_{r_1 i}|^2 + |U_{r_2 i}|^2\right)
    = 1,
\end{equation}
and thus, for any SSG in which process E does not contribute to $\bm{Q}_B$, that is, any SSG with the proper symmetries
\begin{equation}
    \begin{split}
    c_{\text{D}_\inp\rightarrow \bm{Q}_B} \geq \frac{1}{{n_\inp}} \quad \quad \quad \\
    \text{if } m = n_\inp \text{ and no Proc. E contribution}.
    \end{split}
\end{equation}
The condition $m=n_\inp$, and hence the lower bound $c_{\text{D}_\inp\rightarrow \bm{Q}_B} \geq 1/n_\inp$, holds for all of the SSGs in \Cref{sec:example_circuits}.
For a more general interferometer with additional vacuum input modes, $m>n_\inp$, unitarity only implies $\sum_i x_i \leq 1$, so we must use the more general bound of $(\sum_i x_i)^2/n_\inp$ instead of $1/n_\inp$.
The $1/n_\inp$ lower bound when $m=n_\inp$ is saturated if and only if $x_1 = x_2 = \ldots = x_{n_\inp}$, as is the case for all of the SSGs in \Cref{sec:example_circuits} except for the S6 BSG and 4-3DFT 4GHZ/4-line.

We can get an upper bound for the $m = n_\inp$ case by noting
\begin{equation}
    \begin{split}
        x_1^2 + \ldots + x_{n_\inp}^2 &= x_1x_1 + \ldots + x_{n_\inp}x_{n_\inp}\\
        & \leq \max_i(x_i)(x_1 + \ldots + x_{n_\inp})\\
        & = \max_i(x_i).
    \end{split}
\end{equation}

\subsection{The $(BZ, I, \ldots, I)$ Subspace}

The analysis here is almost identical to the previous section.
For the same reasons as before, processes E and F do not contribute to $c_{\text{D}_\inp\rightarrow \bm{Q}_{BZ}}$, where $\bm{Q}_{BZ} = (BZ, I, \ldots, I)$.
The contribution from process D is given by projecting the GHZ state $\ket{\phi^\pm_n}$ from \Cref{eq:proc_d_contribution} onto the flipped GHZ state $\ket{\phi^\mp_n}$, which gives
\begin{equation}\label{eq:bz_response}
    c_{\text{D}_\inp\rightarrow \bm{Q}_{BZ}} = \sum_{i = 1}^{n_\inp}\left(\frac{|U_{r_1 i}|^2 - |U_{r_2 i}|^2}{2}\right)^2.
\end{equation}

\subsection{The $(BX, I, \ldots, I)$ Subspace}

For this response we note that process D no longer contributes to the probability of measuring the output state in $\bm{Q}_{BX} = (BX, I, \ldots, I)$, but that process F now can.
To show that process E still does not contribute, we repeat the analysis from \Cref{sec:q_b} but now, rather than calculating overlaps between the output state given by \Cref{eq:proc_e_bad_state} and \Cref{eq:dist_basis}, we calculate them between \Cref{eq:proc_e_bad_state} and
\begin{equation}
    \left(d_{r_2}^\dagger a_{r_1} + d_{r_1}^\dagger a_{r_2}\right)\ket{\tilde{\Phi}_{\pb^*}}.
\end{equation}

The contribution from process F, and thus the entire value of the response coefficient, is given by
\begin{equation}\label{eq:bx_response}
    c_{\text{D}_\inp\rightarrow \bm{Q}_{BX}} = \sum_{i = 1}^{n_\inp}\left|\frac{U^*_{r_1 i}U_{r_2 i} + U^*_{r_2 i}U_{r_1 i}}{2}\right|^2.
\end{equation}

\subsection{The $(BY, I, \ldots, I)$ Subspace}

Analogously, we find that
\begin{equation}\label{eq:by_response}
    c_{\text{D}_\inp\rightarrow \bm{Q}_{BY}} = \sum_{i = 1}^{n_\inp}\left|\frac{U^*_{r_1 i}U_{r_2 i} - U^*_{r_2 i}U_{r_1 i}}{2}\right|^2.
\end{equation}

Combining the results of \Cref{eq:b_response}, (\ref{eq:bz_response}), (\ref{eq:bx_response}), and (\ref{eq:by_response}), we get the relationship
\begin{equation}
    c_{\text{D}_\inp\rightarrow \bm{Q}_{B}} = c_{\text{D}_\inp\rightarrow \bm{Q}_{BX}} + c_{\text{D}_\inp\rightarrow \bm{Q}_{BY}} + c_{\text{D}_\inp\rightarrow \bm{Q}_{BZ}}.
\end{equation}

\subsection{Success-weighted fidelity}

This subsection does not relate to the error response coefficient framework of the rest of the paper.
However, it is included as it is useful to see how the insights discussed so far can let us derive some simple fidelity expressions.

For each successful pattern $\pb$, let $\ket{\Phi_{\pb}}$ be the normalized ideal output state, with any pattern-dependent Pauli-frame correction understood.
We define the success-weighted fidelity
\begin{equation}
    \mathcal{F}_\text{sw}
    :=
    \sum_{\pb\in S}
    \bra{\Phi_{\pb}}
    \tilde{\rho}_{\out}^{(\pb)}
    \ket{\Phi_{\pb}},
\end{equation}
where $\tilde{\rho}_{\out}^{(\pb)}$ is the unnormalized heralded output state for pattern $\pb$.
This quantity is not conditioned on successful heralding; in the ideal case, $\mathcal{F}_\text{sw}=\Pr(\text{success};\Id)$.
The usual fidelity of the state conditioned on success would require dividing by the noisy success probability.

To calculate the first-order distinguishability contribution to $\mathcal{F}_\text{sw}$, we look at processes A, B, and C in \Cref{fig:dist_process}.

For process A we proceed similarly to the analysis of process D in \Cref{sec:q_b}.
As in \Cref{eq:proc_d_terms}, consider the terms 
\begin{equation}
    \begin{split}
        \ket{\psi_\text{int}^{(i)}} &= \sum_{\pb\in S} c(\pb)\left(\sum_{k \in \mu_\text{det}} |U_{ki}|^2 a_k d_k^\dagger \ket{\pb}\right)\ket{\Phi_{\pb}} + \ldots\\
    \end{split}
\end{equation}
where we now both create and annihilate a photon in the detector mode $k$.
Here we are taking the liberty of identifying detector states which differ only by replacing a good detected photon with a bad detected photon in the same mode.
This is valid in this particular case because the detectors are insensitive to this hidden degree of freedom and the detected modes are traced out.
Under this identification, the output state is the same as in the ideal case, $\ket{\Phi_{\pb}}$, and the corresponding probability weight for input $i$ is
\begin{equation}
    \sum_{k\in\mu_\text{det}} p_k |U_{ki}|^4.
\end{equation}
In the above equation we have used the fact that there are $p_k$ good photons which could be replaced in mode $k$, and
\begin{equation}
    |U_{ki}|^2 = \frac{1}{{n_\inp}}
\end{equation}
for all $k\in\mu_\text{det}$ in all of our SSGs.
This gives a total process A ideal-output weight of $n_\detc / n_\inp$ after summing over all possible distinguishable input photons.

Therefore, for a fixed successful pattern $\pb$, the ideal no-error contribution and process A contribution together give
\begin{equation}
    \bra{\Phi_{\pb}}
    \tilde{\rho}_{\out}^{(\pb)}
    \ket{\Phi_{\pb}}
    =
    \Pr(\pb;\Id)
    \left[
        1
        -
        \left(
            n_\inp
            -
            \frac{n_\detc}{n_\inp}
        \right)\beta_\text{D}
    \right]
    + \bigO(\beta_\text{D}^2).
\end{equation}
Here and below, this expression assumes that processes B and C do not contribute to the target-state overlap, as we show below.
Summing over $\pb\in S$ gives
\begin{equation}
    \begin{split}
    \mathcal{F}^{(\text{A})}_\text{sw}
    & =
    \Pr(\text{success};\Id)
    \left[
        1
        -
        \left(
            n_\inp
            -
            \frac{n_\detc}{n_\inp}
        \right)\beta_\text{D}
    \right]
    + \bigO(\beta_\text{D}^2) \\
    \text{if } & m=n_\inp \text{ and } |U_{ki}|^2=1/n_\inp \ \forall\ k \in \mu_\detc. 
    \end{split}
    \label{eq:proc_a_fidelity}
\end{equation}

For processes B and C we can define two quantities
\begin{equation}\label{eq:proc_bc}
    \begin{split}
        Q_{jk}^{(B)}(\pb) &:= \braket{\psi_{\text{int}}|{\pb}}\!\bra{\pb}a_j^\dagger a_k\ket{\psi_{\text{int}}}\\
        Q_{rs}^{(C)}(\pb) &:= \braket{\psi_{\text{int}}|{\pb}}\!\bra{\pb}a_r^\dagger a_s\ket{\psi_{\text{int}}}
    \end{split}
\end{equation}
where $j, k, r\in\mu_\text{det}$ and $s\in\mu_\text{out}$.
Similar to the analysis of process E, we can prove that these quantities are both 0 for the SSGs in \Cref{sec:example_circuits}, showing that neither of these processes contributes to the success-weighted target overlap.

For process B, we can rewrite $Q_{jk}^{(B)}(\pb)$ as
\begin{equation}
    Q_{jk}^{(B)}(\pb) = \bra{\pb}a_j^\dagger a_k \rho_d\ket{\pb}
\end{equation}
where
\begin{equation}
    \rho_d = \tr_\text{out}\left[\ketbra{\psi_\text{int}}{\psi_\text{int}}\right]
\end{equation}
is the reduced density matrix for the detection modes.

Let $\ket{\pb^-}$ be defined such that
\begin{equation}
    \ket{\pb} = \ket{p_j,p_k}\ket{\pb^-}.
\end{equation}
A symmetry of the form of \Cref{eq:symm} implies that
\begin{align}
    \bra{\pb}a_j^\dagger a_k \rho_d\ket{\pb} &= \bra{\pb^-}\braket{p_j-1,p_k+1|\rho_d|p_j,p_k}\ket{\pb^-}\\
    &= \bra{\pb^-}\braket{p_j-1,p_k+1|\mathcal{D}_\detc\rho_d \mathcal{D}_\detc^\dagger|p_j,p_k}\ket{\pb^-}\\
    &= e^{-i(\alpha_j - \alpha_k)}\bra{\pb}a_j^\dagger a_k \rho_d\ket{\pb},
\end{align}
up to normalization factors.
This says that if there is some symmetry \Cref{eq:symm} with a detector phase matrix $D_\text{det}$ such that $e^{-i(\alpha_j - \alpha_k)} \neq 1$, we must have $Q_{jk}^{(B)}(\pb) = 0$.
Recall that $j\neq k$ for process B.
For the S6 BSG and 2-H4 3GHZ generator, the symmetries listed above already have this property and, while the listed symmetries for the other SSGs do not have this property, it is possible to find ones which do.
Thus, $Q_{jk}^{(B)}(\pb) = 0$ and process B does not contribute.

It is easy to see from the form of \Cref{eq:proc_bc} that $Q_{rs}^{(C)}(\pb)$ has the same form as $Q_{kj}^{(E)}(\pb)$ and thus, using the symmetries defined previously, we can prove that process C also does not contribute.
This leaves \Cref{eq:proc_a_fidelity} as the first-order distinguishability contribution to the success-weighted fidelity.

\section{The Standard n-GHZ generator}
\begin{figure*}
    \includegraphics[width=1.5\columnwidth]{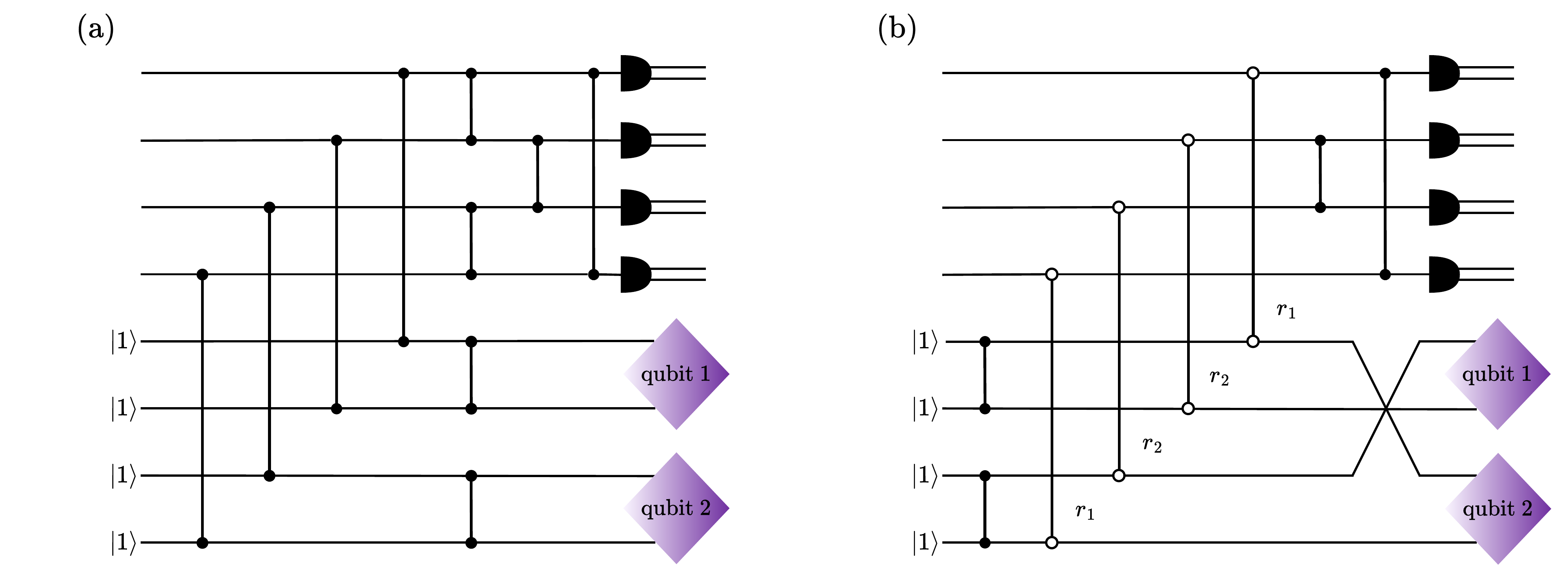}
    \caption{
    The standard (a) and primate-converted with differential coupling down (b) Bell state generators. 
    This can be generalized to an $n$-GHZ generator following construction from ref.~\cite{bartolucci2021creation}.
    \label{fig:pcGHZ}}
\end{figure*}
\subsection{Recap}\label[appendix]{app:n_ghz}
It is well known that it is possible to create an n-GHZ state from $2n$ single photons with probability $1/2^{2n-1}$ \cite{browne2005resource,zou2005generation, zhang2008demonstration,gimeno2016towards,bartolucci2021creation}.
The standard version of this circuit can be seen in \Cref{fig:pcGHZ}(a).
We refer to the beamsplitters which couple the occupied input modes to the unoccupied ones as the ``coupling down'' beamsplitters.

The version of the standard SSG that we consider uses the identity used in \Cref{app:multi_identity} to remove the layer of beamsplitters on the output modes after the coupling-down layer and replace them with 50:50 beamsplitters coupling pairs of the input modes which are occupied by single photons.
This new circuit is referred to as the ``primate'' version of the n-GHZ generator due to its resemblance to the scheme in \cite{bartolucci2021creation} (Section VI therein).

\subsection{Coupling Down}\label[appendix]{app:diff_down_coupling}
It can also be shown that the circuit described above actually produces $n$-GHZs \textit{regardless of the reflectivity of the coupling down beamsplitters}.
Let the reflectivity of each of the coupling down beamsplitters be $r$, making the unitary transformation $U(r)$ a function of the coupling strength.
The standard SSG has the associated unitary $U(r = 1/2)$.
We find that the success probability of the SSG, as well as the probabilities of having certain errors on the output state are a function of $r$.
In particular, we have error response coefficients
\begin{align}
    c_{\text{A}_\inp\to\bm{Q}_M} &= 3(1 - r)\\
    c_{\tilde{\text{A}}_\detc\to\bm{Q}_M} &= \frac{1 - r}{r}\\
    c_{\text{L}_\detc\to\bm{Q}_V} &= \frac{r}{1 - r}
\end{align}
which shows the generator can be tailored to trade-off better responses to some errors, such as having a vacuum qubit in the output state, with worse responses to others, such as having a multiphoton qubit.
The success probability for an $n$-GHZ state is given by
\begin{align}
    \Pr(\text{success};\Id) = 2\left(r(1 - r)\right)^n
\end{align}
and has its maximum at $r = 1/2$.
We refer to this practice of varying $r$ to obtain a better error profile in the SSG as ``variable coupling down''.
Similar tradeoffs between detector errors and success probability were found in refs.~\cite{sparrow2018quantum,czaban2024suppressing} for different SSG circuits.

The fact that $c_{\text{L}_\detc\to\bm{Q}_V} \neq 0$ and $c_{\tilde{\text{A}}_\detc\to\bm{Q}_M} \neq 0$ for the standard n-GHZ generator proves that it does not intrinsically remove any sort of detector error.
It can be made to approach intrinsic removal of one of these errors in the limits $r\to 0$ or $r\to 1$, but in either of these limits the success probability goes to 0.
Also, as we suppress either vacuum or multiphoton error terms, the other increases correspondingly.
Because we have $c_{\text{L}_\detc\to\bm{Q}_V} = 1$ for $r = 1/2$, we know from \Cref{eq:SL_DL} that $c_{\text{L}_\inp\to\bm{Q}_V} = 2$ and so we can say that, to first order, the standard SSG doubles the per-photon vacuum probability from input to output.

It is also possible to tailor the error profile of an individual qubit in these $n$-GHZ generators by just modifying the values of $r$ for the coupling down beamsplitters that correspond to the two output modes for that particular qubit.
In this case, the success probability of the SSG is given by
\begin{align}
    \Pr(\text{success};\Id) = 2\prod_{i=1}^n r_i(1 - r_i),
\end{align}
where the $r_i$ are the values of $r$ corresponding to qubit $i$.
This is a useful tool for when different qubits from a generated seed state are used in different places within the RSG, as it allows the error properties of each qubit to be optimized for its specific role in the architecture.
We refer to this scenario in which there are multiple values of $r$ in a single SSG as ``differential coupling down''.

\subsection{Pauli-to-Leakage Conversion}\label[appendix]{app:pauli_to_leakage}

It is also possible to tailor the error properties of generated seed states by applying some linear optical transformation to the SSG output which leaves the intended output state unchanged but which transforms the error terms.
For a dual-rail encoded $n$-GHZ state, the simplest of these transformations is a set of mode swaps between modes which contain the same photon number for all terms in the state, e.g.\ swapping the logical-1 modes in the Bell states \begin{equation}
\ket{\phi^\pm} = \frac{\ket{1010}\pm\ket{0101}}{\sqrt{2}}.
\end{equation}
If combined with a projection back into the dual-rail subspace, these mode swaps perform pairwise $ZZ$ checks between the swapped modes and so any qubit state which undergoes the operation and is not stabilized by $+ZZ$, say the $\ket{\psi^\pm}$ Bell states, will be converted to a state outside the qubit subspace.
Conversely, swapping modes which correspond to opposite logical states, e.g.\ modes 2 and 3 in the $\ket{\phi^\pm}$ Bell states defined above, will perform a $-ZZ$ check when combined with a dual-rail subspace projection.

This procedure can take any part of the SSG output state which is equivalent to the ideal state followed by a Pauli $X$ or $Y$ error and convert it into leakage.
In doing so, it will create fusion heraldable states, containing correlated multiphoton and vacuum qubits.

Applying this procedure to the standard $n$-GHZ generator described in the previous sections removes almost all of the Pauli $X$ and $Y$ error from the output state, while raising the multiphoton and vacuum probabilities a corresponding amount.
The general effectiveness of this procedure is highly dependent on the structure of the output errors, since outputs which are already in a state where a vacuum qubit is correlated with a multiphoton can be brought back into the qubit subspace, resulting in additional Pauli errors.
For the SSGs described in the main text, we have not seen the same drastic Pauli reductions as for this circuit.

Combining all these ideas together for the $n$-GHZ generator gives us a helpful construction for the realization of this SSG. 
We refer to it as the ``primate-converted n-GHZ generator with differential coupling down'' and give the circuit in \Cref{fig:pcGHZ}(b).

\section{An $n$-GHZ state generator requires $\geq 2n$ input photons}\label[appendix]{app:photon_proof}

To show that an $n$ qubit, dual-rail GHZ state requires at least $2n$ photons, we will use a stronger result. 
To state this result, we first define a state property which we call \textit{mode irreducibility}. 
We say that a state has support on all $m$ modes if, for every mode $i$, the reduced state of that mode is not exactly the vacuum state, i.e.\
\begin{equation}
    \tr \! \left( \bra{0}_i \rho \ket{0}_i \right) \neq 1 .
\end{equation}
An $m$-mode irreducible state is a state with support on all $m$ modes which cannot be transformed using only linear optics into a tensor product of a state with support on fewer than $m$ modes and additional vacuum modes.

The more general result is the following: a state of $n_\inp$ single photons in $n_\inp$ modes, combined with any number of vacuum modes cannot herald a mode irreducible state containing more than $n_\inp$ modes.
Although alternative proofs have been given for this result~\cite{aralov2026photon,chin2024exponentially}, we sketch our proof here as the methods may offer an alternative for intuitive understanding.

We sketch a proof by contradiction for this statement by showing that any SSG which produces states with $m_\out>n_\inp$ modes is not $m_\out$-mode irreducible.
Without loss of generality, we can express any seed state generation circuit as shown in \Cref{fig:general_ssg} i.e. having all nonvacuum inputs coming before all vacuum inputs and all detected output modes before all undetected output modes.
For $n_\inp$ nonvacuum input modes, $m-n_\inp$ vacuum input modes, $m_\detc$ detected modes and $m_\out$ output modes, we can define a unitary matrix, $U$, which describes the mode transformation. 
The dimension of $U$ is $m = m_\detc + m_\out$.

\begin{figure}
    \centering 
    \includegraphics[width=1.1\columnwidth]{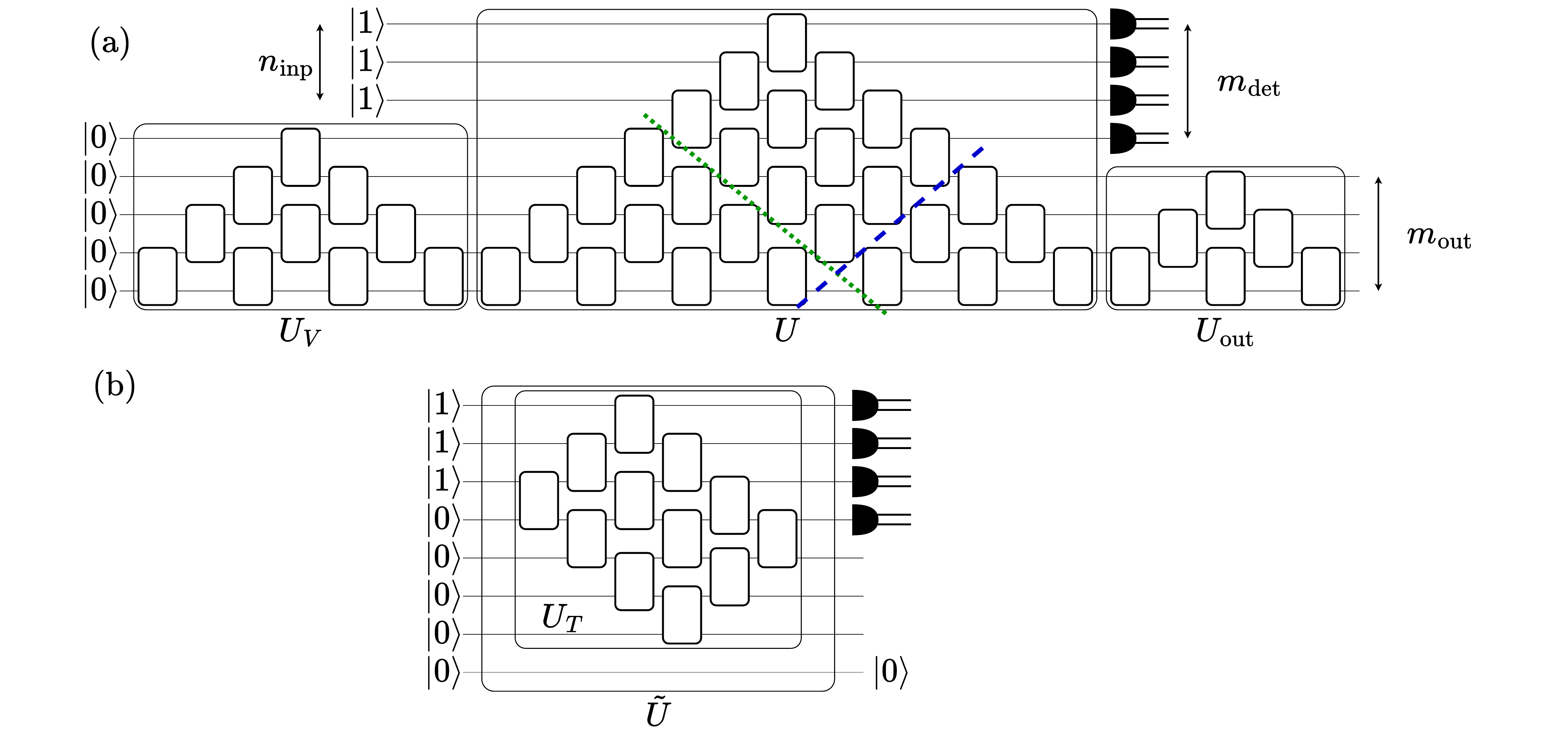}
    \caption{(a) General ballistic seed-state generator, with the additional freedom of $\tilde{U} = \left( \I_{m_\detc} \oplus U_\out \right) \cdot U \cdot \left(\I_{n_\inp} \oplus U_V \right)$ that leaves the mode irreducibility property of the output state unchanged. 
    Gray rounded boxes denote positions of beam splitters in a decomposition of the unitary transformations. 
    A dotted green line shows the beam splitters in $U$ which are able to be compiled into the action of $U_V$. 
    A dashed blue line shows the beam splitters which can be compiled into $U_\out$.
    (b) If we set $U_V$ and $U_\out$ such that all these compiled beam splitters act trivially and when $m_\out > n_\inp$, we end up with a transformation, $\tilde{U}=U_T \oplus \I_{(m-n_\inp-m_\detc)}$ that acts trivially on at least one of the vacuum modes.
    This means that any state produced by this generator will not be $m_\out$-mode irreducible. 
    In this example, $n_\inp=3, m=8, m_\detc=4, m_\out=4$. 
    \label{fig:mode_proof}}
\end{figure}

We now note that vacuum states are invariant under linear optical operations and that the $m$-mode irreducible property of a state is also invariant under linear optical operations.
This means that we can investigate the structure of the matrix 
\begin{align}
    \tilde{U} = \left( \I_{m_\detc} \oplus U_\out \right) \cdot U \cdot \left(\I_{n_\inp} \oplus U_V \right)
\end{align}
as drawn in \Cref{fig:mode_proof}.
Where $U_\out$ and $U_V$ are arbitrary $m_\out$ and $(m-n_\inp)$ dimensional unitary transformations acting on the output modes and vacuum input modes respectively.
The choice of $U_\out$ will change the elements of the last $m_\out$ rows of $\tilde{U}$, while the choice of $U_V$ will change the elements of the last $(m-n_\inp)$ columns of $\tilde{U}$.
We know that for any choice of $U_\out$, $U_V$, we will have a $\tilde{U}$ which is unitary.

If we define $T$ as the upper left $m_\detc \times n_\inp$ submatrix which describes the transformation from the nonvacuum input modes to the detector modes, it can be shown that $\tilde{U}$ is able to be set as any unitary matrix which preserves $T$ as the upper left $m_\detc \times n_\inp$ submatrix.
We will not prove this step explicitly, but it can be seen for example by considering unitary parameterizations of Hurwitz~\cite{hurwitz1897erzeugung,diaconis2017hurwitz} or Reck~\cite{reck1994experimental}, as shown in \Cref{fig:mode_proof}.
One such allowed option is $\tilde{U} = U_T \oplus \I_{(m_\out - n_\inp)}$, where $U_T$ is an $m_\detc+n_\inp$ dimension matrix that is a unitary dilation of $T$.

Because the $\I_{(m_\out - n_\inp)}$ part of $\tilde{U}$ will act trivially on some of the vacuum input states, we know that any state created by such a circuit will not be $m_\out$-mode irreducible.
Since $n$-GHZ dual rail states are $2n$-mode irreducible states (shown in \Cref{app:irr}), if the input ancillary resources are single photon Fock states, this tells us that we cannot use any fewer than $n_\inp=2n$ single photon Fock states to herald an $n$-GHZ state.
It can be similarly shown that any genuinely entangled state of $n$ multi-rail qu$d$its, where all $n$ single-qu$d$it reduced density matrices are rank $d$, requires at least $n d$ photons.

\section{Mode irreducibility of $n$-GHZ states}\label[appendix]{app:irr}
If we write it in first quantization, a pure fixed photon number $m$-mode irreducible state with $n$ photons will have a Schmidt rank of $m$ if we partition the state between one photon and the remaining $n-1$ photons.
To see this, recall that an $n$ photon, $m$ mode state in first quantization is the symmetric subspace of $(\mathbb{C}_m)^{\otimes n}$ and that linear optics will act locally on the $\mathbb{C}_m$ Hilbert space of each photon.
Therefore, linear optical transformations will preserve the Schmidt rank of the state between the $\mathbb{C}_m$ Hilbert space of one photon and the $(\mathbb{C}_m)^{\otimes (n-1)}$ space of the remaining photons.
From this, we see that a Schmidt rank of $m$ will imply that a state is $m$-mode irreducible.

Before proving the mode irreducibility of the dual-rail $n$-GHZ state, we first recall that the state of $n$ single photons is written like $\ket{1}^{\otimes n}$ in second quantization and like $\sum_{\sigma \in S_n} \otimes_{i=1}^n \ket{\sigma(i)}/\sqrt{n!}$ in first quantization, where $S_n$ is the symmetric group of degree $n$.

A dual-rail $n$-GHZ state in second quantization is written as 
\begin{align}
    \ket{n\text{-GHZ}} = \frac{1}{\sqrt{2}} \left( \ket{01}^{\otimes n} + \ket{10}^{\otimes n}\right)
\end{align}
which in first quantization becomes
\begin{multline}
    \ket{n\text{-GHZ}} = \frac{1}{\sqrt{2 n!}} \Biggl( \left[ \sum_{\sigma \in S_n} \bigotimes_{i=1}^n \ket{2 \sigma(i)} \right] \\ 
    + \left[ \sum_{\sigma \in S_n} \bigotimes_{i=1}^n \ket{2 \sigma(i)-1} \right]\Biggr) .
\end{multline}
We can then rearrange this to more clearly show the $\mathbb{C}_m \otimes (\mathbb{C}_m)^{\otimes (n-1)}$ tensor product structure for the Schmidt rank.
For each $i\in[1,n]$, define normalized $(n-1)$-photon symmetric states $\ket{E_i}$ and $\ket{O_i}$ as the states with one photon in each even mode $\{2,4,\ldots,2n\}\setminus\{2i\}$ and one photon in each odd mode $\{1,3,\ldots,2n-1\}\setminus\{2i-1\}$, respectively.
We can then write
\begin{align}
    \ket{n\text{-GHZ}}
    =
    \frac{1}{\sqrt{2n}}
    \sum_{i=1}^n
    \left(
        \ket{2i}\otimes\ket{E_i}
        +
        \ket{2i-1}\otimes\ket{O_i}
    \right).
\end{align}
The $2n$ single-photon states $\{\ket{2i},\ket{2i-1}\}_{i=1}^n$ are orthonormal.
The corresponding $(n-1)$-photon states $\{\ket{E_i},\ket{O_i}\}_{i=1}^n$ are also mutually orthonormal, since any two of them have different occupied mode sets.
This is therefore a Schmidt decomposition with $2n$ nonzero Schmidt coefficients, all equal to $1/\sqrt{2n}$.
So we see that the state has a Schmidt rank of $2n$, proving that the dual-rail $n$-GHZ state is $2n$-mode irreducible.
 
\section{5DFT Bell State Generator Variants}\label[appendix]{app:5dft_variants}
We prove here that the 5DFT BSG circuit of \Cref{sec:5dft} produces a Bell state with any input of the form $\ket{x, x, x, x, x}, x\geq 1$ upon the detection of $5x - 2$ photons in the detected mode, and with any permutation of $\ket{x + 1, x, x, x, x}, x\geq 1$  upon the detection of $5x - 1$ photons.
First consider the polynomial
\begin{equation}
    P_x(z_1, \ldots, z_5) = \frac{\left(\sum_{k=1}^5 z_k U_{k1}\right)^x}{\sqrt{x!}}\times\ldots\times\frac{\left(\sum_{k=1}^5 z_kU_{k5}\right)^x}{\sqrt{x!}}
\end{equation}
where $U = \text{DFT}_5$ is the circuit's transfer matrix.
We constructed this polynomial so that the state of the photons after the interferometer is $P_x(a_1^\dagger, \ldots, a_5^\dagger)\ket{0}$.
For $x = 1$, the scenario which corresponds to the original 5DFT Bell state generator of \Cref{sec:5dft}, one can find that
\begin{equation}
    P_1(z_1, \ldots, z_5) = \alpha z_1^5 + \beta p_{\psi}(z_2, z_3, z_4, z_5)z_1^3 + \ldots,
    \label{eq:5dft_poly_output}
\end{equation}
where $\alpha = 1/(25\sqrt{5}), \beta = -1/(5\sqrt{5})$ and the polynomial $p_{\psi}(z_2, z_3, z_4, z_5) = z_3z_4 + z_2z_5$.
Note that we are only showing the two highest powers in $z_1$ and the term proportional to $z_1^4$ is missing because of the destructive interference within the 5DFT.
\Cref{eq:5dft_poly_output} shows that if we herald 3 photons in the first output mode we project the remaining modes onto a Bell pair.

Notice now that for any $x$ we have
\begin{equation}
    P_x(z_1, \ldots, z_5) = \frac{1}{\sqrt{x!^5}}P_1(z_1, \ldots, z_5)^x.
    \label{eq:5dft_pm}
\end{equation}
Therefore, for any $x$ the polynomial $P_x(z_1, \ldots, z_5)$ is of the form
\begin{equation}
    P_x(z_1, \ldots, z_5) = \alpha_x z_1^{5x} + \beta_x p_{\psi}(z_2, z_3, z_4, z_5)z_1^{5x - 2} + \ldots
\end{equation}
Thus, if we herald $5x - 2$ photons in the first mode we get a Bell pair with probability
\begin{equation}
    p_{5x-2}(x) = 2|\beta_x|^2(5x - 2)!.
\end{equation}
From \Cref{eq:5dft_pm} we know that $\beta_x = x\alpha^{x-1}\beta/\sqrt{x!^5}$ and so
\begin{equation}
    p_{5x-2}(x) = \frac{50x^2}{x!^5}\frac{(5x-2)!}{3125^x}.
\end{equation}

We also observe that this SSG produces a Bell state if the input is of the form $(2, 1, 1, 1, 1)$.
Then, the corresponding polynomial is related to \Cref{eq:5dft_poly_output} as
\begin{equation}
    \begin{split}
        P_{+1}(z_1, \ldots, z_5) &= \frac{1}{\sqrt{2}}\left(z_1 U_{11} + \ldots\right)P_1(z_1, \ldots, z_5)\\
        & = \alpha'z_1^6 + \beta'p_{\psi}(z_2, z_3, z_4, z_5)z_1^4 + \ldots
    \end{split}
\end{equation}
Thus, if the input of the 5DFT BSG is of the form (2, 1, 1, 1, 1) we get a Bell pair with probability
\begin{equation}
    p_{+1}(x=1) = \frac{24}{625}.
\end{equation}
This observation holds for any 5-mode input of the form $(x + 1, x, x, x, x), x\geq 1$ with probability of success
\begin{equation}
    p_{+1}(x) = \frac{5 x - 1}{5 (x+1)} p_{5 x - 2}(x).
\end{equation}

\section{Detector pattern error dependence}\label[appendix]{app:det_pattern_error_dependence}
Many SSGs with multiple detector patterns have different valued error response coefficients $c_{E\to\bm{Q}}^{(\bm{p})}$ for different success patterns $\bm{p}$.
In particular, the Pauli error level vectors
\begin{equation}
    \bm{Q}_P = (P,\ideal,\ldots \ideal) \quad P\in\{X, Y, Z\}
\end{equation}
and some correlated leakage output subspaces like the $(V, V, M)$ subspace discussed in \Cref{sec:2-H4} can show large pattern-dependent variations.
Occasionally, it can be favorable to treat some nominally successful detector patterns as failures in order to trade off lower success probability for a higher quality of output state.
Deciding how to make such a trade will depend on the specific architecture being considered.

All of the SSGs from \Cref{sec:example_circuits} with more than one detector pattern, that is the 2-3DFT BSG, H8 BSG, 2-H4 3GHZ generator, and the 4-3DFT 4GHZ generator, have variations in errors on the output state between the different detector patterns.
For example, \Cref{tab:pattern_dist_z_coeffs} gives the minimum, maximum, and average error response coefficients $c_{\text{D}_\inp\to\bm{Q}_Z}^{(\bm{p})}$ relating source distinguishability to Pauli $Z$ error for these SSGs.

\begin{table}
    \begin{tabular}{c|c|c|c}
        & $c_{\text{D}_\inp\to\bm{Q}_Z, \text{min}}^{(\bm{p})}$ & $c_{\text{D}_\inp\to\bm{Q}_Z, \text{max}}^{(\bm{p})}$ & $c_{\text{D}_\inp\to\bm{Q}_Z, \text{avg}}^{(\bm{p})}$\\ \hline
        $\text{2-3DFT}$ & 1/12 & 1/3 & 5/24 \\
        $\text{H8}$ & 1/24 & 11/8 & 3/8 \\
        $\text{2-H4}$ & 3/40 & 7/8 & 5/24 \\
        $\text{4-3DFT}$ & 7/96 & 7/6 & 71/480 \\
    \end{tabular}
    \caption{Detector pattern variation in SSG source distinguishability to Pauli $Z$ response coefficients.
    Columns show the minimum, maximum and average error response over all patterns.}\label{tab:pattern_dist_z_coeffs}
\end{table}

Other error response coefficients, such as those relating input loss to the single qubit $V$ subspace, don't vary between detector patterns.
A comprehensive explanation for these variations is not yet understood.

\section{Error responses}\label[appendix]{app:error_responses}

Error response coefficients, averaged over all successful heralding detector patterns $\bm{p}$, for the SSGs in \Cref{sec:example_circuits} and \Cref{fig:ssg_circuits}, plus the 4-line generator mentioned in \Cref{sec:4-3dft}, are given in \Cref{tab:5dft_coeffs,tab:2-3dft_coeffs,tab:s6_coeffs,tab:h8_coeffs,tab:h4_coeffs,tab:4x3dft_coeffs,tab:4line_coeffs}.
Each row corresponds to some error channel $E$ that takes place during seed state generation, while each column corresponds to an error level $Q_i$.
The values in the table give the error response coefficients $c_{E\to\bm{Q}_{Q_i}}$, where $\bm{Q}_{Q_i}$ is the error level vector
\begin{equation}
\bm{Q}_{Q_i} = (Q_i, \ideal, \ldots, \ideal).
\end{equation}
These column choices are not intended to be exhaustive, just to highlight some important differences between the different SSGs.
We use a dash to indicate that a coefficient is zero via conservation of photon number.
When a coefficient is given as ``$0$'', this is due to its error suppressing properties.

We assume that errors which can be equivalently expressed as occurring on multiple qubits, such as a Pauli $Z$ error on a GHZ state, are symmetrized so as to have an equal probability of occurring in any configuration which leads to the same noisy state.
However, if multiple errors happen simultaneously, such as when a distinguishable qubit is accompanied by a Pauli error, if possible we choose to minimize the overall error weight by representing the state as if the errors had occurred on the same qubit.

In \Cref{tab:4x3dft_coeffs} there are two extra Pauli levels which cannot be described by a weight-1 Pauli string: $XX$ and $XY$ which refer to the correlated Pauli flip error level vectors
\begin{align}
    \bm{Q}_{XX} &= (X, X, \ideal, \ideal) \\ 
    \bm{Q}_{XY} &= (X, Y, \ideal, \ideal)  
\end{align}
which are the only subspaces labeled with correlated Pauli flips which have non-zero probability to leading order.
The coefficients given in \Cref{tab:4x3dft_coeffs} relate the input error to the probability of \textit{each} of the valid correlated Pauli flips, i.e. $X_1X_2$ and $X_3X_4$ for $XX$ and $Y_1X_2$, $X_1Y_2$, $Y_3X_4$ and $X_3Y_4$ for $XY$.

In \Cref{tab:4line_coeffs} the only Pauli errors which occur above the limit given by \Cref{eq:pauli_bound} project the state into the spaces labeled by $(I, Z, I, I), (I, I, Z, I)$ and $(I, Z, Z, I)$.
The column heading $Z$ refers to the coefficients for \textit{either} of the first two spaces with a single $Z$ label, which are equal, and the label $ZZ$ refers to the final one.
Each of these spaces can also be reached via other combinations of Pauli errors but we choose here to label them via these representations.

Both the 4GHZ and 4-line versions of the SSG are full-fusion-heraldable (FFH) for detector loss and $k=2$ fusion heraldable (FH) for detector noise. 
This can be found by showing that all the relevant error response coefficients are zero.

\begin{table*}
    \begin{tabular}{c|c|c|c|c|c|c|c|c|c}
        & $V$ & $M$ & $B$ & $X$ & $Y$ & $Z$ & ${BX}$ & ${BY}$ & ${BZ}$ \\ \hline
        $\text{D}_\inp$ & - & - & 1/5 & 1/15 & 1/15 & 0 & 1/10 & 1/10 & 0 \\
        $\text{A}_\inp$ & - & 2 & - & - & - & - & - & - & - \\
        $\text{L}_\detc$ & 0 & - & - & - & - & - & - & - & - \\
        $\tilde{\text{A}}_\detc$ & - & 1/3 & - & - & - & - & - & - & - \\
        $(\text{A}_\inp, \text{L}_\inp)$ & - & - & - & 3/4 & 3/4 & 1/4 & - & - & - \\
        $(\text{A}_\inp, \text{L}_\detc)$ & - & - & - & 0 & 0 & 0 & - & - & - \\
        $(\text{L}_\inp, \tilde{\text{A}}_\detc)$ & - & - & - & 1/3 & 1/3 & 0 & - & - & - \\
        $(\text{L}_\detc, \tilde{\text{A}}_\detc)$ & - & - & - & 0 & 0 & 0 & - & - & - \\
    \end{tabular}
    \caption{5DFT error response coefficients. 
    }\label{tab:5dft_coeffs}
\end{table*}

\begin{table*}
    \begin{tabular}{c|c|c|c|c|c|c|c|c|c}
        & $V$ & $M$ & $B$ & $X$ & $Y$ & $Z$ & ${BX}$ & ${BY}$ & ${BZ}$ \\ \hline
        $\text{D}_\inp$ & - & - & 1/6 & 0 & 0 & 5/24 & 0 & 0 & 1/6 \\
        $\text{A}_\inp$ & - & 3/2 & - & - & - & - & - & - & - \\
        $\text{L}_\detc$ & 0 & - & - & - & - & - & - & - & - \\
        $\tilde{\text{A}}_\detc$ & - & 0 & - & - & - & - & - & - & - \\
        $(\text{A}_\inp, \text{L}_\inp)$ & - & - & - & 1/4 & 1/4 & 5/4 & - & - & - \\
        $(\text{A}_\inp, \text{L}_\detc)$ & - & - & - & 0 & 0 & 1 & - & - & - \\
        $(\text{L}_\inp, \tilde{\text{A}}_\detc)$ & - & - & - & 0 & 0 & 5/4 & - & - & - \\
        $(\text{L}_\detc, \tilde{\text{A}}_\detc)$ & - & - & - & 0 & 0 & 5/4 & - & - & - \\
    \end{tabular}
    \caption{2-3DFT error response coefficients.} \label{tab:2-3dft_coeffs}
\end{table*}

\begin{table*}
    \begin{tabular}{c|c|c|c|c|c|c|c|c|c}
        & $V$ & $M$ & $B$ & $X$ & $Y$ & $Z$ & ${BX}$ & ${BY}$ & ${BZ}$ \\ \hline
        $\text{D}_\inp$ & - & - & 5/24 & 0 & 1/12 & 0 & 1/24 & 1/8 & 1/24 \\
        $\text{A}_\inp$ & - & 3/2 & - & - & - & - & - & - & - \\
        $\text{L}_\detc$ & 0 & - & - & - & - & - & - & - & - \\
        $\tilde{\text{A}}_\detc$ & - & 0 & - & - & - & - & - & - & - \\
        $(\text{A}_\inp, \text{L}_\inp)$ & - & - & - & 1/4 & 3/4 & 1/4 & - & - & - \\
        $(\text{A}_\inp, \text{L}_\detc)$ & - & - & - & 0 & 1/2 & 0 & - & - & - \\
        $(\text{L}_\inp, \tilde{\text{A}}_\detc)$ & - & - & - & 0 & 1/2 & 0 & - & - & - \\
        $(\text{L}_\detc, \tilde{\text{A}}_\detc)$ & - & - & - & 0 & 1/2 & 0 & - & - & - \\
    \end{tabular}
    \caption{S6 error response coefficients.} \label{tab:s6_coeffs}
\end{table*}

\begin{table*}
    \begin{tabular}{c|c|c|c|c|c|c|c|c|c}
        & $V$ & $M$ & $B$ & $X$ & $Y$ & $Z$ & ${BX}$ & ${BY}$ & ${BZ}$ \\ \hline
         $\text{D}_\inp$ & - & - & 1/8 & 0 & 0 & 3/8 & 0 & 0 & 1/8 \\
         $\text{A}_\inp$ & - & 3/2 & - & - & - & - & - & - & - \\
         $\text{L}_\detc$ & 0 & - & - & - & - & - & - & - & - \\
         $\tilde{\text{A}}_\detc$ & - & 0 & - & - & - & - & - & - & - \\
         $(\text{A}_\inp, \text{L}_\inp)$ & - & - & - & 1/4 & 1/4 & 3 & - & - & - \\
         $(\text{A}_\inp, \text{L}_\detc)$ & - & - & - & 0 & 0 & 11/4 & - & - & - \\
         $(\text{L}_\inp, \tilde{\text{A}}_\detc)$ & - & - & - & 0 & 0 & 3 & - & - & - \\
         $(\text{L}_\detc, \tilde{\text{A}}_\detc)$ & - & - & - & 0 & 0 & 3 & - & - & - \\
    \end{tabular}
    \caption{H8 error response coefficients.} \label{tab:h8_coeffs}
\end{table*}

\begin{table*}
    \begin{tabular}{c|c|c|c|c|c|c|c|c|c}
        & $V$ & $M$ & $B$ & $X$ & $Y$ & $Z$ & ${BX}$ & ${BY}$ & ${BZ}$ \\ \hline
        $\text{D}_\inp$ & - & - & 1/8 & 0 & 0 & 5/24 & 0 & 0 & 1/8 \\
        $\text{A}_\inp$ & - & 3/2 & - & - & - & - & - & - & - \\
        $\text{L}_\detc$ & 0 & - & - & - & - & - & - & - & - \\
        $\tilde{\text{A}}_\detc$ & - & 0 & - & - & - & - & - & - & - \\
        $(\text{A}_\inp, \text{L}_\inp)$ & - & - & - & 1/4 & 1/4 & 25/12 & - & - & - \\
        $(\text{A}_\inp, \text{L}_\detc)$ & - & - & - & 0 & 0 & 11/6 & - & - & - \\
        $(\text{L}_\inp, \tilde{\text{A}}_\detc)$ & - & - & - & 0 & 0 & 5/3 & - & - & - \\
        $(\text{L}_\detc, \tilde{\text{A}}_\detc)$ & - & - & - & 0 & 0 & 5/3 & - & - & - \\
    \end{tabular}
    \caption{2-H4 error response coefficients.} \label{tab:h4_coeffs}
\end{table*}

\begin{table*}
    \begin{tabular}{c|c|c|c|c|c|c|c|c|c|c|c}
        & $V$ & $M$ & $B$ & $X$ & $Y$ & $Z$ & ${BX}$ & ${BY}$ & ${BZ}$ & ${XX^*}$ & ${XY^*}$\\ \hline
        $\text{D}_\inp$ & - & - & 1/6 & 0 & 0 & 71/480 & 0 & 0 & 1/6 & 4/15 & 67/480\\
        $\text{A}_\inp$ & - & 3/2 & - & - & - & - & - & - & - & - & -\\
        $\text{L}_\detc$ & 0 & - & - & - & - & - & - & - & - & - & -\\
        $\tilde{\text{A}}_\detc$ & - & 0 & - & - & - & - & - & - & - & - & -\\
        $(\text{A}_\inp, \text{L}_\inp)$ & - & - & - & 1/4 & 1/4 & 417/160 & - & - & - & 17/5 & 281/160\\
        $(\text{A}_\inp, \text{L}_\detc)$ & - & - & - & 0 & 0 & 377/160 & - & - & - & 17/5 & 281/160\\
        $(\text{L}_\inp, \tilde{\text{A}}_\detc)$ & - & - & - & 0 & 0 & 71/40 & - & - & - & 16/5 & 67/40\\
        $(\text{L}_\detc, \tilde{\text{A}}_\detc)$ & - & - & - & 0 & 0 & 71/40 & - & - & - & 16/5 & 67/40\\
    \end{tabular}
    \caption{4-3DFT 4GHZ error response coefficients. Asterisks denote Pauli responses which are not symmetric over all subspace labels. See \Cref{app:error_responses} for details.}\label{tab:4x3dft_coeffs} 
\end{table*}

\begin{table*}
    \begin{tabular}{c|c|c|c|c|c|c|c|c|c|c|c}
        & $V$ & $M$ & $B$ & $X$ & $Y$ & $Z^*$ & ${BX}$ & ${BY}$ & ${BZ^*}$ & ${ZZ^*}$\\ \hline
        $\text{D}_\inp$ & - & - & 1/6 & 0 & 0 & 2/3 & 0 & 0 & 1/6 & 2/3\\
        $\text{A}_\inp$ & - & 3/2 & - & - & - & - & - & - & - & -\\
        $\text{L}_\detc$ & 0 & - & - & - & - & - & - & - & - & -\\
        $\tilde{\text{A}}_\detc$ & - & 0 & - & - & - & - & - & - & - & -\\
        $(\text{A}_\inp, \text{L}_\inp)$ & - & - & - & 1/4 & 1/4 & 10 & - & - & - & 17/2\\
        $(\text{A}_\inp, \text{L}_\detc)$ & - & - & - & 0 & 0 & 19/2 & - & - & - & 17/2\\
        $(\text{L}_\inp, \tilde{\text{A}}_\detc)$ & - & - & - & 0 & 0 & 8 & - & - & - & 8\\
        $(\text{L}_\detc, \tilde{\text{A}}_\detc)$ & - & - & - & 0 & 0 & 8 & - & - & - & 8\\
    \end{tabular}
    \caption{4-3DFT 4-line error response coefficients. Asterisks denote Pauli responses which are not symmetric over all subspace labels. See \Cref{app:error_responses} for details.}\label{tab:4line_coeffs}
\end{table*}

\end{document}